%% file: DDBF.tex
\documentclass{aa}
\usepackage[varg]{txfonts}
\usepackage[utf8]{inputenc}
\usepackage[english]{babel}
\usepackage{indentfirst}
\usepackage{graphicx}
\usepackage[]{natbib}
\bibpunct{(}{)}{;}{a}{}{,} 
\usepackage{lineno}

\newcommand{\rem}[1]{}
\newcommand{\acos}{\mathrm{acos\,}}
\begin{document}

\title{Direction dependent background fitting for the Fermi GBM data}

\author{Dorottya Sz\'ecsi\inst{1,2,3}
  \and Zsolt Bagoly\inst{1,4}
  \and J\'ozsef  K\'obori\inst{1}
  \and Istv\'an Horv\'ath\inst{4}
  \and Lajos G. Bal\'azs\inst{1,2}     
  }

\offprints{Z. Bagoly, \email{zsolt.bagoly@elte.hu}}

\institute{E\"otv\"os  University, Budapest, Hungary,
  \and MTA CSFK Konkoly Observatory, Budapest, Hungary,
  \and Argelander-Institute f\"ur Astronomie der Universit\"at Bonn, Germany
  \and Bolyai Military University, Budapest, Hungary
  }

\date{Received 8 January, 2013 / Accepted June 6, 2013}

\abstract
{
We present a method for determining the background of the gamma-ray bursts (GRBs) of the Fermi Gamma-ray Burst Monitor (GBM) using the satellite positional information and a physical model. Since the polynomial fitting method typically used for GRBs is generally only indicative of the background over relatively short timescales, this method is particularly useful in the cases of long GRBs or those that have autonomous repoint request (ARR) and a background with much variability on short timescales.
} 
{Modern space instruments, like Fermi, have some specific motion to survey the sky and catch gamma-ray bursts in the most effective way.
However, GBM bursts sometimes have highly varying backgrounds (with or without ARR),
and modelling them with a polynomial function of time is not
efficient --  one needs more complex, Fermi-specific methods. This article
presents a new 
direction dependent background fitting 
method and shows how it
can be used for filtering the lightcurves. } 
{First, we investigate how the celestial position of the satellite may have
influence on the background and define three underlying variables with
physical meaning: 
celestial distance of the burst and the detector's orientation, the contribution of the Sun and the contribution of the Earth. 
Then, we use multi-dimensional general least square fitting
and 
Akaike model selection criterion 
for the background fitting of the GBM lightcurves. 
Eight bursts are presented as examples, of which we computed the duration using background fitted cumulative lightcurves.} 
{We give a direction dependent background fitting (DDBF) method for separating
the motion effects from the real data and calculate the duration (T$_{90}$, T$_{50}$, and confidence intervals) of the 
nine example bursts, from which two resulted an ARR. We also summarize the 
features 
of our method and compare it qualitatively with the official GBM Catalogue.
} 
{
Our background filtering method uses a model based on the physical information of the satellite position. Therefore, it has many advantages compared to previous methods. It can fit long background intervals, remove all the features caused by the rocking behaviour of the satellite, and search for long emissions or not-triggered events. Furthermore, many parts of the fitting have now been automatised, and the method has been shown to work for both Sky Survey mode and ARR mode data. Future work will provide a burst catalogue with DDBF.
} 

\maketitle 

\section{Introduction}\label{sec:intro}
\input{intro}

\section{Difficulties with the Fermi background}\label{sec:problem}
\input{problem}

\section{Investigation of possible background sources}\label{sec:sources}
\input{sources}
   \subsection{Earth}\label{sec:earth}
   \input{earth}

   \subsection{Sun}\label{sec:sun}
   \input{sun}

   \subsection{Other gamma sources}\label{sec:other}
   \input{other}

\section{Background subtraction}\label{sec:subtrac}
\input{subtrac}

   \subsection{General Least Square}\label{sec:gls}
   \input{gls}

   \subsection{Multidimensional fit}\label{sec:multidim}
   \input{multidim}
   \subsection{Singular Value Decomposition}\label{sec:svd}
   \input{svd}

   \subsection{Model selection}\label{sec:modelsel}
   \input{modelsel}

   \subsection{Features of DDBF}\label{sec:features}
   \input{features}

\section{Results}\label{sec:result}
   \subsection{Direction dependent fit and T$_{90}$ for GRB 091030.613}
   \input{result}

   \subsection{Examples}\label{sec:example}
   \input{examples}

\section{Confidence intervals}\label{sec:error}
\input{error}

	\subsection{Comparison with the Fermi GBM Catalogue}\label{sec:catal}
	\input{catal}

\section{Summary and conclusion}
\input{discus}

\begin{acknowledgements} 
This study was supported by the Hungarian OTKA-77795 grant, by OTKA/NKTH
A08-77719 and A08-77815 grants (Z.B.). 
D. Sz. is grateful to P\'eter Veres
for the introduction to the field of the GRB data analysis and for all the useful
explanations and to \'Aron Szab\'o for highlighting the mathematical
basics of the statistical methods and for his patience and advices. 
We would like to express our gratitude to William Paciesas for his generous help 
with the ARR cases. Additionally, thanks to David Gruber for his comments concerning GRB~091024 and for all the discussions and inspirations. Special thanks to Gerard Fitzpatrick for the language editing. We also thank the anonymous Referee for the especially constructive remarks and suggestions.
D.Sz. has been supported by the "Lend\"ulet-2009" Young Researchers's
Program of the Hungarian Academy of Sciences and the OTKA-NIH Grant
MB0C 81013.
\end{acknowledgements}

\bibliographystyle{aa} 
\bibliography{Yourfile} 

\appendix
\section{Earth in the FoV}\label{sec:earthapp}
\input{appendix}

\end{document}

%% file: intro.tex
NASA's \textsl{Fermi} Gamma-ray Space Telescope has an 
orbit of altitude $\sim$\,$565\mathrm{~km}$ 
and period of $\sim$\,$96$ minutes. It carries two
main instruments on board. 
The \textsl{Large Area Telescope}'s (LAT) energy range
($20\mathrm{~MeV}-300\mathrm{~GeV}$) overlaps the energy range of the
\textsl{Gamma-ray Burst Monitor} (GBM, $8\mathrm{~keV}-40\mathrm{~MeV}$). 
GBM consists of two types of detectors: 12 Sodium Iodide (NaI) and 2 Bismuth
Germanate-Oxide (BGO) detectors \citep{Meegan}. 

The primary observation mode of \textsl{Fermi} is sky-survey mode. This enables the
LAT to monitor the sky systematically, whilst maintaining an uniform
exposure. In this mode, the entire sky is observed for $\sim$\,$30$ minutes per 2
orbits. If a sufficiently bright GRB is detected by GBM, an autonomous
repoint request (ARR) may be issued. This will cause the satellite to slew,
so that the burst's coordinates (calculated by the GBM) stay within the
field of view of the LAT for $\sim$\,$2$ hours \citep{Fitzpatrick}. 
However, this repositioning right after the trigger results in rapid and high background rate variations of the GBM 
lightcurves -- sometimes even 
during the burst, which is the most important time of the observation. 
Therefore, it is crucial
to have a filtering method, which is capable of correcting for the
background variations caused by the ARR.

To date, GBM has triggered on 1000 GRBs \citep{GCN}, \citep{ARR}. Only a small fraction 
($\sim$70 GRBs) 
resulted an ARR \citep{Paciesas}. 
The relatively low rate of ARR's is due to the  GBM trigger that has to meet certain criteria (such as high peak flux) before an ARR occurs.
When we started to analyse GRBs
detected by GBM, we found that several non-ARR bursts have 
a background
variation of the same order of magnitude as the burst itself. 
As we will show, one can find connection between these background rates and the actual position and orientation of the satellite. 
Therefore it is necessary to use 
the directional information to filter the background not only for ARR but also for many non-ARR cases.

Here, we present the effect of the slew and how it is represented in the measured data of the GBM.
We summarize why the usual background subtraction methods are
inefficient in most cases, especially for the long bursts, as seen in Sec.~\ref{sec:problem}. 
Then, we introduce
variables based on the position of the satellite related to the Earth and
the Sun (Sec.~\ref{sec:sources}) and use them with the time variable to fit a
general multi-dimensional linear function to the background
(Sec.~\ref{sec:subtrac}). Our method is called
\textsl{direction dependent background fitting} (DDBF). 

We also present examples where we compute the duration (T$_{90}$ and T$_{50}$)
from our background-filtered lightcurves and show that 
the DDBF method can be used for both the Sky Survey and ARR observations 
(Sec.~\ref{sec:result}).
Confidence levels and a comparison to the GBM catalogue are given in Sec.~\ref{sec:error}.

%% file: problem.tex
\subsection{Lightcurves with unpredictably varying background}

\begin{figure}[h!]
  \resizebox{\hsize}{!}{\includegraphics[angle=270]{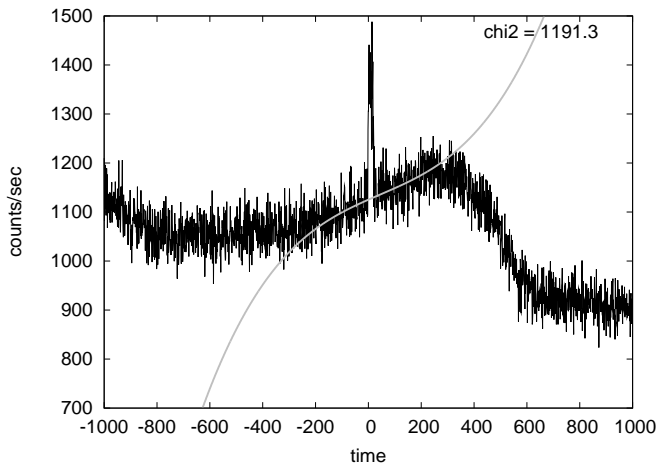}}
  \caption{Lightcurve of the \textsl{Fermi} burst 091030.613 measured by the 3rd
  GBM-detector without any background filtering with 1-second bins. The grey
  line is a fitted polynomial function of time of order 3
  for the ranges of [-200:-20] and
  [38:200] seconds, 
  which does not seem to be a correct model for this whole
  background. Reduced chi-square statistics are given in the top right
  corner \citep{Acta}.}
  \label{fig:lc.eps}
\end{figure}

The lightcurve for GBM trigger 091030.613 is shown in Fig.~\ref{fig:lc.eps} in the energy range $\sim$\,11--980~keV.
This burst did not result in an ARR \citep{GCN}.
We decided to use the sum of the channels except for the highest and lowest, where the
detector's efficiency drops, so the signal is statistically stronger. 
Since we are only interested in the duration information of the bursts, we
use the high time resolution data (CTIME, see Sec.~\ref{sec:sources} for the detailed
description)
and sum of the channels.
We note that, however, the analysis can be done using
either different channels or the 
high 
spectral resolution data files (CSPEC), so spectral information can be obtained \citep[see][]{Munchen}.

In Fig.~\ref{fig:lc.eps},
the burst is clearly visible above the background, but the background is
varying so rapidly and to such an extent that one can question the
usefulness of fitting and subtracting a simple polynomial function of order
3
(grey line in Fig.~\ref{fig:lc.eps}). 
This situation is typical in the case of \textsl{Fermi}, 
as can be seen
in the examples in 
Sec.~\ref{sec:example}. 
Especially when a long burst occurs, the background rate can change too quickly for analyses without some knowledge about the satellite position and the
gamma sources on the sky. 
In the following, we are investigating for possible background sources. 
We will see that one can find a correspondence between the gamma background and the celestial orientation of the satellite. Furthermore, both the Sun and the Earth limb have a contribution, given that they move in and out of the field of view because of the rocking motion of the satellite. 
Based on these physical conditions, we are constructing a background model and a fitting algorithm, both of
which give us a more effective method for filtering the motion effects. 
Since the method is based on the actual directional information of the satellite, it is possible to analyse 
bursts for which an ARR was issued.

\subsection{Previous methods}

In the BATSE era, it was sufficient to fit a low-order polynomial in the
function of time for most cases. 
It was because BATSE has had a 
fixed 
orientation and has not been able to change it during a burst. 
As a result, sources moving in and out of the field of view could not play an important role on a shorter timescale, and all the backgrounds could be subtracted by fitting a time-dependent low-order (up to 3) polynomial \citep{Koshut,Sakamoto,Varga}. 

In the \textsl{Fermi} era, this situation has however fundamentally changed. To
present this on our example above, we fitted a simple 3rd-order
polynomial function of time shown with a grey line in Fig.~\ref{fig:lc.eps}. 
The fitting was done by using only a selected short time interval around the burst, which is a common method of the BATSE era. 
This fit may be sufficient around the burst prompt emission, but is sufficient only there. It is clear that the background cannot be well modelled with this simple function over a long timescale. Moreover, an incidental longtime emission 
would be overlooked. 

Fitting higher order polynomials of time could be suggested. We rule out this solution because of two reasons. First, these fittings show polynomial instabilities in the burst interval, as we have seen it in our early experiments; namely, we got high order, low amplitude oscillations of these fittings during the interval of the burst.
Second, we wanted to take into consideration that the main cause of the complicated background is well known (namely the rocking motion of the satellite). Indeed, we use physically defined underlying variables, as we will show in Sec.~\ref{sec:subtrac}, and with them, we fit higher order multidimensional functions. As a conclusion, time-dependent polynomial fittings may have been 
sufficient for the BATSE data but \textsl{Fermi}-data
cannot be analysed that way due to the rapid motion of the satellite: we
need a \textsl{Fermi} specific method. 

Such a method was presented by \citet{Fitzpatrick}. They
estimated the background successfully with the rates from adjacent days, when
the satellite was at the same geographical coordinates. This solution is only applicable when the satellite is in Sky
Survey Mode and cannot be used if an ARR occurred. If an
ARR is accepted, this technique cannot be employed.

%% file: sources.tex
\subsection{Orientation of NaI detectors}

As we mentioned above, \textsl{Fermi} uses a complex algorithm to optimize the
observation of the Gamma-Ray Sky. In Sky Survey Mode, the satellite rocks
around the zenith within $\pm 50^{\circ}$, and the pointing alternates between
the northern and southern hemispheres each orbit \citep{Meegan, Fitzpatrick}.

\begin{figure}[h!]
  \resizebox{\hsize}{!}{\includegraphics{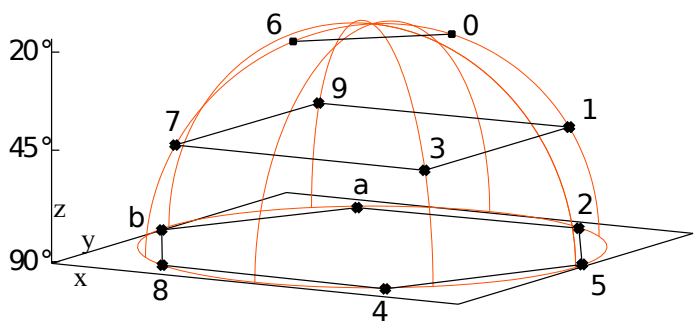}}
  \caption{\small{
  Setup of the 12 NaI detectors of GBM given in the Spacecraft 
  Coordinates \citep[see][]{Meegan}. The zenith angle of the
  detectors in degrees is marked. This design is built in order to cover the
  whole visible part of the sky with the GBM. (The figure is based on Table 1. of
  \citet{Meegan}.   
  Notations 'a' and ’b' mean the 10th and 11th NaI detectors, respectively.)
  }}
  \label{fig:detekt}
\end{figure}

The set-up of the instruments on-board is well known from the literature
\citep{Meegan}. 
The 12 NaI detectors are placed in such a way that the entire
unocculted sky is observable with them at the same time,
as seen in Fig.~\ref{fig:detekt}. \textsl{Fermi} has a proper coordinate system, whose $Z$
axis is given by the LAT main axis. 
From now on, we only analyse
the data of the NaI detectors; the BGO detectors will be considered in a future work.

The \textsl{Fermi} data set is available from the web for the GBM's 12 NaI
detectors\footnote{~The High Energy Astrophysics Science Archive Research
Center (HEASARC): \textsl{legacy.gsfc.nasa.gov}}.
The positional
information of the spacecraft is contained in the LAT data (called
Spacecraft Data\footnote{~LAT Photon, Event, and Spacecraft Data Query:
\textsl{http://\textsl{Fermi}.gsfc.nasa.gov/cgi-bin/ssc/LAT/LATDataQuery.cgi}}). 
The GBM data, which we use in our analysis (called CTIME), are
available at 8 energy channels with $0.064$-second and $0.256$-second
resolution (for triggered and non-triggered mode, respectively). 
The position
data is available in $30$-second resolution. 

\begin{figure}[h!]
  \resizebox{\hsize}{!}{\includegraphics[angle=0]{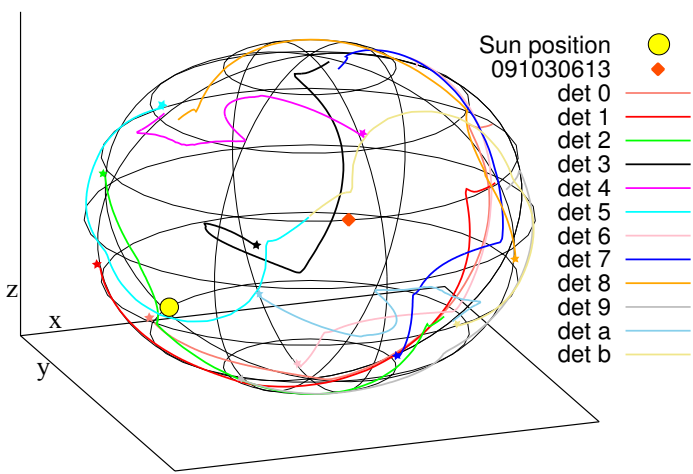}}
  \caption{\small{ 
  Orientation of the 12 NaI detectors on the
  sky (in the second equatorial system), during the pre- and post-1000 seconds
  around the burst 091030.613. To show the direction with time, we marked the 
  starting points of every line with a small star. The Sun's position is marked with big 
  sphere. The burst position is marked with diamond. 
}}
  \label{fig:091030613all}
\end{figure}

The $30$-second Spacecraft Data were evenly proportioned to $0.256$-second and
$0.064$-second bins using linear interpolation, to correspond to the
CTIME data of non-triggered and triggered mode, respectively.
We created a 3D-plot from this data using the known orientation of the 12 NaI
detectors given in the Spacecraft coordinate 
system. Fig.~\ref{fig:091030613all} shows the 
detectors' orientation (path) on
the sky during the pre- and post-1000 seconds around the trigger of
091030.613 (lightcurve was shown in Fig. \ref{fig:lc.eps}). 

\begin{figure}[h!]
  \resizebox{\hsize}{!}{\includegraphics[angle=270]{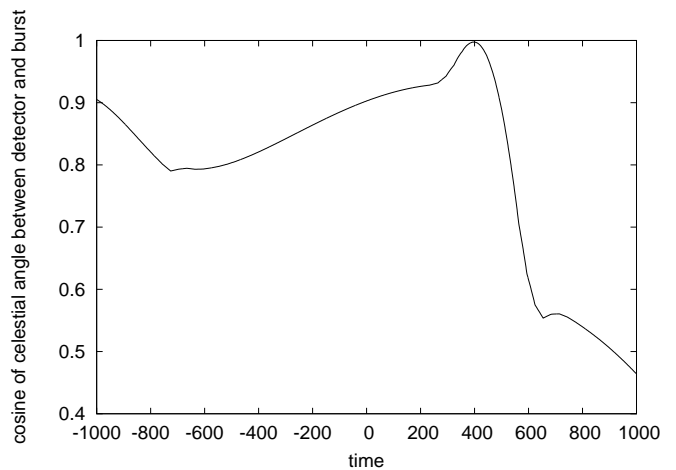}}
  \caption{The celestial distance of the 3rd GBM detector and the \textsl{Fermi}-burst
  091030.613 as a function of time. It is worth comparing this figure to
  Fig.\,\ref{fig:lc.eps}.}
  \label{fig:angle}
\end{figure}

The catalogue location for the GRB is shown with a diamond ($\alpha=260.72^{\circ}$, $\delta=22.67^{\circ}$, see \citet{catalogue}). 
Since we
wanted to know the position of the detectors on the sky, we needed to transform
the proper coordinate system of the \textsl{Fermi} shown in Fig.~\ref{fig:detekt} to
the general (second) equatorial system, since the burst's position was given in
the latter.
In addition, we plot the celestial angle between the 3rd detector (black line
in 
Fig.~\ref{fig:091030613all}) and the burst 091030.613 
(marked with a diamond 
in 
Fig.~\ref{fig:091030613all})
against time 
in Fig.~\ref{fig:angle}. 

At this point, we have to mention the effect of the NaI detectors' characteristics. Figure\,12 from \citet{Meegan} shows the angular dependence of a NaI detector 
effective area: The angular response 
for the flat crystal is approximately cosine. For this reason, we define our first underlying variable as the \textsl{cosine} of the celestial angle between the detector and the burst (as it is shown in Fig.~\ref{fig:angle}). We will find further underlying variables in Secs.~\ref{sec:earth} and \ref{sec:sun}. 

However, the NaI characteristics 
are 
also energy dependent: The dependence of the transmissivity on the angle of incidence is more important at higher than at lower energies. Furthermore, a detector has two small sensitivity peaks around -150 and 150 degrees, which means that they can detect photons coming under the plane of the crystal. We consider these features by allowing higher orders when performing the fits seen in Sec.~\ref{sec:subtrac}.

If we compare Fig.~\ref{fig:angle} to Fig.\,\ref{fig:lc.eps}, it is clear that the unpredictable variation in the background is
connected to the orientation of the detector in question. We can also
examine other bursts (see Sec.~\ref{sec:example}. for more examples).
However, we cannot state a clear relation between the angle and the
lightcurve.

%% file: earth.tex
The satellite's $Z$ axis 
(the direction of the LAT) 
is pointing to
the opposite direction of the Earth, when it is possible. 
Due to the rocking behavior, 
GBM detectors' orientation are, however, towards the
Earth-limb from time to time. 

The Earth-limb is notable from the board of \textsl{Fermi}. At an
altitude of $\sim$\,$565\mathrm{~km}$, it corresponds to an aperture of $\sim$\,$134^{\circ}$ when fully in
the FoV.
Therefore, we have to 
consider
the effect of the
Earth-limb when analysing the data of the GBM detectors. 
There are terrestrial
gamma-ray flashes (brief bursts of gamma-radiation that are thought to be associated with lightning in the upper atmosphere); furthermore,
gamma-rays of the GRB's scatter on the atmosphere. 
The main contributor in our background model is the latter. Terrestrial
gamma-ray flashes have a duration of only tens of
milliseconds \citep{Briggs} and are too
short to have a significant effect.

\begin{figure}[h!]
  \resizebox{\hsize}{!}{\includegraphics[angle=270]{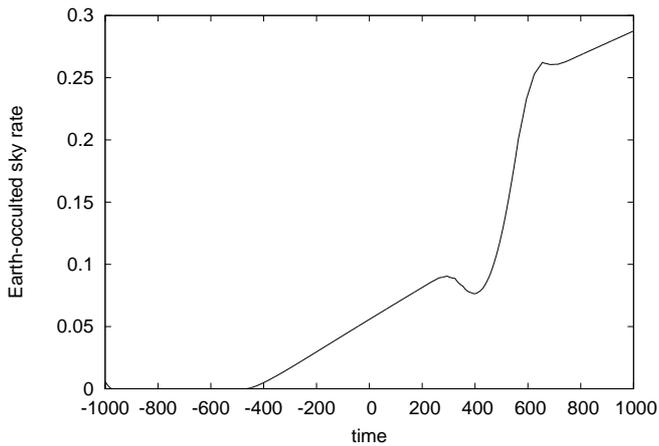}}
  \caption{The Earth-occulted sky rate for the 3rd GBM detector
  as a function of time during the GRB 091030.613. (The Earth-occulted sky rate is zero, if the 
  Earth-limb is out of the FoV.)}
  \label{fig:limb}
\end{figure}

We presume therefore that the detected background also depends on how much sky the
Earth-limb shields from the detector's FoV. To measure
this, we define the \textsl{Earth-occulted sky rate} as 
the rate of the Earth-covered sky correlated to the size of the FoV. 
As \textsl{Fermi} has a proper motion, the Earth-occulted sky rate is a function of time, satellite position, and orientation. 
Based on spherical geometrical computations given in Appendix
\ref{sec:earthapp}, we can get the Earth-occulted sky rate as
a function of the aperture of the Earth-limb and the maximum altitude of the
Earth seen from the \textsl{Fermi}.  The Earth-occulted sky rate is plotted in Fig.~\ref{fig:limb} as a function of time.

We can see the same effect like above: There is some noticeable connection
between the lightcurve in Fig.~\ref{fig:lc.eps} and the Earth-occulted sky rate
in Fig.~\ref{fig:limb}.

%% file: sun.tex
One of the main contributors of the gamma-ray sky is the Sun. Flares and other eruptive solar events produce gamma rays 
in addition to those 
created by cosmic rays striking the Sun’s gas.
If we are
looking for a 
complete 
model of the background, we need to consider the
contribution of the Sun as well.

The Sun's position is known from ephemeris tables for the day of the burst.
We do not need more precise data than one day, because the time interval
around the burst is only $2000\mathrm{~s}$ in our 
analysis, and the position of the Sun
does not change significantly during that time.

\begin{figure}[h!]
  \resizebox{\hsize}{!}{\includegraphics[angle=270]{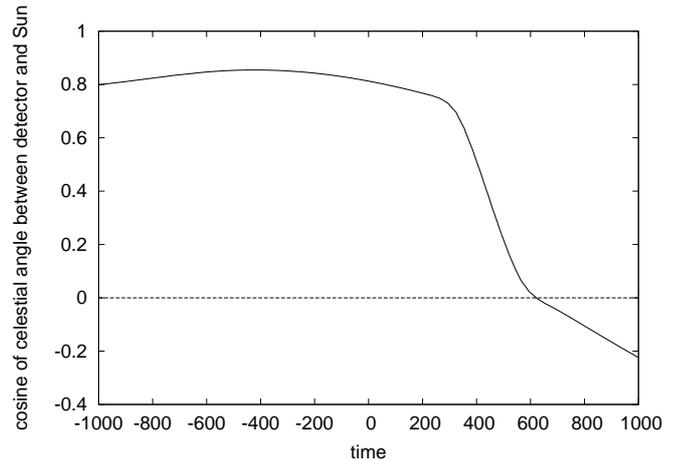}}
  \caption{The celestial distance of the 3rd GBM detector and the Sun as a
  function of time during the GRB 091030.613. 
The dashed line shows the 0 level (under this the Sun and the detector close in an angle larger than 90$^{\circ}$). 
It is worth comparing this figure to Fig.~\ref{fig:lc.eps}. }
  \label{fig:sun}
\end{figure}

We compute the celestial distance (i.e. the angle) between the detector's
direction and the Sun's position. This parameter is shown in
Fig.~\ref{fig:sun}. 
The Sun's position is also shown in the
Fig.~\ref{fig:091030613all} with a yellow circle.

Comparing Fig.~\ref{fig:lc.eps} to Fig.~\ref{fig:sun}, one can see a
connection between them. 	
It is interesting to take notice of the fact that when the Sun's angle is larger then 90$^{\circ}$ (the cosine is lower that 0) around 600\,sec, the background rate in Fig.\,\ref{fig:lc.eps} drops. 
It shows a further correspondence 
of the background and the direction of the satellite towards to the Sun.

%% file: other.tex
It is known today that the gamma-ray sky is not dark \citep{gammasky}.
Apart from the gamma-ray bursts, the terrestrial 
flashes,
and the Sun's activity, there are also additional gamma-ray sources. 
Some examples include the
gamma-rays produced when cosmic rays collide with gas in the Milky Way and
the contribution from individual galactic sources, such as pulsars and other
transient sources. 
As an extragalactic counterpart, we see collective radiation from galaxies that we are not detecting directly and gamma-rays from jets of active galaxies.
	
All this gamma-background has to be paid respect to. 
Rather than consider each contributing source
individually, we introduce them into our model by allowing higher order	
terms when constructing 
the basis function 
of the general least square problem in Secs.~\ref{sec:gls}~and~\ref{sec:multidim}. Furthermore, we use the method of singular value decomposition and Akaike model selection described in Secs.~\ref{sec:svd}~and~\ref{sec:modelsel} for choosing the contributing ones, since the net effect of all these sources is hard to compute at every second.

%% file: subtrac.tex
In Sec.~\ref{sec:sources}, we have found three variables, 
which contribute to the variation 
in the background (see Figs.~\ref{fig:angle},
\ref{fig:limb} and \ref{fig:sun}). They may help extend the polynomials of
time that are only usable in some short intervals around the bursts. These
three variables contain physical information of the background, because they
are suggested by the actual position and orientation of the satellite. 

However, we cannot
quantify the contribution from the various sources at any given time. 
As we know that they have an influence on the
background, we can fit a theoretical function of these physical underlying
variables. Therefore, we fit and subtract the background using the three defined
variables 
(burst position, Sun, and Earth) 
and the time variable, on a higher degree.

At this point, the following question may arise: why is the burst location needed?
If a curve contains no burst for sure,
there is no sense of using the burst position
as an underlying variable. In that case, we would probably need to use
only the Sun and the Earth (maybe implement the position of some other
gamma-sources as well). 

The reason why we use the burst position when there is a burst in the
data is that the burst itself is a gamma source. Of course, it does not
produce gamma photons at a constant level, but transiently. It is
possible, nevertheless, that a not yet identified long emission would be
enhanced (or weakened) because the satellite moved toward (or away of) the
burst. To analyse (or sometimes even detect) emission coming from
the astrophysical source outside of the main burst interval, it is needed to identify the
fluctuations of the background rate caused by the change in the distance between the
detector and the burst.

Next, we summarize the method of general least square for multidimensional
fits, the algorithm of singular value decomposition, its numerical
solution, and the Akaike model selection criterion for choosing the best model.
Since we use underlying variables, which are calculated based on the actual
direction and orientation of the satellite, we call this method
direction dependent background fitting (DDBF).

%% file: gls.tex
For simplifying the explanation, we will use the following notation:
\begin{flushleft}
$y_i$=counts per bin;\\
$x^{(1)}_i$=celestial distance between burst and detector orientation (Fig.~\ref{fig:angle});\\
$x^{(2)}_i$=celestial distance between Sun and detector orientation (Fig.~\ref{fig:sun});\\
$x^{(3)}_i$=rate of the Earth-uncovered sky (Fig.~\ref{fig:limb});\\
$x^{(4)}_i$=time.
\end{flushleft}
We have
a set of datapoints $(\mathbf{x}_i,y_i)$, where the
components of $\mathbf{x}_i$ are $\mathbf{x}_i=(x^{(1)}_i, x^{(2)}_i,
x^{(3)}_i, x^{(4)}_i)$, while $i=1\ldots N$.

We use the general least square method \citep{NR} for a multidimensional fit
(since $\mathbf{x}_i$ has more than one component). The theoretical value of
$y(\mathbf{x}_i)$ can be expressed with functions of $\mathbf{x}_i$,
known as the \textsl{basis functions} $X_k(\mathbf{x}_i)$:
\begin{equation}
\displaystyle y(\mathbf{x}_i)=\Sigma_{k=1}^{M}a_k X_k(\mathbf{x}_i),
\label{eq:yXx}
\end{equation}
where the weights $a_k$ are the model parameters that we need to estimate from
the data ($k=1...M$).  Note that the basis functions $X_k(\mathbf{x}_i)$ can
be nonlinear functions of $\mathbf{x}_i$ (this is why the method is called
generalized), but the model depends only linearly on its parameters $a_k$.

The maximum likelihood estimate of the model parameters $a_k$ is obtained by
minimizing the quantity
\begin{equation}
\chi^2 = \Sigma_{i=1}^N \left( \frac{y_i- \Sigma_{k=1}^{M}a_k X_k(\mathbf{x}_i) }{\sigma_i} \right)^2,
\label{eq:chi2}
\end{equation}
which is known as the chi-square statistics or chi-square function. 

One can write the chi-square function in a matrix equation form as well.
For that, it is useful for defining the \textsl{design matrix} $\mathbf{A}$
($N\times M$, $N\geq M$) of the fitting problem. Since the measured values of
the dependent variable do not enter the design matrix, we may also define the
vector $\mathbf{b}$. The components of $\mathbf{A}$ and $\mathbf b$ are defined
to be the following:
\begin{equation}
\displaystyle
A_{ij} = \frac{X_j(\mathbf{x}_i)}{\sigma_i}, \hspace{20pt} b_{i} = \frac{y_i}{\sigma_i}.
\label{eq:design}
\end{equation}
From now, we set $\sigma_i=$constant.

In terms of the design matrix $\mathbf{A}$ and the vector $\mathbf{b}$, the
chi-square function can be written as
\begin{equation}
\chi^2 = (\mathbf{A}\cdot \mathbf{a} - \mathbf{b})^2,
\label{eq:chi2matrix}
\end{equation}
and we need an $\mathbf{a}$ that minimizes this function, so the derivatives
of $\chi^2$ with respect of the components of $[\mathbf{a}]_k=a_k$ are zeros.
That leads us to the equation for $\mathbf{a}$:
\begin{equation}
\mathbf{a} = (\mathbf{A}^T \mathbf{A})^{-1} \mathbf{A}^T \mathbf{b},
\label{eq:solution}
\end{equation}
where $\mathbf{A}^T$ means the transpose of $\mathbf{A}$, and the expression
$(\mathbf{A}^T \mathbf{A})^{-1} \mathbf{A}^T$ are called \textsl{generalized
inverse} or \textsl{pseudoinverse} of $\mathbf{A}$. The best technique of
computing pseudoinverse is based on \textsl{singular value decomposition} (SVD),
which we describe in Sec.~\ref{sec:svd}. 
We first specify 
the general method written above for the case of the \textsl{Fermi} GBM lightcurves in the following section.

%% file: multidim.tex
Equation~(\ref{eq:yXx}) describes a hypersurface, and it is a generalization of
fitting a straight line to the data. 
Very simple 
backgrounds may be fitted well
with first degree hypersurface (hyperplane) of the four variables described as
$\mathbf{x}_i=(x^{(1)}_i,x^{(2)}_i,x^{(3)}_i,x^{(4)}_i)$:
\begin{equation}
y(\mathbf{x}_i) = a_1\cdot x^{(1)}_i + a_2\cdot x^{(2)}_i + a_3\cdot x^{(3)}_i + a_4\cdot x^{(4)}_i, \label{eq:1st}
\end{equation}
where the basis functions are $X_l(\mathbf{x}_i)=x^{(l)}_i$, respectively,
and the design matrix simply consists of the components of $\mathbf{x}_i$ with $A_{ij}=x^{(j)}_i$.

\begin{figure}[h!]
  \resizebox{\hsize}{!}{\includegraphics{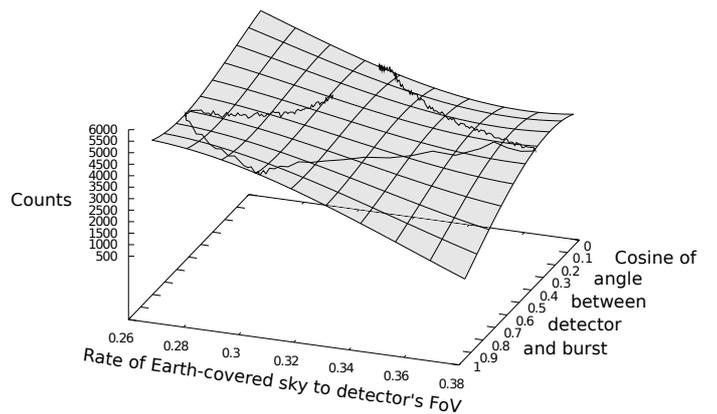}}
  \caption{\small{
  The 2-dimensional hypersurface of a 3rd degree fitting to a \textsl{Fermi} lightcurve
  is shown. The fitted variables ($x^{(1)}_i$, $x^{(3)}_i$) are along the
  horizontal axes, while vertical axis represents the counts of the lightcurve
  $y_i$ (shown by the black curve on the fitted grey plane). 
  }}
  \label{fig:surf}
\end{figure}

For the most complicated 
\textsl{Fermi} backgrounds, higher degree
of the variables are needed, however. One can illustrate the lightcurve data $y_i$ and
the fitted hypersurface $y(\mathbf{x}_i)$ using the two variables $x^{(1)}_i$ and
$x^{(3)}_i$, which are both of 3rd degree on a 3D plot, as seen in Fig.~\ref{fig:surf}. The
design matrix of this problem is 
\begin{equation}
\mathbf{A}=
\left(\begin{array}{cccccc} 
x^{(1)}_1 & (x^{(1)}_1)^2 & x^{(3)}_1 & (x^{(3)}_1)^2 & x^{(1)}_1\cdot x^{(3)}_1 & 1 \\ 
x^{(1)}_2 & (x^{(1)}_2)^2 & x^{(3)}_2 & (x^{(3)}_2)^2 & x^{(1)}_2 \cdot x^{(3)}_2 & 1 \\ 
... & & & & & \\
x^{(1)}_N & (x^{(1)}_N)^2 & x^{(3)}_N & (x^{(3)}_N)^2 & x^{(1)}_N \cdot x^{(3)}_N & 1 
\end{array}\right).
\label{eq:design2}
\end{equation}

Since we would like to have a method for all the cases of \textsl{Fermi}-bursts
(whether it is simple, complicated, non-ARR, or ARR), 
we define our model to be comprehensive. Let
us have $y(\mathbf{x}_i)$ as the function of
$\mathbf{x}_i=(x^{(1)}_i,x^{(2)}_i,x^{(3)}_i,x^{(4)}_i)$ of order 3, so the
basis functions $X_k(\mathbf{x}_i)$ (and columns of the design matrix) consist
of every possible products of the components $x^{(l)}_i$ up to order 3. That
means that we have $M=k_{max}=35$ basis functions and $a_1$, $a_2$...$a_{35}$ as
free parameters. We are sure that we do not need so many free parameters to
describe a
simple background, and although a complicated or ARR background may require more free parameters, 35 is too much in every practical case.
Therefore, we decrease the number of free parameters using SVD in the next section.

%% file: svd.tex
In Sec.~\ref{sec:gls}, we showed that the least square problem can be solved
by computing the pseudoinverse of the design matrix $\mathbf{A}$.  For this
purpose, we used Singular Value Decomposition (SVD), since SVD is
robust and very stable numerically \citep{Octave} \citep{NR}.

The SVD takes an $N \times M$ matrix $\mathbf{A}$ and factors it into
$\mathbf{A} = \mathbf{USV}^T $.  In this expression, $\mathbf{U}$ and
$\mathbf{V}$ are $N\times N$ and $M\times M$ orthogonal matrices, respectively,
and $\mathbf{S}$ is an $N \times M$ diagonal matrix. The columns of $\mathbf
U$ and  $\mathbf V$ are the eigenvectors of $\mathbf{AA}^T$ and
$\mathbf{A}^T\mathbf{A}$, respectively.  Furthermore, $\mathbf{S}$ contains
the square roots of the eigenvalues of $\mathbf{AA}^T$ and
$\mathbf{A}^T\mathbf{A}$ (both have the same eigenvalues, but different
eigenvectors). These eigenvalues (diagonal elements in $\mathbf{S}$) are
called the \textsl{singular values}, $s_i$. 

In overdetermined cases ($N\geq M$), the last $N-M$ singular values, however,
are zeros, so we may consider only $\mathbf{U}$ as an $N\times M$
matrix, $\mathbf{V}$ as an $M\times M$ matrix, and $\mathbf{S}$ as
$M\times M$ (it is called \textsl{economic} SVD).

If $\mathbf{U}$ and $\mathbf{V}$ enter the SVD decomposition of $\mathbf{A}$
as described above, one can show easily (using the orthogonalithy of
$\mathbf{U}$ and $\mathbf{V}$) that the pseudoinverse of $\mathbf{A}$ can be
obtained as \begin{equation} pinv(\mathbf{A}) = (\mathbf{A}^T \mathbf{A})^{-1}
	\mathbf{A}^T = \mathbf{VS}^{-1}\mathbf{U}^T.  \label{eq:pseudo}
\end{equation}

SVD is implemented in several numerical software. In our work, we used
\textsc{Octave}'s SVD function\footnote{~GNU
\textsc{Octave}: http://www.gnu.org/software/octave/}, known as the \texttt{svd}, and
pseudoinversion function, known as the \texttt{pinv} \citep{Octave}. 

Computing the pseudoinverse, we need the reciprocal of the singular values in
the diagonals of $\mathbf{S}^{-1}$, and there is a problem with this.
The size of a singular value tells you exactly how much influence the
corresponding rows and columns of $\mathbf{U}$ and $\mathbf{V}$ have over the
original matrix $\mathbf{A}$.  We can find the exact value of $\mathbf{A}$ by
multiplying $\mathbf{USV}^T$. If we, however, remove (for example) the last columns
of $\mathbf{U}$ and $\mathbf{V}$ and the final singular value, we are removing
the least important data. If we then multiplied these simpler matrices, we
would only get an approximation to $\mathbf{A}$ but one which still contains
all but the most insignificant information. This means that SVD allows us to
identify linear combinations of variables that do not contribute
much to reducing the chi-square function of our data set.

The singular values are usually arranged in the order of size with the first
being the largest and most significant. The corresponding columns of
$\mathbf{U}$ and $\mathbf{V}$ are therefore also arranged in importance.  If a
singular value is tiny, very little of the corresponding rows and columns get
added into the matrix $\mathbf{A}$ when it is reconstructed by SVD. If we
compute the pseudoinverse of $\mathbf{A}$, the reciprocals of the tiny and not
important singular values will be unreasonably huge and enhance the numerical
roundoff errors as well. 

This problem can be solved defining a \textsl{limit} value, below which
reciprocals of singular values are set to zero. It means that the resulted
matrix is an approximation of the real pseudoinverse, but we only omit
information of the less interest.

With equation (\ref{eq:yXx}), we can define models of any number of variables
and of arbitrary degree. In our case, we define models with four underlying
variables of degree 3. Therefore, we have $M=35$ free fitting parameters, as
described above in Sec.~\ref{sec:multidim}. We do not know how many and
which ones of these parameters have real importance in the variation in the
background, but SVD can give us the answer trivially: Pseudoinverse should be
done by omitting the singular values which do not contribute so much. 

The only question that remains is where
this limit should be when singular values are not so important.
We find an answer to that question in Sec.~\ref{sec:modelsel} using model selection criteria.

%% file: modelsel.tex
Model selection is usually based on some information criterion. 
We use the Akaike information criterion (AIC) method 
to distinguish between different models to the data \citep{AIC}. 
However, we note here that AIC has to be used with caution, especially in the most 
complicated cases of backgrounds (see examples in Sec.~\ref{sec:example}).
	
We first assume that we have $M$ models so that the
$k$th model has $k$ free parameters ($k=1...M$). When the deviations of the
observed values from the model are normally and independently distributed,
every model has a value AIC$_k$ so that
\begin{equation}
\mathrm{AIC}_k=N\cdot \log{\frac{RSS_k}{N}} + 2\cdot k,
\label{eq:AIC}
\end{equation}
where $RSS_k$ is the residual sum of squares from the estimated model
($RSS=\Sigma_{i=1}^N \left( y_i- y(\mathbf{x}_i,k) \right)^2$), $N$ is the
sample size, and $k$ is the number of free parameters to be estimated.  
The first term of equation (\ref{eq:AIC}) measures the \textsl{goodness of fit} (discrepancy between observed values and the values expected under the model in question), the second term penalizes the free parameters. 
Given any two estimated models, the model
with the lower value of AIC$_k$ is the one to be preferred. Given many models, the
one with lowest AIC$_k$ will be the best choice: It has as many free
parameters as needed but not more. 
Note that we do not use AIC for deciding how good the fit is but only for choosing one model over the another. The goodness of fit is given by the chi-square statistics defined by equation (\ref{eq:chi2}).

So far, we defined a complex model with 35 free parameters and,
therefore, the design matrix $\mathbf{A}$ has 35 singular values (see Sec.~\ref{sec:multidim}). However, we know that we can omit some of the tiny
singular values when computing the pseudoinverse of $\mathbf{A}$ -- the ones,
which are not necessary to the best fit of the gamma background. Thus, we take a
loop over the pseudoinverse operation and decrease the omitted number (that
is, increase the used number) of singular values in every step. Furthermore,
we also compute the AIC$_k$ in every step with $k$ being the number of
singular values not omitted. In that way, the number of singular values, which
minimize the AIC$_k$ as a function of $k$ will be the best choice when
calculating the pseudoinverse, so we get the most useful estimation of the
model parameters $\mathbf{a}$ (let us remember that singular values are sorted
in decreasing order, so the last and not important ones will be penalized by
the second term of AIC).

\begin{figure}[h!]
  \resizebox{\hsize}{!}{\includegraphics[angle=270]{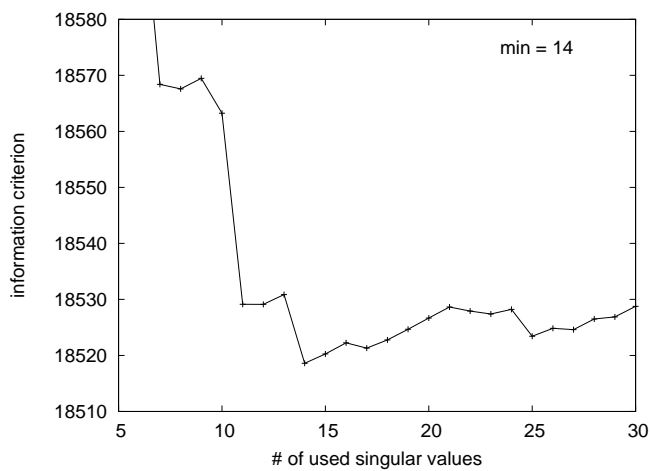}}
  \caption{\small{The Akaike Information Criterion for model selection. Model
  with 14 singular values is selected. (First and last five singular values
  are usually too high, so we do not show them.)
  }}
  \label{fig:ic.eps}
\end{figure}

At this point, we return to the \textsl{Fermi}'s GRB 091030.613 presented in Sec.~\ref{sec:problem} and \ref{sec:sources} and follow the method of general
least square, as described above. We compute AIC$_k$ for every $k=1...35$. This
function is shown in Fig.~\ref{fig:ic.eps}. 

Based on the Akaike information criterion, the model with 14 singular values is the
best choice. We present the result of the fitting with this model in
Sec.~\ref{sec:result}.

%% file: features.tex
	
One cornerstone of the fitting algorithm DDBF described above is the definition of
the boundaries that decide the interval of the burst and the intervals of the
background. In this work, we follow the common method of using user-selected
time intervals \citep{catalogue}. 

Unlike in \citet{catalogue}, usage of the position data gives us the possibility of fitting the whole background of the CTIME file instead of selecting two or three small fractions around the burst. This notable feature has two important consequences. 

First, the user has to select only the two boundaries before and after the burst; the other boundaries of the background intervals are inherently at the beginning and at end of the CTIME datafile. This reduces the error factor put into the DDBF method by the user compared to the method of \citet{catalogue}.

Second, one can easily detect a possible long emission coming from the astrophysical source. Since this emission has nothing to do with the direction and orientation of the satellite, the signal consequently has to be present in the lightcurve after the background filtering. (The opposite is also true: a signal after the burst could be considered a long emission when the user defines two short background intervals, although it was caused by the motion of the satellite. One example for this case is presented in Sec.~\ref{sec:Cormley}.)

In the case of the GRB 091030.613, we used a
burst-interval between $-20$ and $38$\,seconds before and after the burst,
respectively (see Sec.~\ref{sec:result}, Figure\,\ref{fig:fit.eps}). This means that the data of this
time interval were omitted when fitting to the background. Other than that, the whole CTIME lightcurve were fitted.

It is one of our future plans to create a self-consistent method, which
can automatically define these intervals based on a self-consistent iteration
algorithm, so the user's presence would be unnecessary and the method would be totally automatic.

%% file: result.tex
In this section, we present the result of the DDBF for the GRB 091030.613 (the
one that we showed in Fig.~\ref{fig:lc.eps} and noted that there are difficulties
with its background fitting). 

\begin{figure}[h!]
  \resizebox{\hsize}{!}{\includegraphics[angle=270]{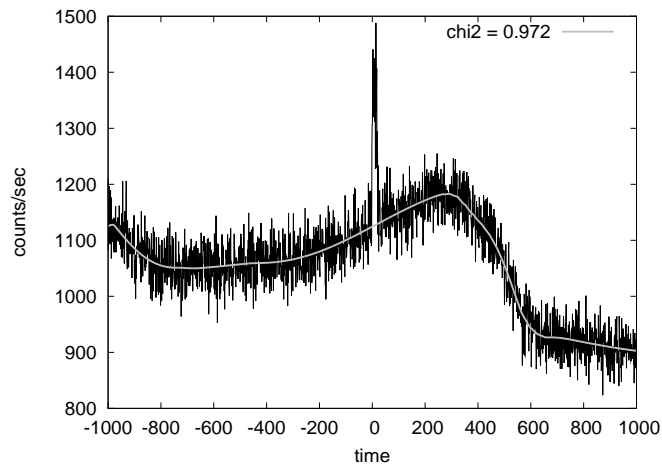}}
  \caption{\small{Fitted background of the lightcurve of the \textsl{Fermi} burst
  091030.613 measured by the 3rd GBM detector. Fitting was done by DDBF method
  \citep{Acta,Monterey},   
  using 14 non-zero singular values according to AIC. Reduced
  chi-square statistics is shown in the top right corner.}}
  \label{fig:fit.eps}
\end{figure}

The DDBF method is a good alternative for the polynomial fitting of time for
two reasons. 
First, the background model consists of astrometric computations of
astrophysical objects, and the fitting variables have physical meanings. This
property is missing when one uses simple polynomial fitting of time; however,
\textsl{Fermi}'s complex motion prefers to have a more detailed model for the background
sources.

Second, using the polynomial fitting of time,
one has to define two
short time intervals before and after the burst, which can be well described by a polynomial function (see Sec.~\ref{sec:features}).
Usually, these intervals have 
to be short
enough and defined precisely to get a correct fit. DDBF can fit all the 2000-sec data of the CTIME (and CSPEC) files. Therefore, we are also able to
study long emissions or precursors. 

\begin{figure}[h!]
  \resizebox{\hsize}{!}{\includegraphics[angle=270]{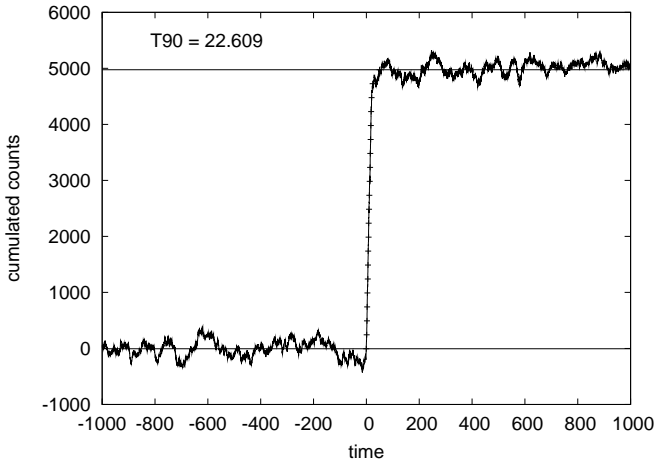}}
  \caption{\small{Cumulative lightcurve of the \textsl{Fermi} burst 091030.613 by the 3rd
  GBM detector. Horizontal lines are drawn at 0\% and 100\% of total cumulated
  counts; dots mark every 5\%. (Model with 14 singular values was selected, as seen in Fig.~\ref{fig:ic.eps}.)
  }}
  \label{fig:int.eps}
\end{figure}

Fig.~\ref{fig:int.eps} shows the cumulative lightcurve from which we computed
the durations \citep{Munchen}. Horizontal lines were computed by averaging the
cumulated background levels before and after the burst: These are the levels
of 0\% and 100\% of total cumulated counts. 

We note that these levels were selected
by the user for the \textsl{Fermi} GBM Catalogue. 
Since they only fitted some short intervals around the burst using time-dependent polynomials, this step could not been automatised \citep{catalogue}. With DDBF, however, we fit all the 2000\,seconds of the CTIME file (except for the burst in the middle) using direction dependent underlying variables. Our method gives us cumulative lightcurves, where the resulting levels are tightly distributed around a constant value, and therefore, the automation (calculating the average of the levels) is possible.

Between the levels of 0\% and 100\%, 19 equally
heightened points mark every 5\% of their cumulated counts (the first and last are fixed where the lightcurves step over and below the levels before and after).
T$_{90}$ is computed by subtracting the value corresponding to 5\% from the
value corresponding to 95\%.

The \textsl{Fermi} GBM Catalogue reports T$_{90}^{cat}$=19.200$\pm$0.871 secs.  Our result is
T$_{90}$=22.609$^{+13.518}_{-4.522}$ seconds.  We always give confidence
intervals instead of error bars with the T$_{90}$ values, 
since the DDBF method is complicated: The error estimation needs further considerations. See Sec.~\ref{sec:error} for details.

This result does not depend on the spectrum or the detector response matrix,
because we summed up the channels of the CTIME files. However, the DDBF can be
used for every channel separately (as it was done in \citet{Munchen}) 
and can also be used with CSPEC data to obtain spectral
information.

%% file: examples.tex
We began with the observation that many of \textsl{Fermi} bursts (even in non-ARR cases) have a
varying background corresponding to the actual direction of the satellite. Thus, our idea was to use this directional information in the filtering algorithm.
We created a method, which is able to
separate this background from the lightcurves. Now, we want to demonstrate the
effectiveness of our method, so we present examples here, with each having an
extreme background.

These examples were purposely chosen to demonstrate how powerful DDBF can be and to give an overall impression about the cases for which it can be used in and the advantages and the difficulties it carries. Two of the examples below are ARR bursts (Secs.~\ref{sec:Meyving} and \ref{sec:Niley}). In general, we would like to draw attention to the connection between the direction dependent underlying variables and the variability features of the lightcurve: the correlation between them are undeniable in every single case (even in no-ARR cases).

In each example, we present figures of the original lightcurves for one of the
triggered detectors, summarizing 
the counts of the effective range of channels of CTIME file.
On these lightcurves, we plot the fitted theoretical background with a solid
line and the reduced chi-square statistics in the top right corner. Then, we
show the \textsl{absolute value} of the direction dependent underlying variables 
(in one graph), 
and the AIC$_k$ as a function of used singular values. 

As a final result, we show the cumulative lightcurves, which we used to compute
the T$_{90}$ values. We also give the preliminary T$_{90}^{pre}$ from the
gamma-ray coordinates network \citep{GCN}, and the T$_{90}^{cat}$ from the
catalogue computed and published by the GBM team \citep{catalogue}.
We give confidence intervals of the computed T$_{90}$s (and T$_{50}$s as well). The description of how these confidence intervals were computed is in Sec.~\ref{sec:error}.

It is important to note, however, that only long GRBs were analysed here. The
reason of this is that short bursts usually are not influenced 
by the fast motions of the satellite.
During one short burst, the background does not change so much that DDBF should be
used. Furthermore, short bursts are better analysed using the time tagged
events (TTE) data type instead of CTIME (and CSPEC), and therefore, they are not presented here.

Since we want to present how effective our method is, we show the detector having the highest background variability without filtering in every case. However, it is possible to combine the same analysis for a number of bright detectors for each burst to reduce the error. It will be a part of a future work to create a catalogue of the durations of the \textsl{Fermi} bursts using DDBF, in which we will use more than one detector's data. Here, we present the method with only one triggered detector for each case.


\subsubsection{GRB~090102.122}\label{sec:Kyron} 
GRB~090102.122 is an example where no fast motion was carried out, and therefore, no high background rate variation is taken place. This burst had no ARR. The lightcurve is simple in the 
sense that a time dependent polynomial function could possibly be used to fit it properly. However, we present DDBF results only to show that the method works in these simple cases as well. The AIC chose 9 singular values, and one can see in the information criterion plot that more values than this are punished by the AIC: Too many free parameters would cause the fitted curve to have unnecessary loops fitted to the noise of the background.
The \textsl{Fermi} catalogue reports T$_{90}^{cat}$=26.624$\pm$0.810\,sec \citep{catalogue}.
Detector 'a' was analysed here.

\begin{figure}[h!]\begin{center}
  \resizebox{.8\hsize}{!}{\includegraphics[angle=270]{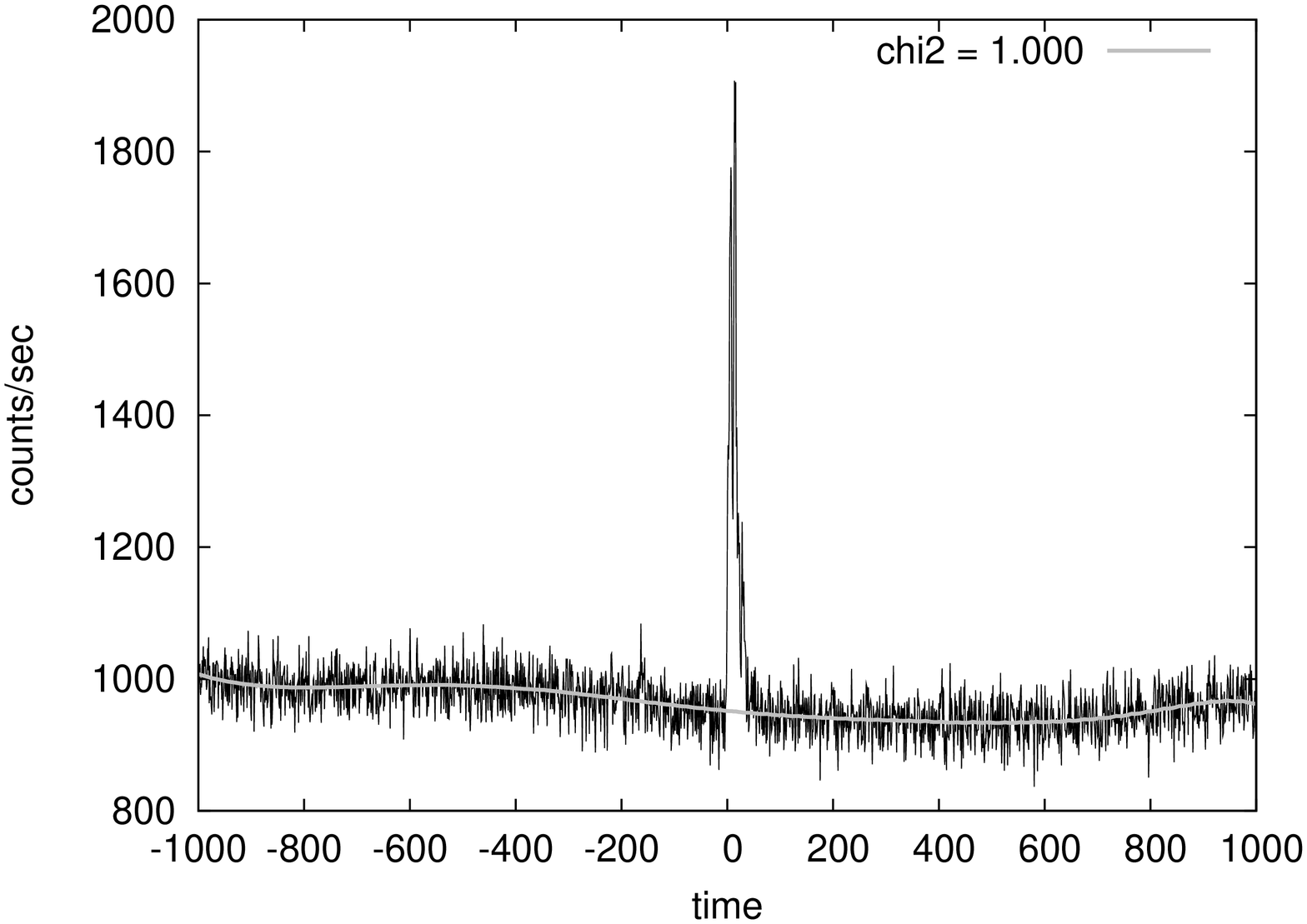}}
  \resizebox{\hsize}{!}{\includegraphics[angle=270]{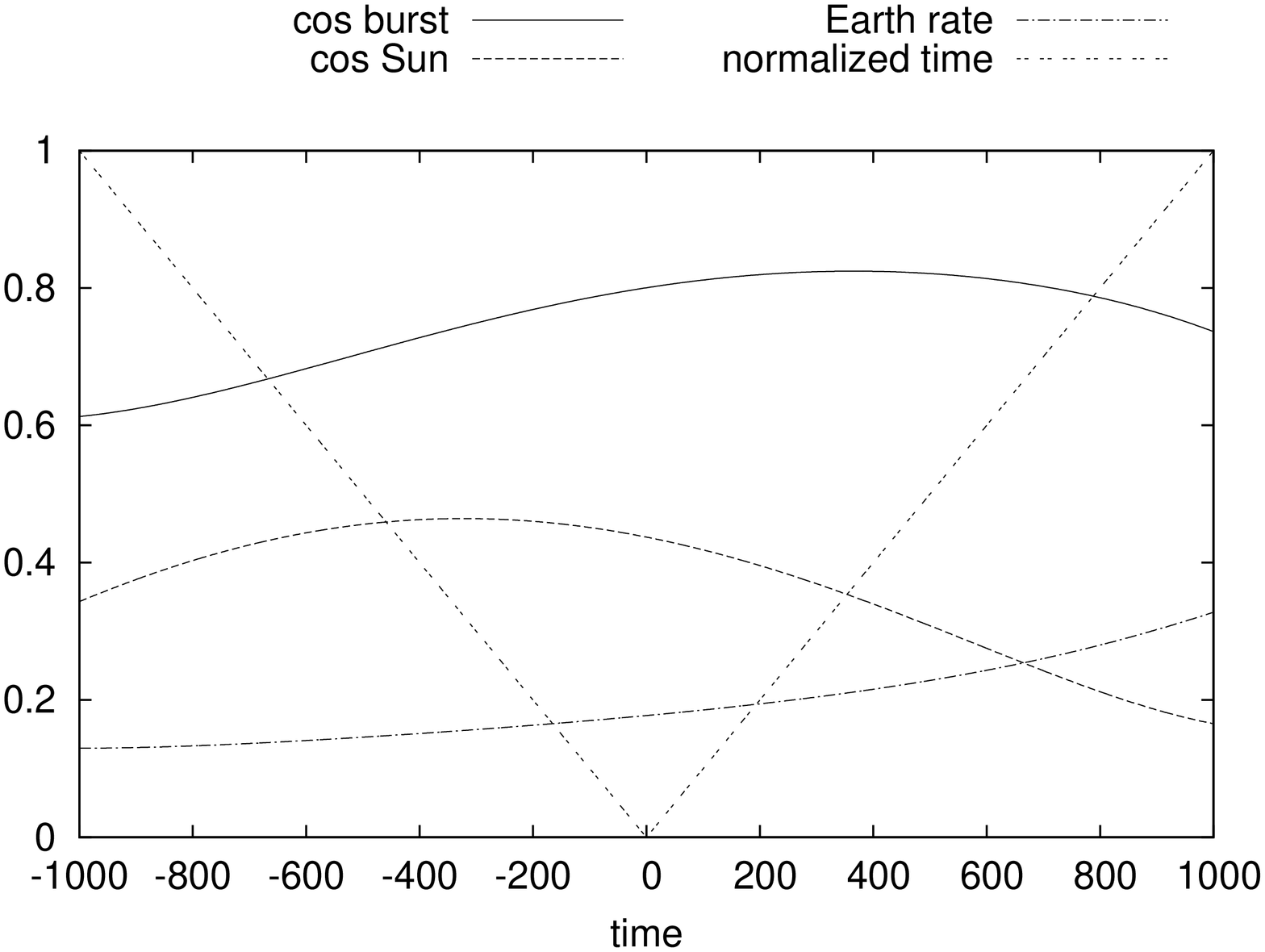}
  			\includegraphics[angle=270]{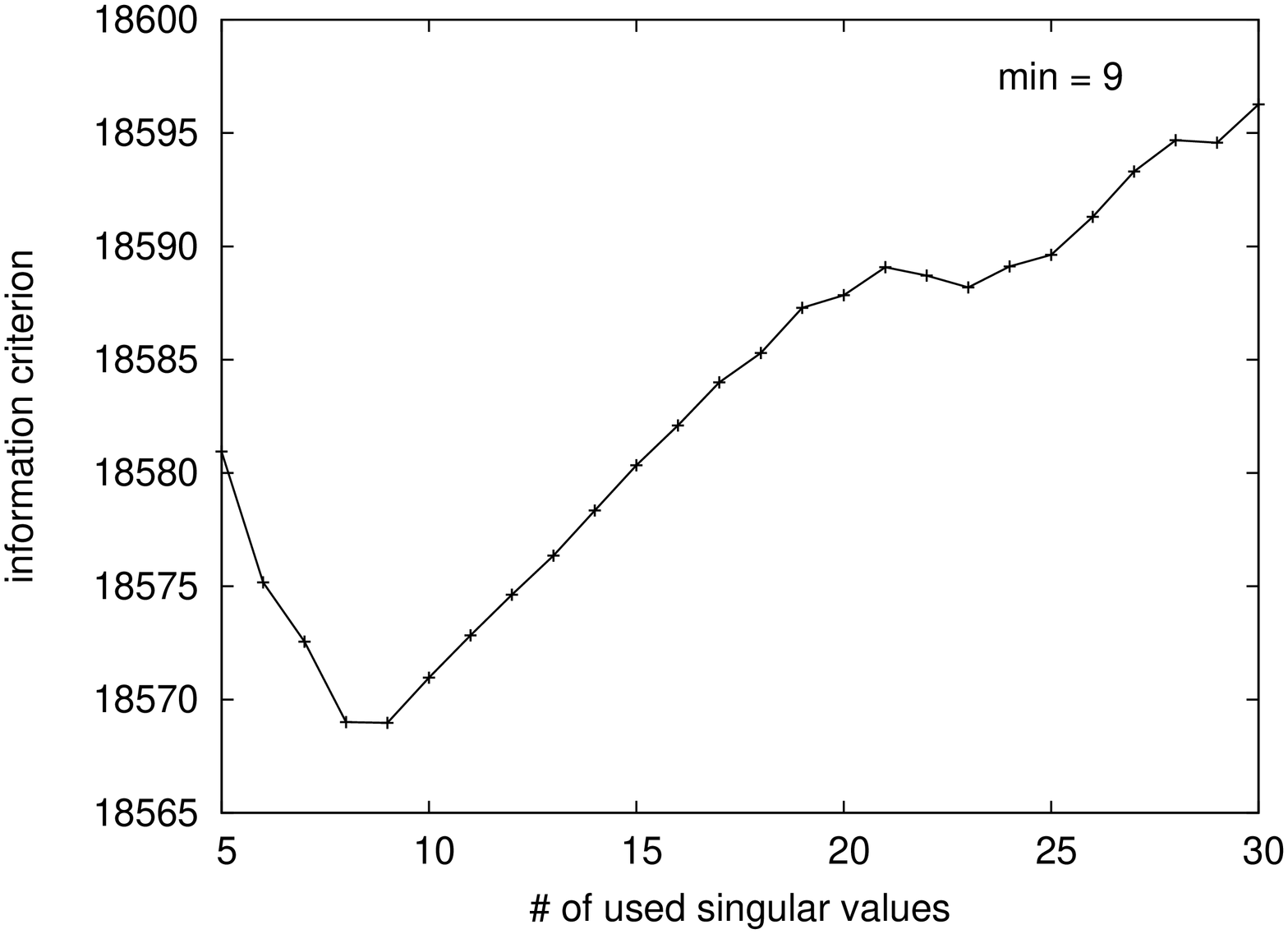}  }
  \caption{\small{
		\textsl{Top:}} Lightcurve of the \textsl{Fermi} GRB~090102.122 as measured
		by the triggered GBM detector\,'a' and the fitted background
		with a grey line. Burst interval (secs): [-5:35].
				\textsl{Bottom left:} Underlying variables (absolute values). See Sec.~\ref{sec:sources}.
				\textsl{Bottom right:} Akaike Information Criterion. See Sec.~\ref{sec:modelsel}.  
  }
  \label{fig:090102122}
  \resizebox{.8\hsize}{!}{\includegraphics[angle=270]{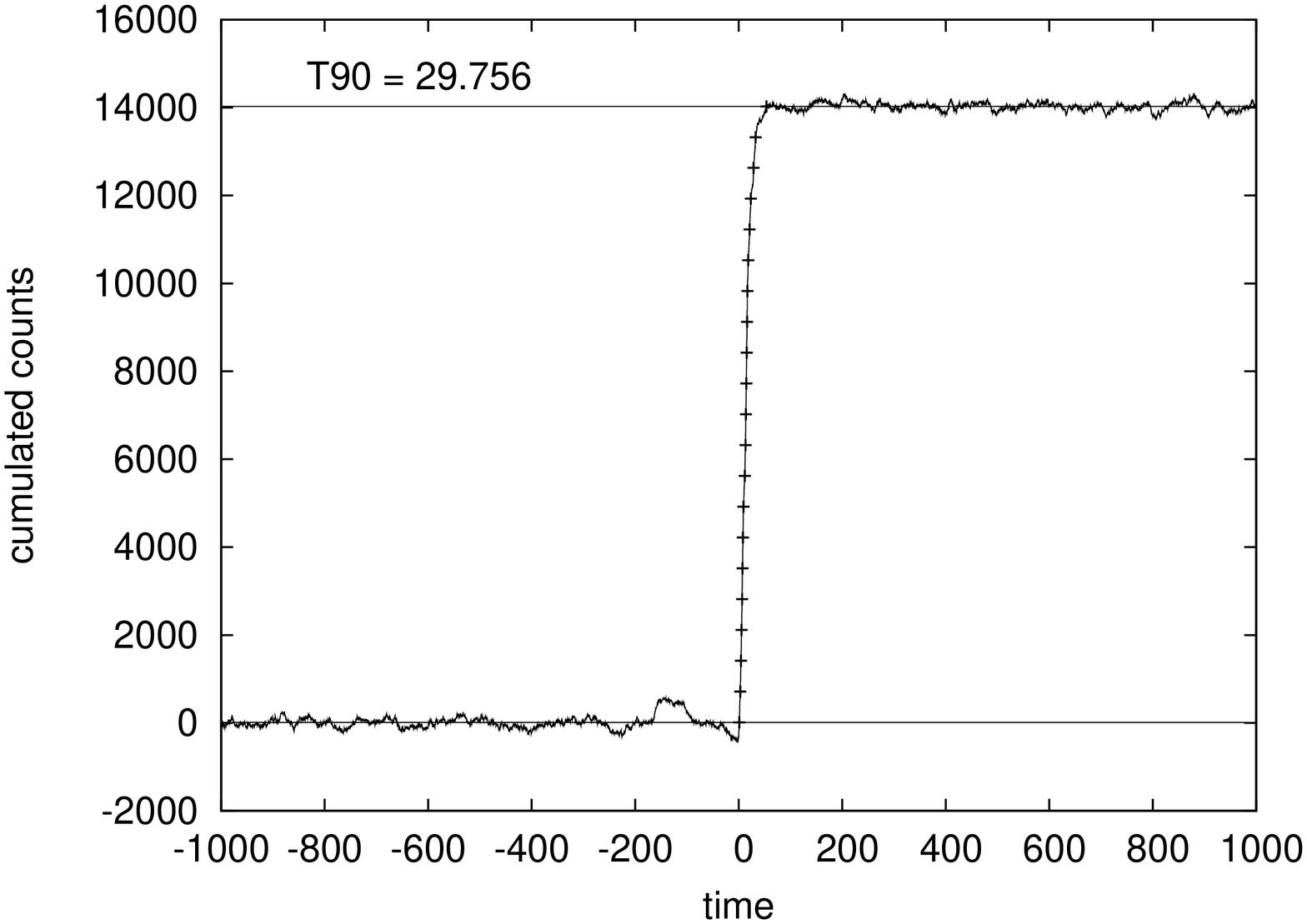} }
  \caption{\small{Cumulative lightcurve of GRB~090102.122. Horizontal lines are
  drawn at 0\% and 100\% of total cumulated counts; dots mark every 5\%. }}
  \label{fig:090102122INT}
\end{center}
\end{figure}

Around -150\,secs in the lightcurve, there is a peak, which cannot be explained
by the physical underlying variables. This causes a little hump in the
cumulative lightcurve in Fig.~\ref{fig:090102122INT}. (Furthermore, the same peak can be seen 
in 
the lightcurves of the other triggered detector.) It is out of the
scope of this article to decide whether it is a pre-burst or another
instrumental effect, however, we emphasize again that DDBF can also be used for finding
pre-bursts or long emissions. 
We measured T$_{90}$=29.756$^{+2.971}_{-1.198}$\,seconds. 


\subsubsection{GRB~090113.778}\label{sec:Cormley}	
The \textsl{Fermi} catalogue reports T$_{90}^{cat}$=17.408$\pm$3.238\,sec \citep{catalogue} and this is a no-ARR case.
Detector\,'0' was analysed here \citep{Monterey}.
This lightcurve in Fig.~\ref{fig:100130777} has some extra counts around 400
and 600\,seconds. 
Both of them can be explained with the variation in the underlying variables: 
Around 400~sec, the Earth limb was out of the FoV and then it came back and peaked at 600~sec until the Sun's position changed significantly. Both of these could cause the extra counts. The best chosen model with 12 singular values could fit these peaks (see the big and small loops in the fitted lightcurve at 400 and 600~sec).
Since the underlying variables are based on the motion of the satellite, it follows that these two peaks are probably not astrophysical effects. They do not come from the GRB but from the combined effect of the background sources in the surroundings: the Earth and the Sun. It is important to note that a statement like that could not be made using the traditional method of polynomial fitting of time.

\begin{figure}[h!]\begin{center}
		\resizebox{.8\hsize}{!}{\includegraphics[angle=270]{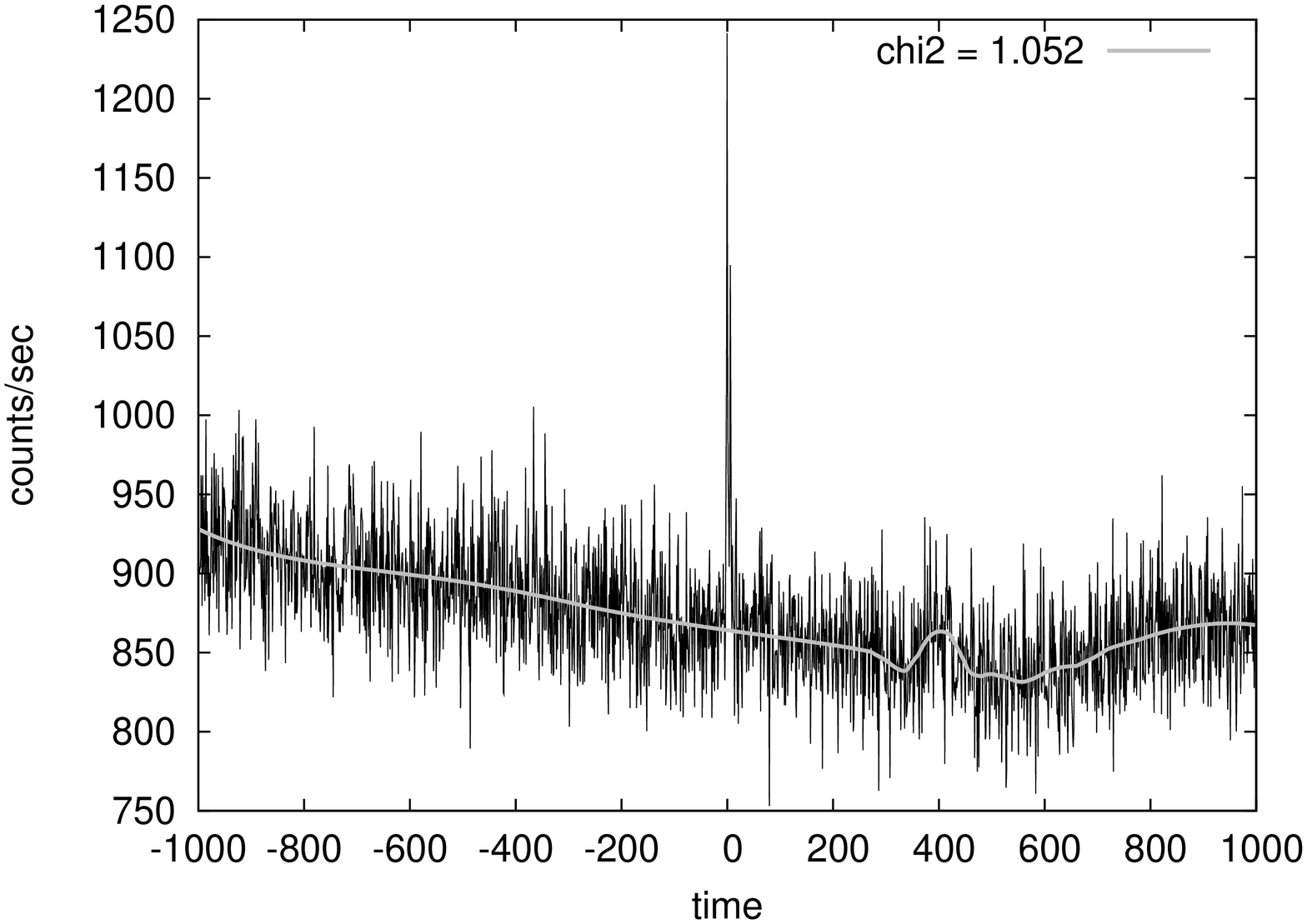}}
		\resizebox{\hsize}{!}{	\includegraphics[angle=270]{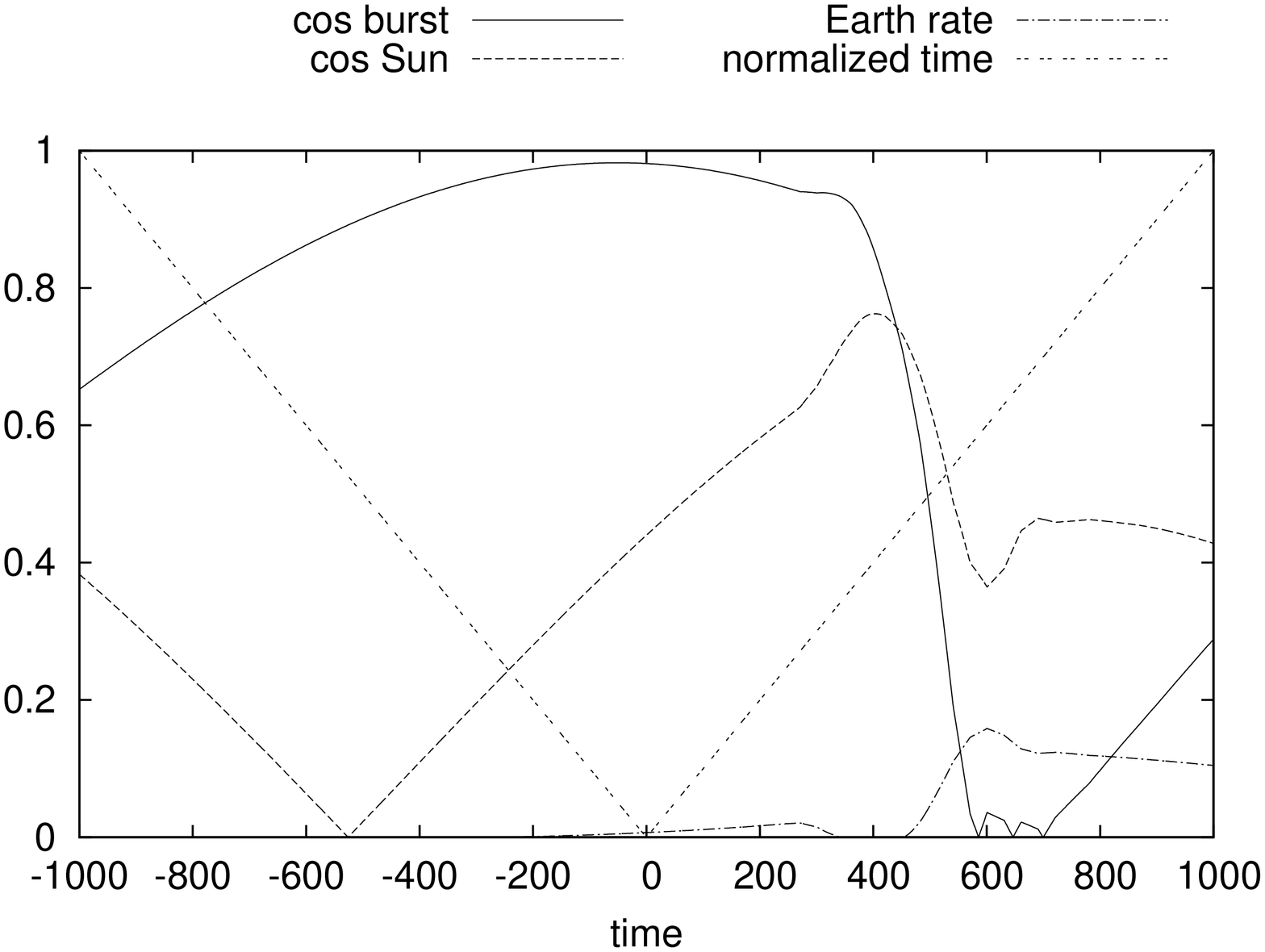}
								\includegraphics[angle=270]{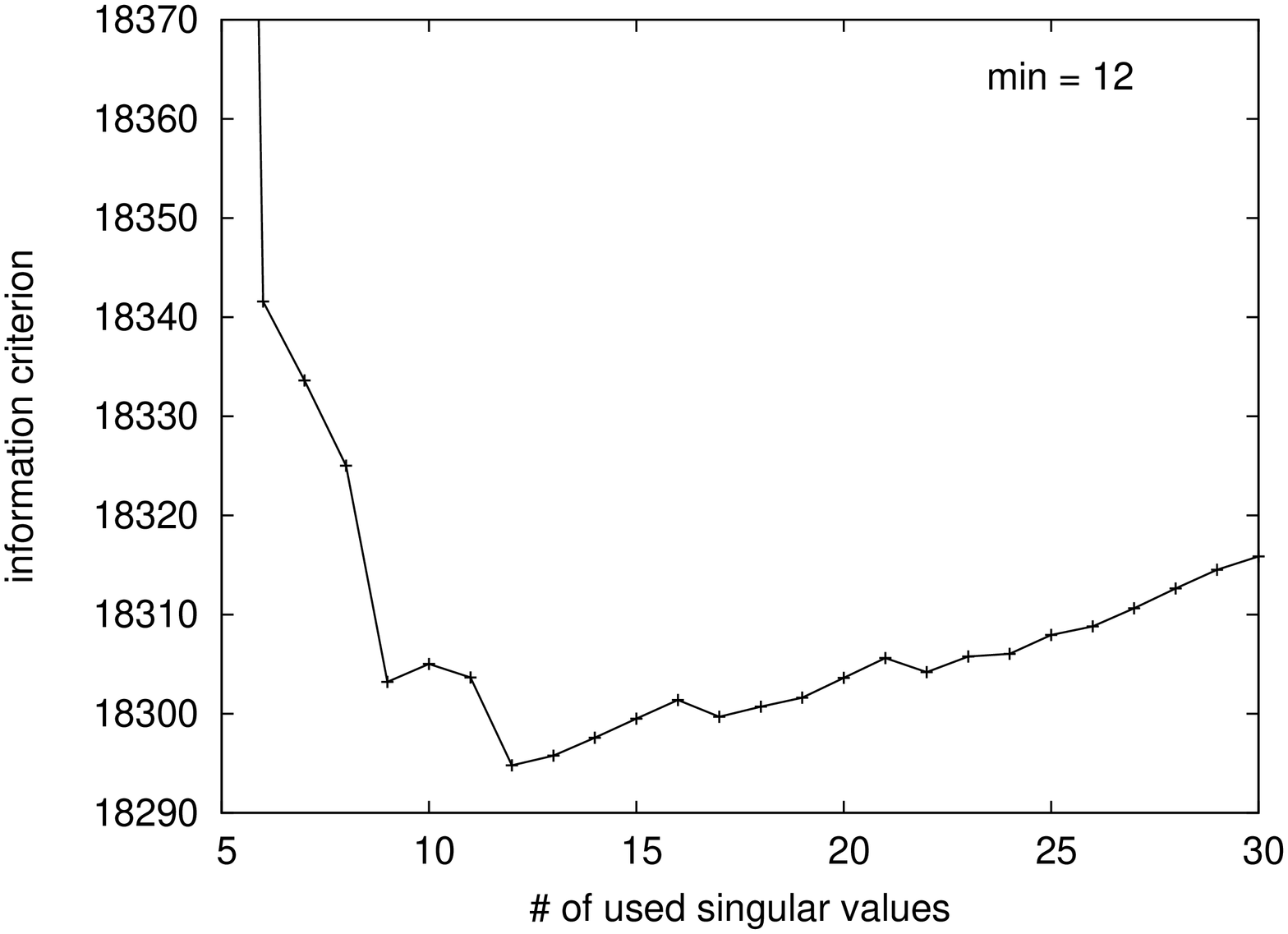}  }
		\caption{\small{
				\textsl{Top:} Lightcurve of the \textsl{Fermi} GRB~090113.778 as measured by the triggered
				GBM detector\,0' and the fitted background
				with a grey line. Burst interval: [-20:40].
				\textsl{Bottom left:} Underlying variables (absolute values). See Sec.~\ref{sec:sources}.
				\textsl{Bottom right:} Akaike Information Criterion. See Sec.~\ref{sec:modelsel}.  
				}}
		\label{fig:090113778}
		\resizebox{.8\hsize}{!}{	\includegraphics[angle=270]{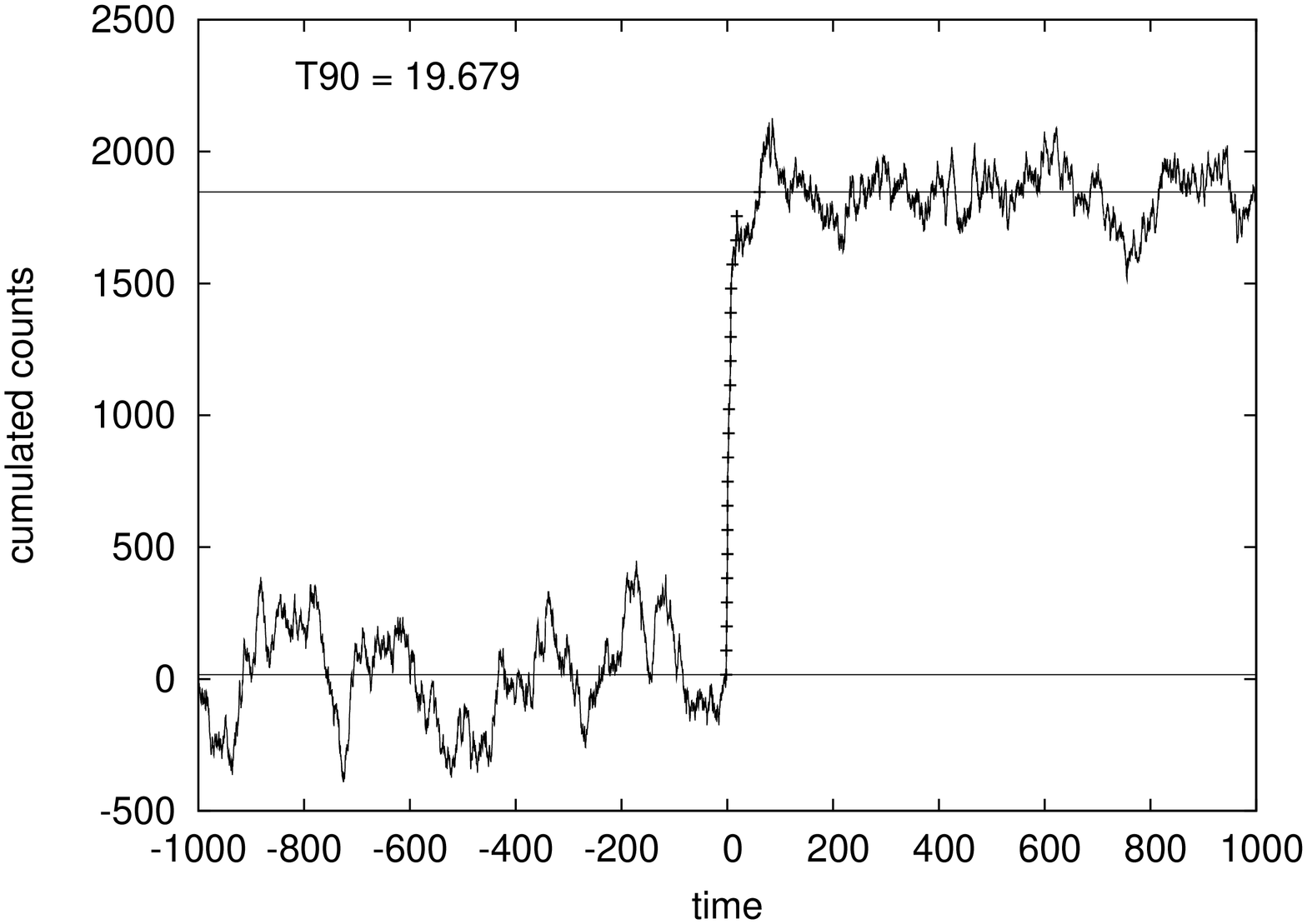}}
		\caption{\small{
				Cumulative lightcurve of GRB~090113.778.
				Horizontal lines are drawn at 0\% and 100\% of
				total cumulated counts; dots mark every 5\%.
  }}
		\label{fig:090113778INT}
\end{center}
\end{figure}

After the background subtraction, the cumulative lightcurve
(Fig.~\ref{fig:090113778INT}) is noisy because this burst was not so
intense with only $\sim$\,1800\,counts, while other examples have 10000-20000
counts. Our result is T$_{90}$=19.679$^{+10.883}_{-6.421}$\,sec.


\subsubsection{GRB~090618.353}\label{sec:Dynold} 
The \textsl{Fermi} catalogue reports T$_{90}^{cat}$=112.386$\pm$1.086 sec
\citep{catalogue}. No ARR was taken. 

The data from detector\,'7' were analysed here. 
Nevertheless, we should
note that detector\,'4' 
has so many counts that almost any kind of
background model seems to be good enough to compute T$_{90}$ when using
detector\,'4'. We still choose to present detector\,'7' here, because
we can show our method working in a more complicated case.

\begin{figure}[h!]\begin{center}
  \resizebox{.8\hsize}{!}{\includegraphics[angle=270]{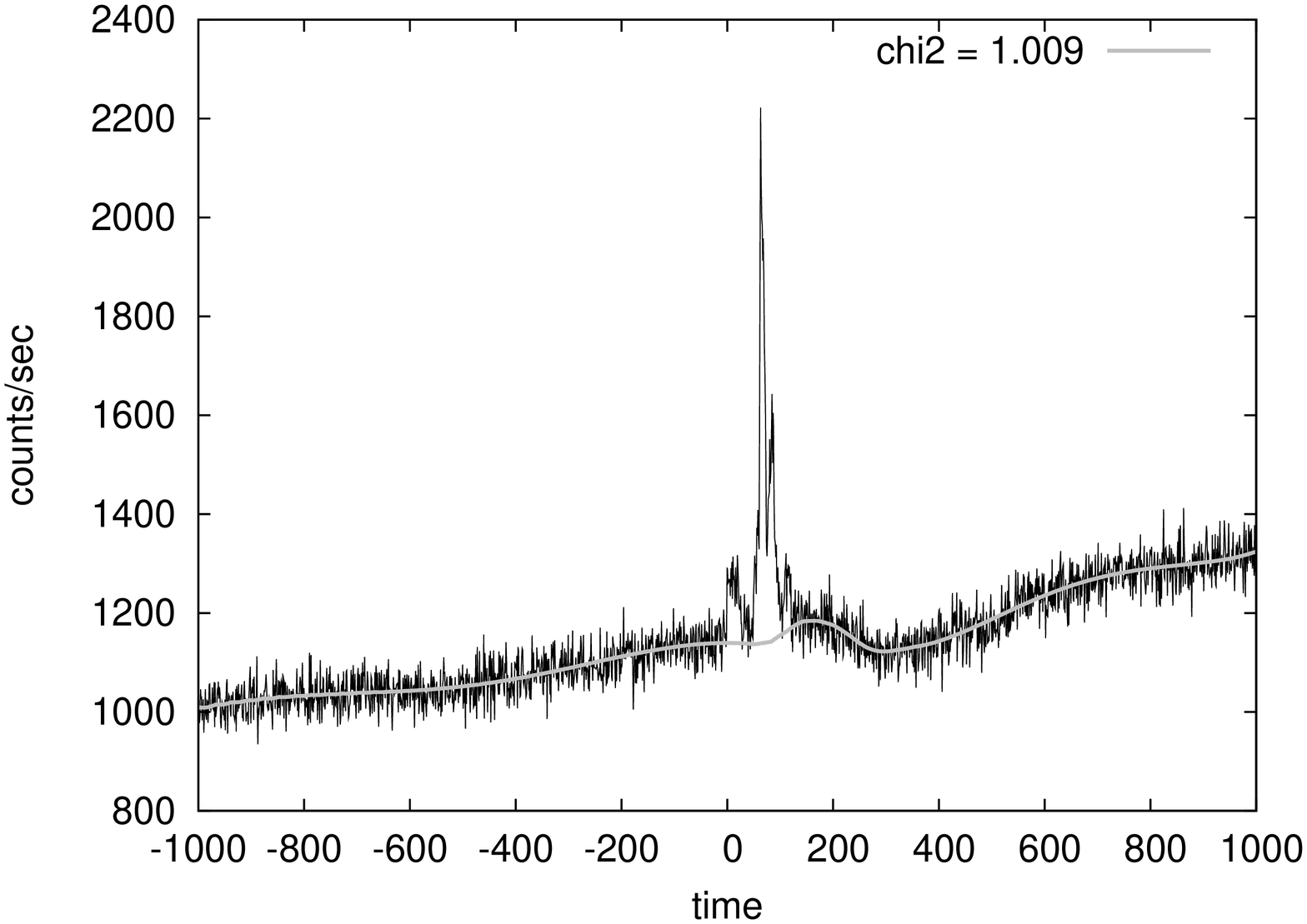}}
  \resizebox{\hsize}{!}{\includegraphics[angle=270]{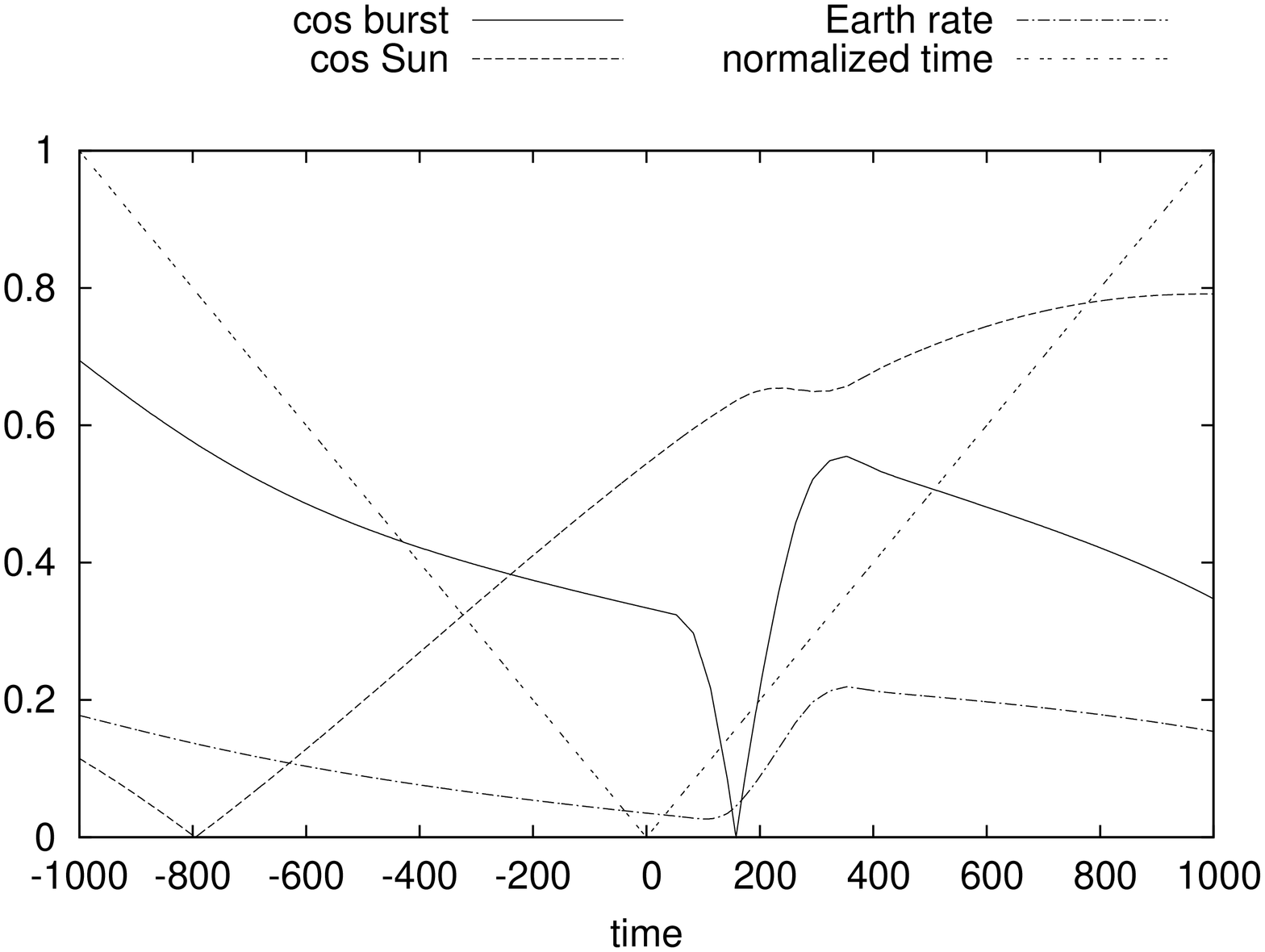}
  			\includegraphics[angle=270]{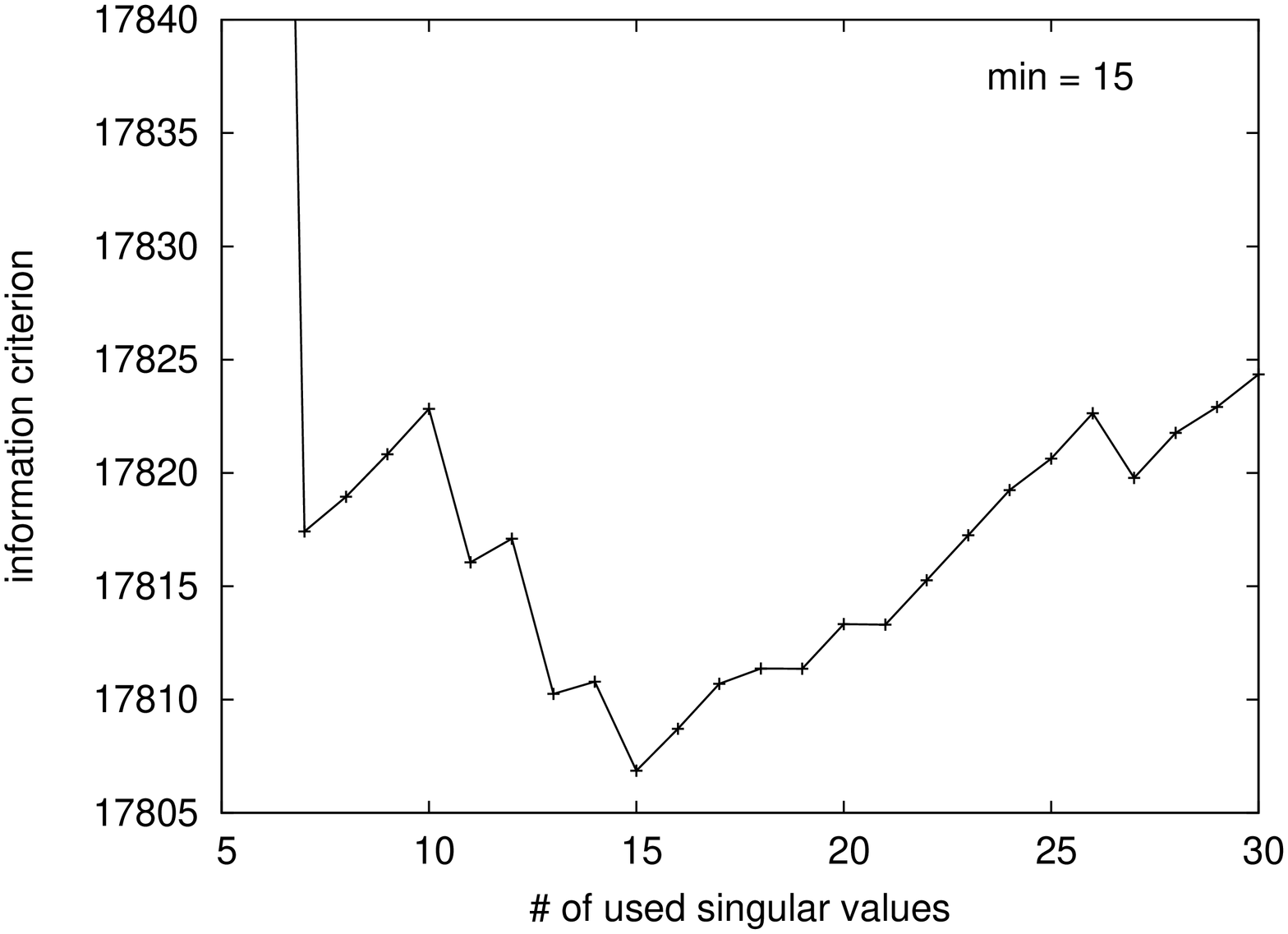}  }
  \caption{\small{
		\textsl{Top:} Lightcurve of the \textsl{Fermi} GRB~090618.353 as measured
		by the triggered GBM detector\,'7' and the fitted background
		with a grey line. Burst interval: [-20:130].
				\textsl{Bottom left:} Underlying variables (absolute values). See Sec.~\ref{sec:sources}.
				\textsl{Bottom right:} Akaike Information Criterion. See Sec.~\ref{sec:modelsel}.  
  }}
  \label{fig:090618353}
  \resizebox{.8\hsize}{!}{\includegraphics[angle=270]{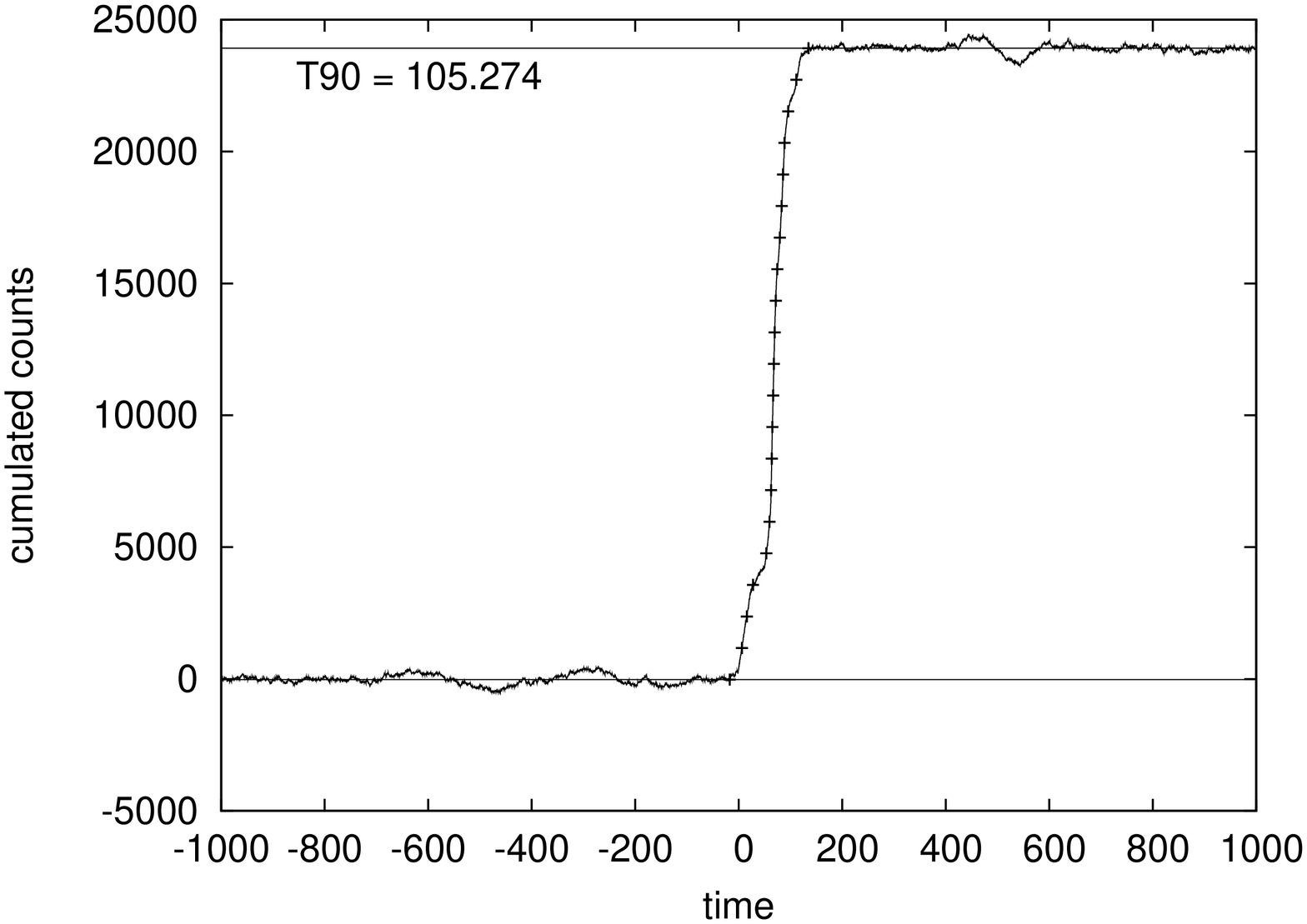} }
  \caption{\small{Cumulative lightcurve of GRB~090618.353. Horizontal lines are
  drawn at 0\% and 100\% of total cumulated counts; dots mark every 5\%. }}
  \label{fig:090618353INT}
\end{center}
\end{figure}

Now we may be used to the fact that quickly varying underlying variables
(which correspond to fast motion of the satellite) cause a quick change in the
lightcurve background at the same time. 
This burst had no ARR, but the satellite started to rotate according to the fast change of the underlying variables after
the trigger. At this point, the
lightcurve is changing more quickly than before. 
The fitted grey line
(chi-square statistics are 1.009) pursue this change, and results in a
duration of T$_{90}$=103.338$^{+3.842}_{-6.725}$\,seconds.


\subsubsection{GRB~090828.099}\label{sec:Lyokari} 
GRB~090828.099 was detected by the GBM on 28 August 2009 at 02:22:48.20\,UT
\citep[9844]{GCN}. The first
GBM catalogue reported T$_{90}^{cat}$=68.417$\pm$3.167\,sec \citep{catalogue}. This is a non-ARR case. The data from detector\,'5' was analysed here.

\begin{figure}[h!]\begin{center}
  \resizebox{.8\hsize}{!}{\includegraphics[angle=270]{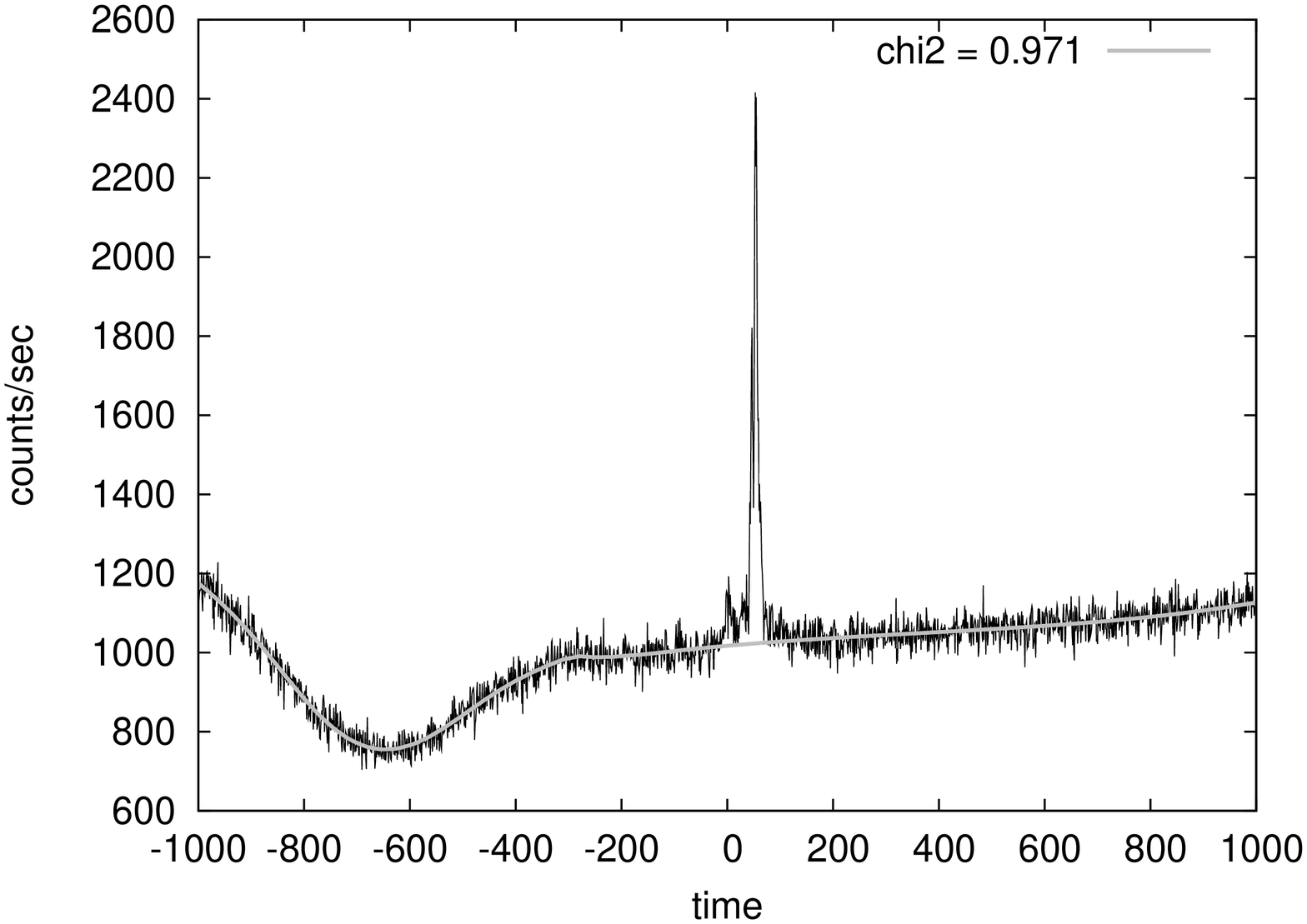}}
  \resizebox{\hsize}{!}{\includegraphics[angle=270]{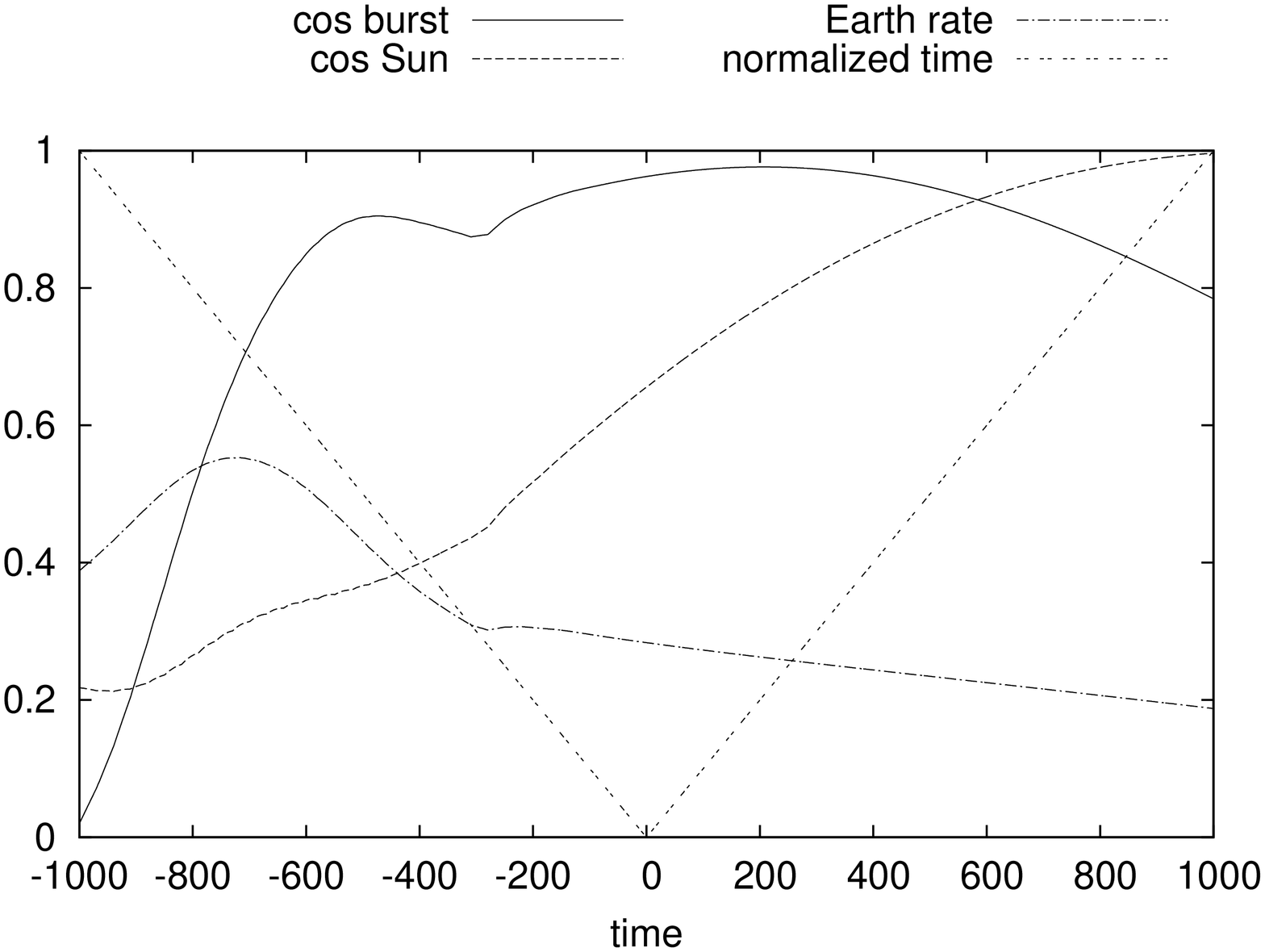}
  			\includegraphics[angle=270]{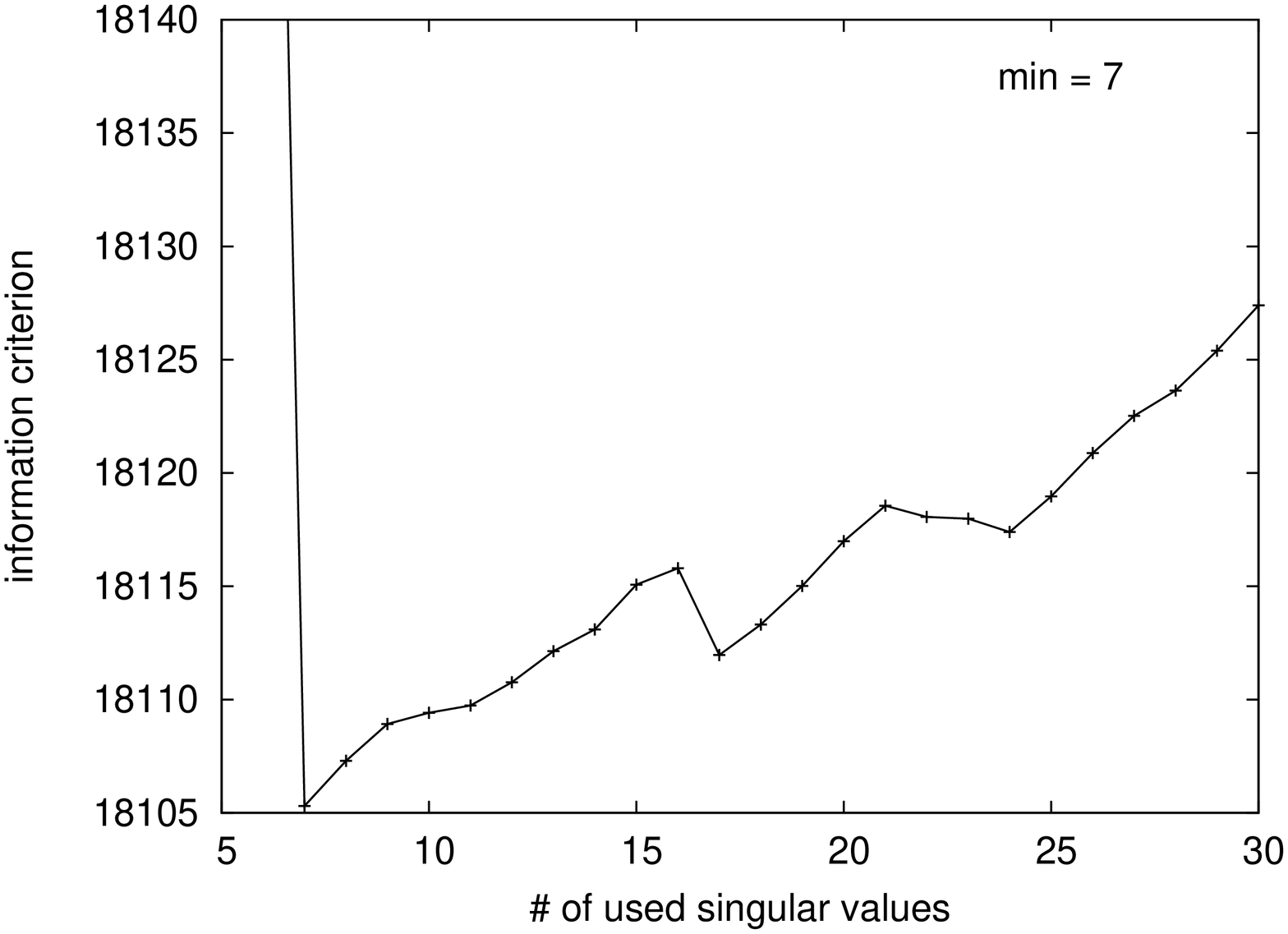}  }
  \caption{\small{
		\textsl{Top:} Lightcurve of the \textsl{Fermi} GRB~090828.099 as measured
		by the triggered GBM detector\,'5' and the fitted background
		with a grey line. Burst interval: [-10:80].
				\textsl{Bottom left:} Underlying variables (absolute values). See Sec.~\ref{sec:sources}.
				\textsl{Bottom right:} Akaike Information Criterion. See Sec.~\ref{sec:modelsel}.   
  }}
  \label{fig:090828099}
  \resizebox{.8\hsize}{!}{\includegraphics[angle=270]{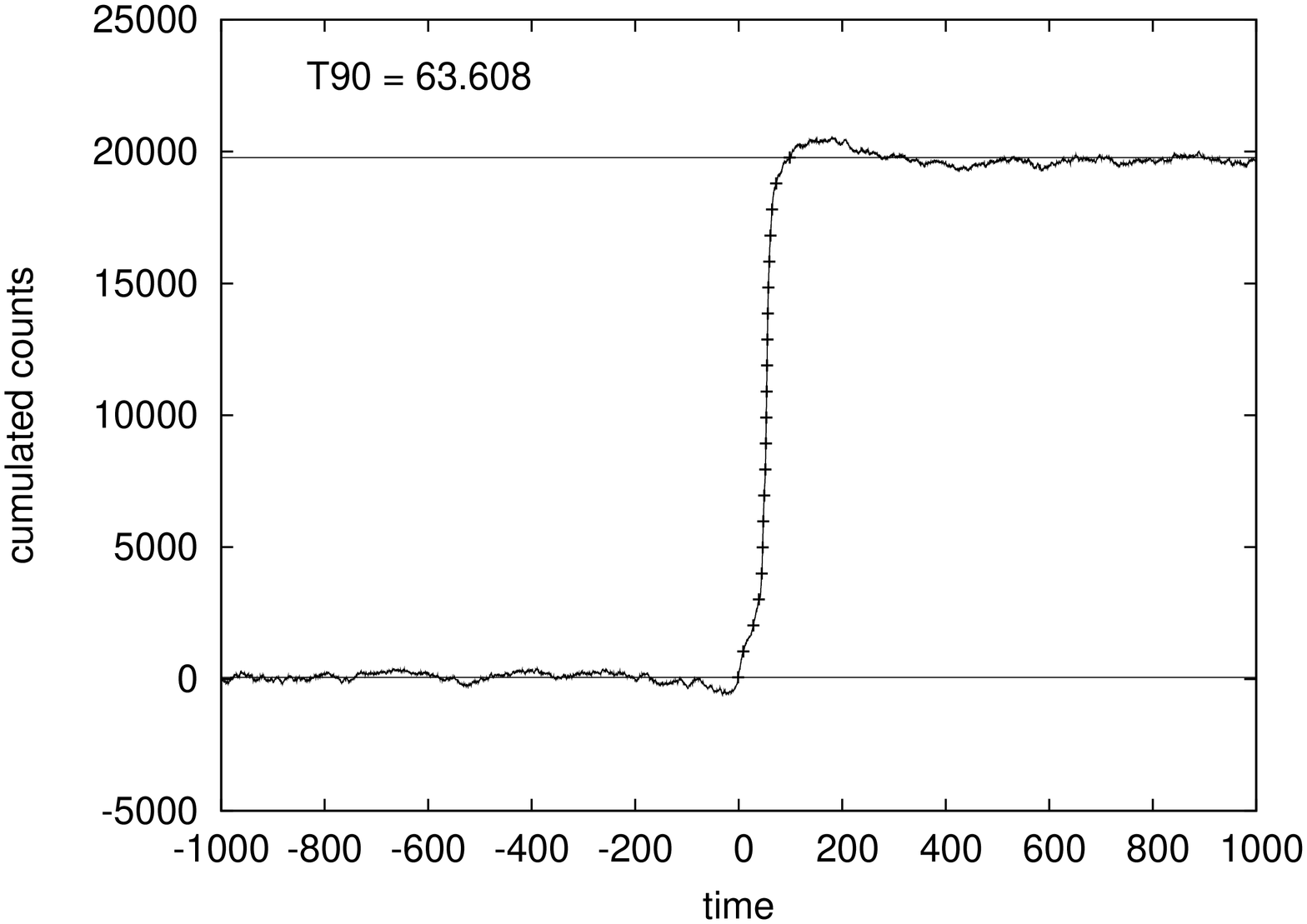} }
  \caption{\small{Cumulative lightcurve of GRB~090828.099. Horizontal lines are
  drawn at 0\% and 100\% of total cumulated counts; dots mark every 5\%. }}
  \label{fig:090828099INT}
\end{center}
\end{figure}

The AIC gives us the model with 7 singular values. 
This is also a simple background. Only the first 300-400~sec are influenced by the fast motion, but DDBF could filter this effect.
The duration computed with the DDBF is T$_{90}$=63.608$^{+1.467}_{-1.652}$\,seconds.


\subsubsection{GRB~091024.372 and .380}\label{sec:Meyving} %
This case deserves attention because an ARR was caused by this burst.
The GBM was triggered twice on GRB~091024: the first time at
08:55:58.47~UT (GRB~091024.372) and the second time at 09:06:29.36~UT (GRB~091024.380).
The GCN 10114 reports: \textsl{'This burst was detected by Swift and the \textsl{Fermi} Gamma-ray  
Burst Monitor with a first emission interval lasting $\sim$50 sec and a  
second emission interval starting $\sim$630~sec after trigger and lasting  
more than 400~sec. The spacecraft performed a repointing  
maneuver for this burst which resulted in pointed observation for 5~hours starting $\sim$350 sec after} 
[the second] 
\textsl{trigger.'} \citep[10114]{GCN} 

Additionally, \citet{Gruber} performed a detailed analysis of this burst and its optical afterglow. Here, we show DDBF duration results separately for the two triggers. Further investigation is needed to analyse the total $\sim$1020 seconds of this extreme long burst as a whole with DDBF. This will be provided in a future work.

\begin{figure}[h!]\begin{center}
  \resizebox{.8\hsize}{!}{\includegraphics[angle=270]{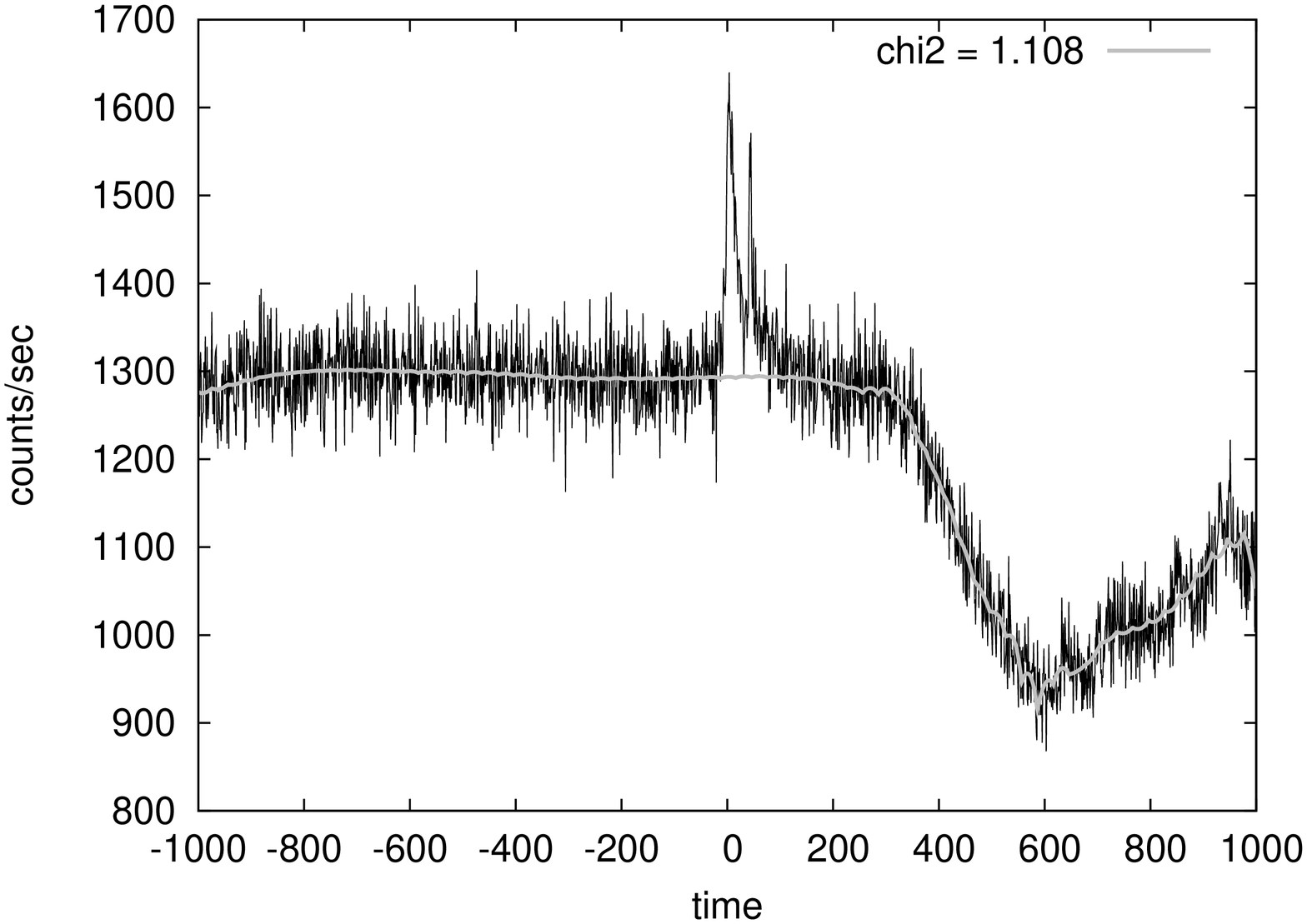}}
  \resizebox{\hsize}{!}{\includegraphics[angle=270]{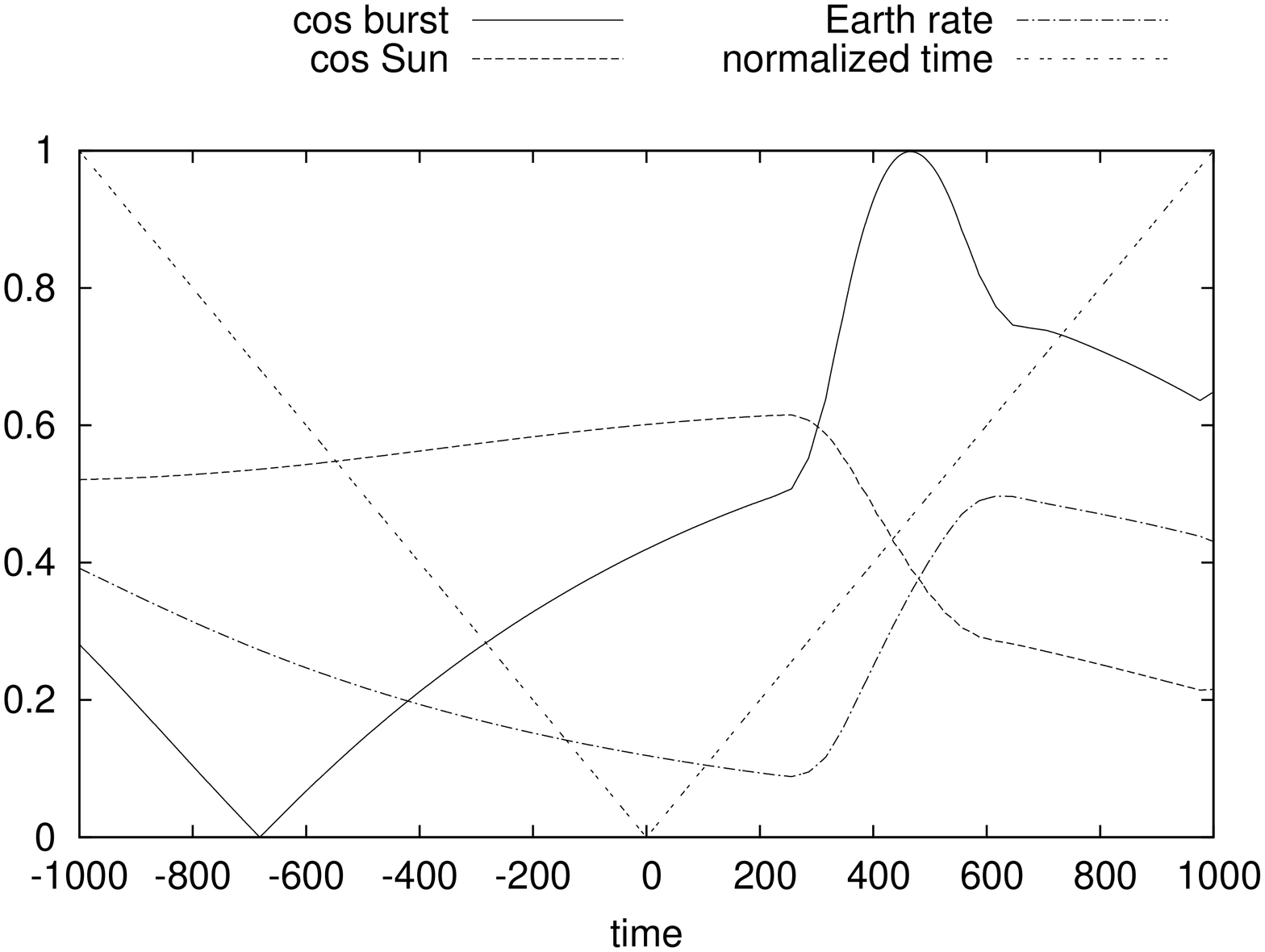}
  			\includegraphics[angle=270]{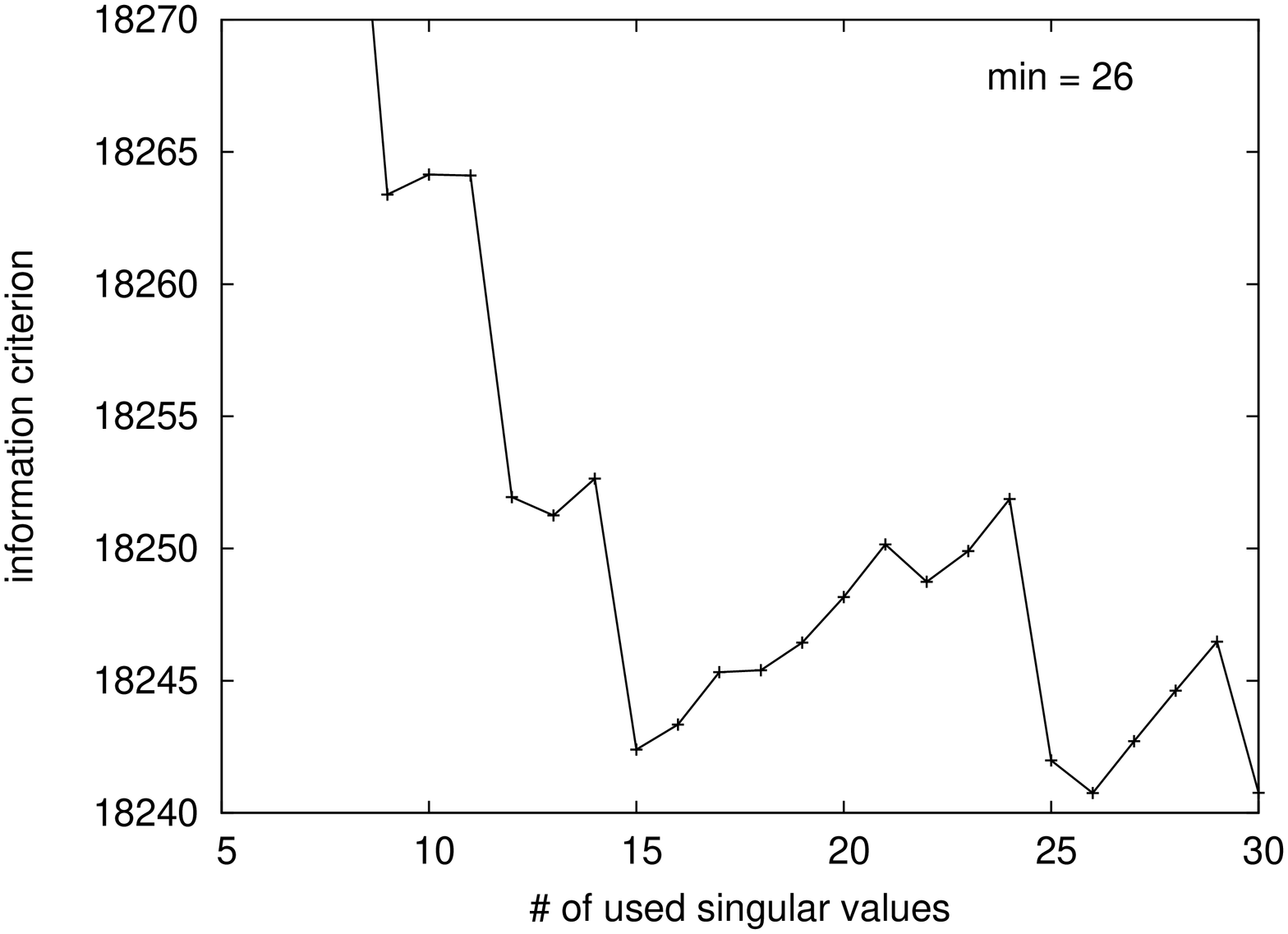}  }
  \caption{\small{
		\textsl{Top:} Lightcurve of the \textsl{Fermi} GRB~091024.372 as measured
		by the triggered GBM detector\,'8' and the fitted background
		with a grey line. Burst interval: [-19:119].
				\textsl{Bottom left:} Underlying variables (absolute values). See Sec.~\ref{sec:sources}.
				\textsl{Bottom right:} Akaike Information Criterion. See Sec.~\ref{sec:modelsel}.   
  }}
  \label{fig:091024372}
  \resizebox{.8\hsize}{!}{\includegraphics[angle=270]{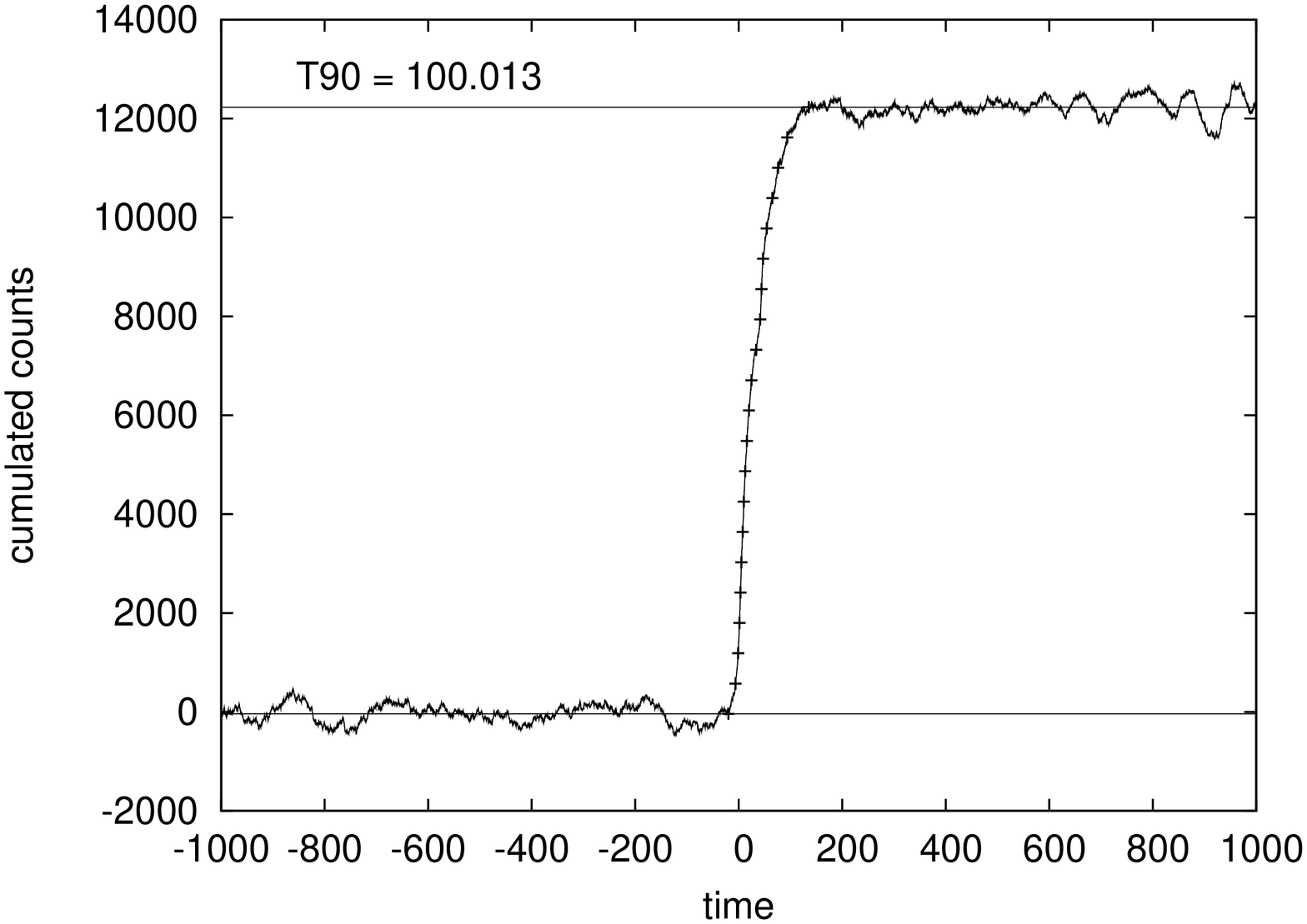} }
  \caption{\small{Cumulative lightcurve of GRB~091024.372. Horizontal lines are
  drawn at 0\% and 100\% of total cumulated counts; dots mark every 5\%. }}
  \label{fig:091024372INT}
\end{center}
\end{figure}

Figure~\ref{fig:091024372} shows the CTIME data of the first trigger (.372) using the triggered detector '8'. 
The second burst episode after 630~sec can also be recognized in the lightcurve by the naked eye 
(however, the satellite changed its position at the time of this second trigger, so this emission looks less intensive here in detector '8'). 
On the other hand, one can notice that the underlying variables do not show any variability at this time interval. Qualitatively this means that something is happening there which is not coming from our modelled sources (Earth or Sun). This can be shown more quantitatively, if one considers that another local minimum can be seen at 15 which are close to the global minimum at 20, which AIC determines for this fit. Here the models with too many free parameters considered the second burst as a background noise and tried to filter it with these polynomial loops. Indeed, the fitted curve shows several loops, especially at the interval of the second burst. 

We can draw two lessons from all of this. First, one has to use AIC with caution. Sometimes, the preferred singular value is not the one AIC gives, if there is another one close enough. In the case of the first emission (.372), there are no loops on the fitted curve, when one uses only 15 singular values (the second local minimum of the AIC). Fortunately, the final T$_{90}$ result does not change much (less than 1\% in this case).
Second, one needs to pay more attention to too many singular values (we would say more than 20, based on our other examples), especially if there is an additional local minimum in AIC close to the chosen one. This can mean that something is happening that cannot be well modelled and may be an astrophysical process. We already mentioned that DDBF can be used to detect long emissions: This is clearly such a case.
Our final result for the first emission (.372) is T$_{90}$=100.013$^{+7.908}_{-4.156}$\,seconds.

\begin{figure}[h!]\begin{center}
  \resizebox{.8\hsize}{!}{\includegraphics[angle=270]{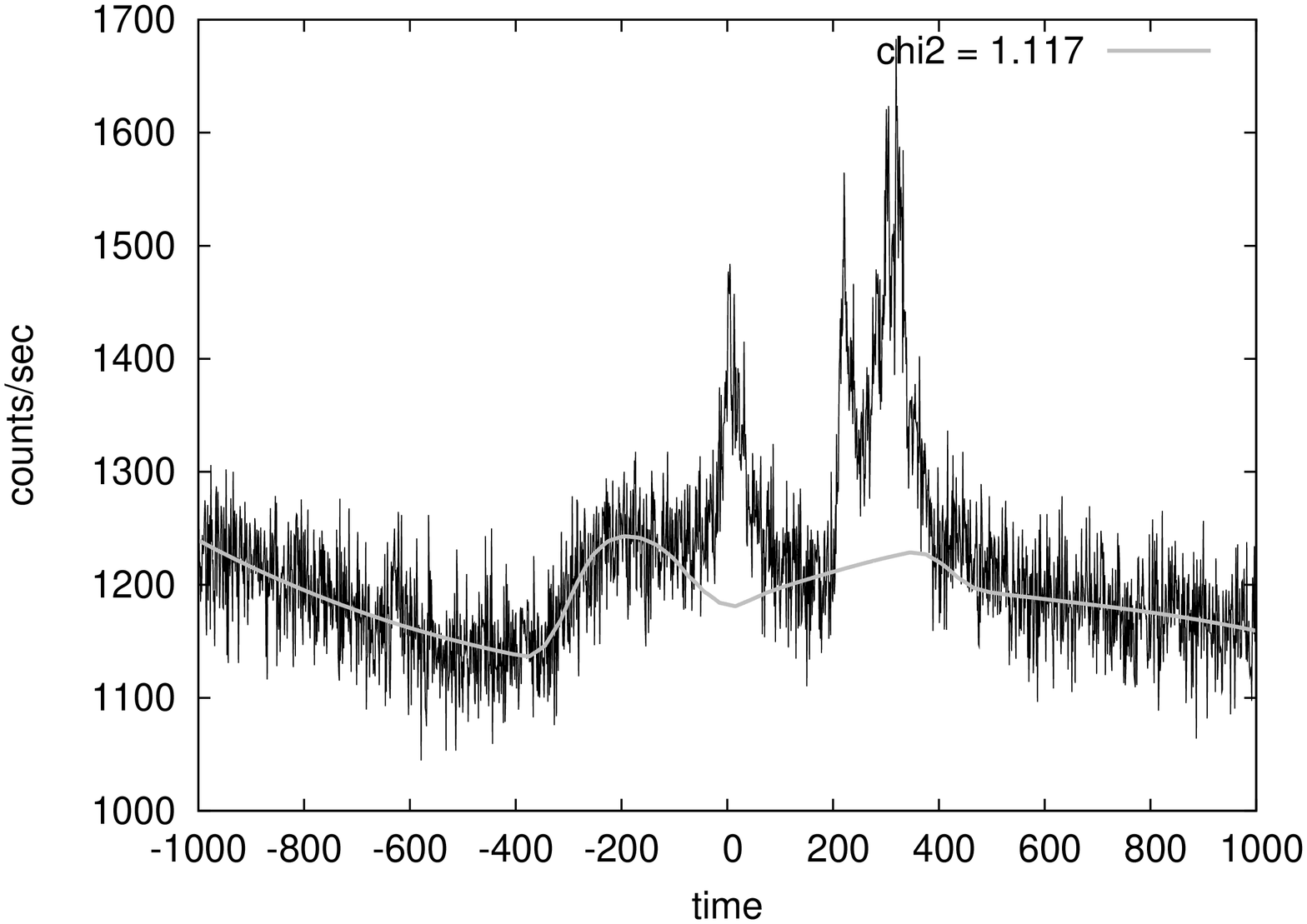}}
  \resizebox{\hsize}{!}{\includegraphics[angle=270]{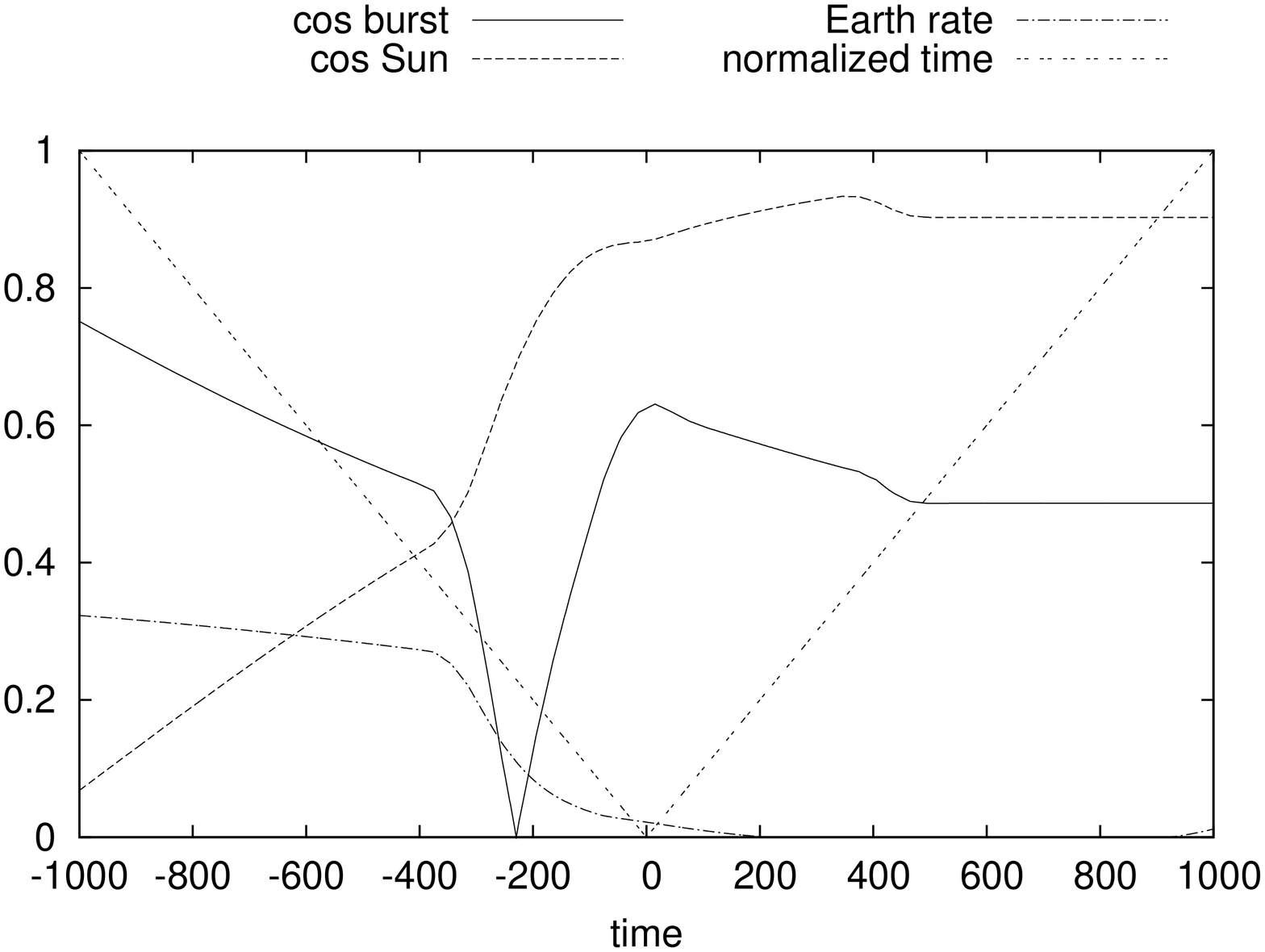}
  			\includegraphics[angle=270]{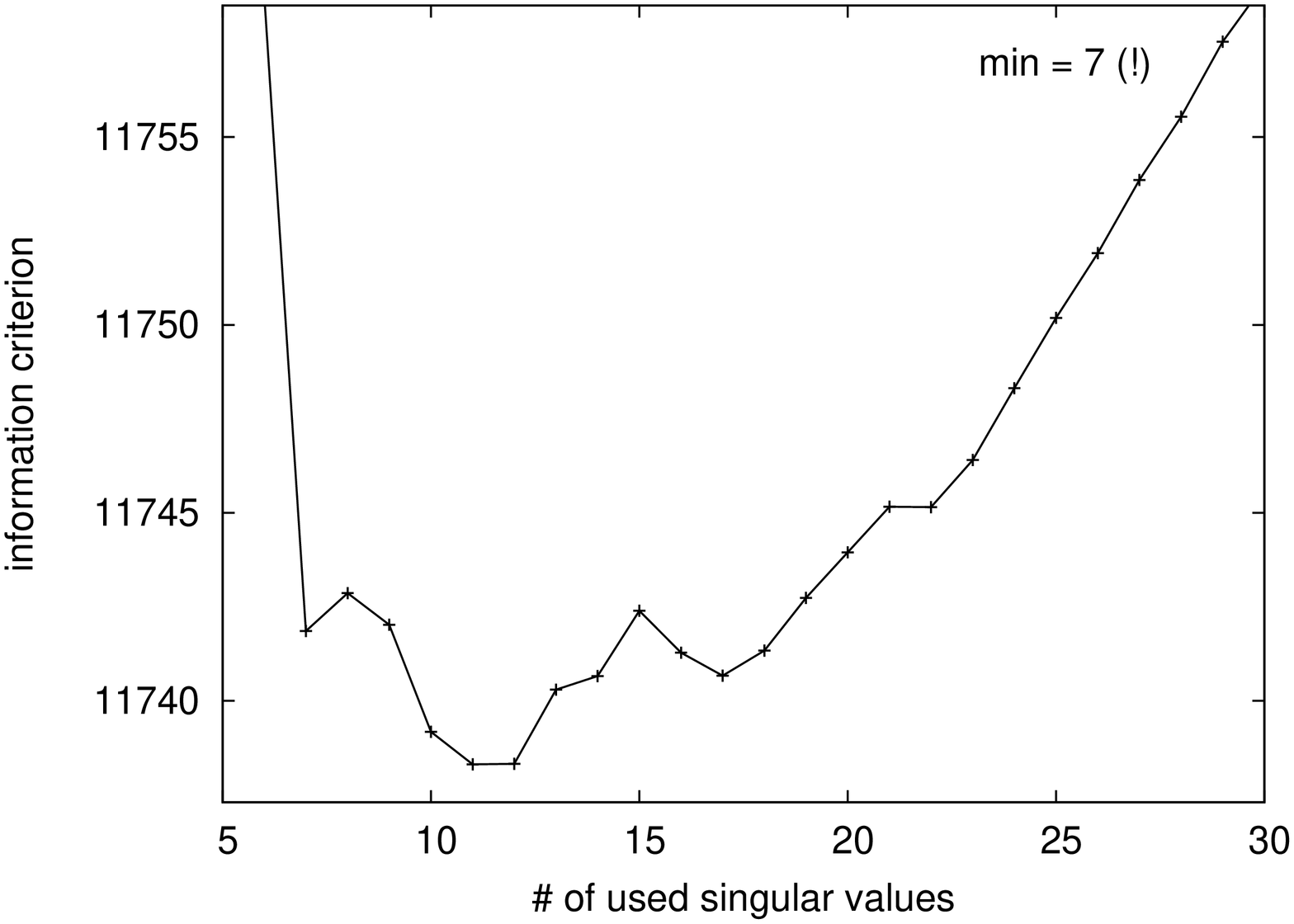}  }
  \caption{\small{
		\textsl{Top:} Lightcurve of the \textsl{Fermi} GRB~091024.380 as measured
		by the triggered GBM detector\,'9' and the fitted background
		with a grey line. Burst interval: [-200:600].
				\textsl{Bottom left:} Underlying variables (absolute values). See Sec.~\ref{sec:sources}.
				\textsl{Bottom right:} Akaike Information Criterion, the smallest local minimum of 7 singular values is used here. See Sec.~\ref{sec:modelsel}.   
  }}
  \label{fig:091024380}
  \resizebox{.8\hsize}{!}{\includegraphics[angle=270]{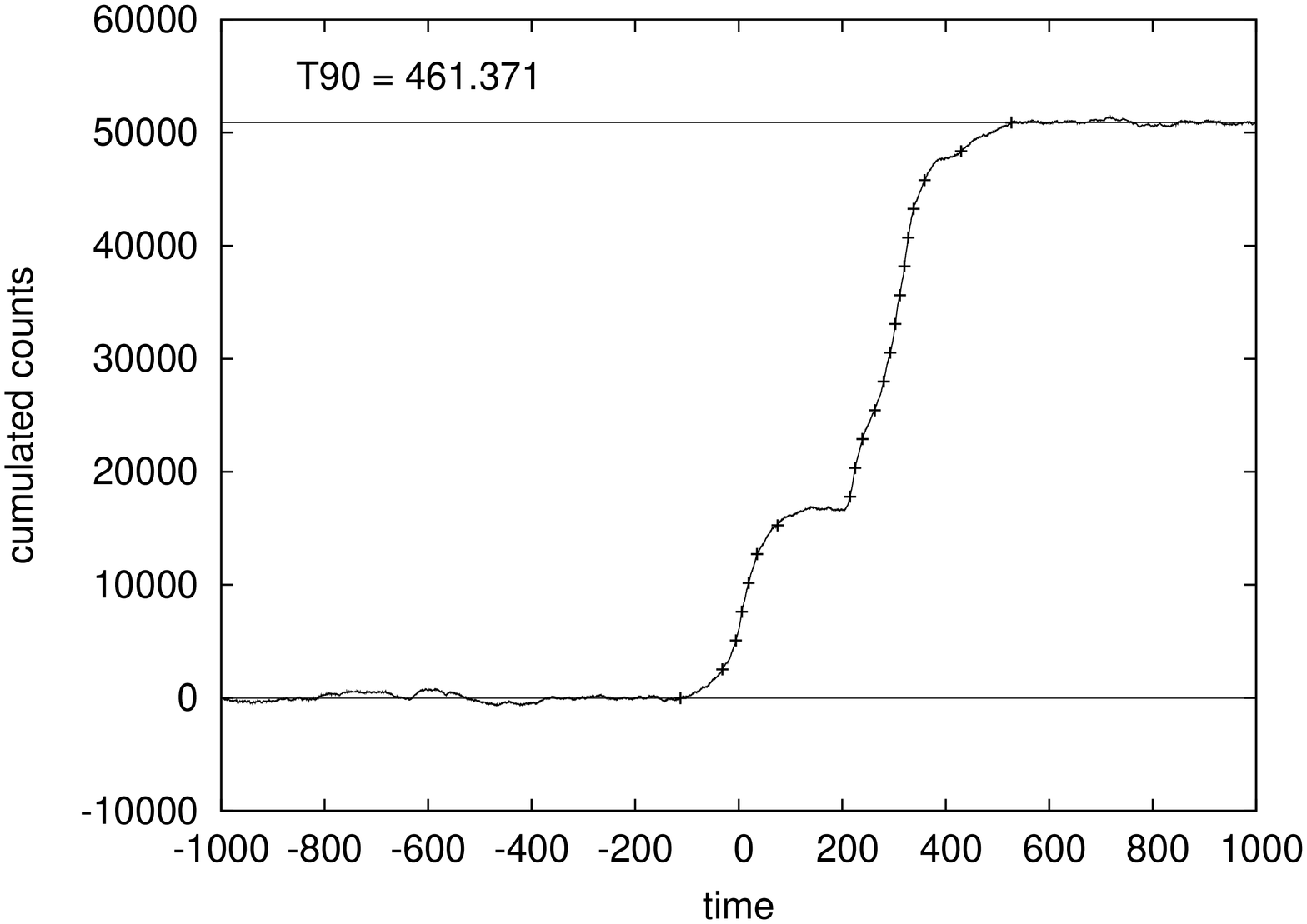} }
  \caption{\small{Cumulative lightcurve of GRB~091024.380. Horizontal lines are
  drawn at 0\% and 100\% of total cumulated counts; dots mark every 5\%. }}
  \label{fig:091024380INT}
\end{center}
\end{figure}

The second burst emission is after 630~sec in Fig.~\ref{fig:091024372}. As we already mentioned, this second emission resulted a second trigger from the GBM (.380), which is shown in Fig.~\ref{fig:091024380} using the data of the triggered detector '9'. Here, the first trigger is visible at -630~sec. However, it is less intensive, since detector '9' was not triggered with the first emission. 	
	
This second burst was so long (GBM Catalogue reported T$_{90}^{cat}$=450.569\,sec \citep{catalogue}) that we needed to reconsider the best model given by AIC. The minimum of AIC as a function of the used singular values is at 11, but this model has a large polynomial loop in the burst interval and is, therefore, useless. Although this is understandable, longer burst intervals lead to shorter fitted backgrounds (and thus, a large amount of information can be lost), it implies that the information criterion has to be used with caution, especially in extreme cases. In this case, we chose the model with 7 singular values. This model fits the background considerably well according to our experience, and is supported by the information criterion: The smallest local minimum is at 7. 
	
The ARR was issued at 09:12:14.28 UT, $\sim$970~sec after the first trigger (.372) and $\sim$350~sec after the second trigger (.380) \citep{Gruber}. A small change in the underlying variables at 350~sec in Fig.~\ref{fig:091024380} can be seen, but the ARR slew was not too large, since the source was already at 15~degrees from the LAT boresight. Nonetheless, the effect of the ARR is represented by the fitted model, as seen by the small knot of the grey line at 350-400~sec in Fig.\ref{fig:091024380}. As for the cumulative lightcurve in Fig.~\ref{fig:091024380INT}, the first emission at -630~sec is present with a non-significant sign, otherwise our result of T$_{90}$=461.371$^{+48.575}_{-71.535}$\,seconds agrees with the GBM Catalogue. 
	
\subsubsection{GRB~100130.777}\label{sec:Belryn}		
The \textsl{Fermi} GRB~100130B was detected by the GBM on 10 January 2010 at 18:38:35.46\,UT. The GBM GRB Catalogue presented T$_{90}^{cat}$=86,018$\pm$6,988\,sec
\citep{catalogue}.  We analyse the data of triggered NaI detector\,'8' using
DDBF.

\begin{figure}[h!]\begin{center}
		\resizebox{.8\hsize}{!}{\includegraphics[angle=270]{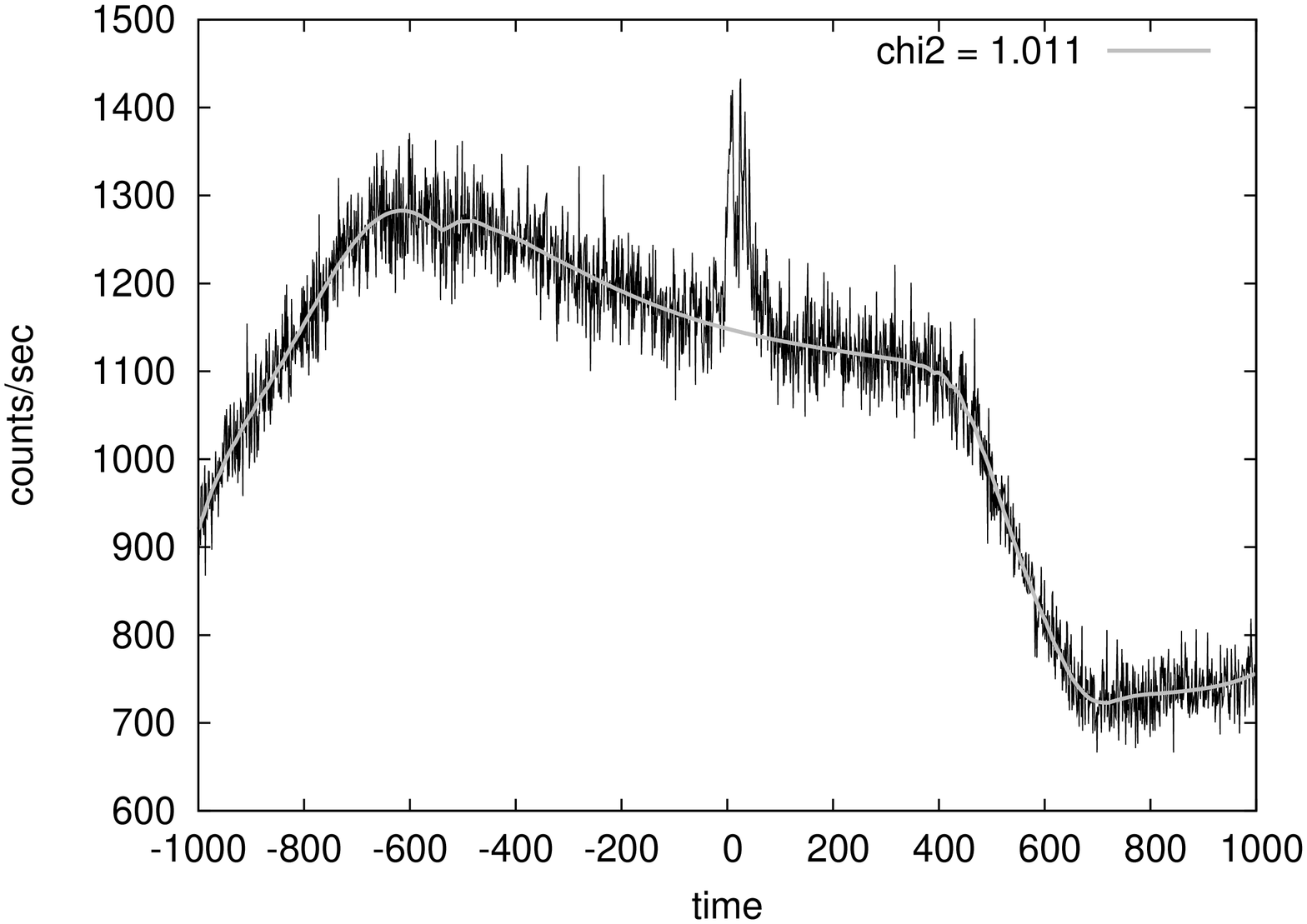}}
		\resizebox{\hsize}{!}{	\includegraphics[angle=270]{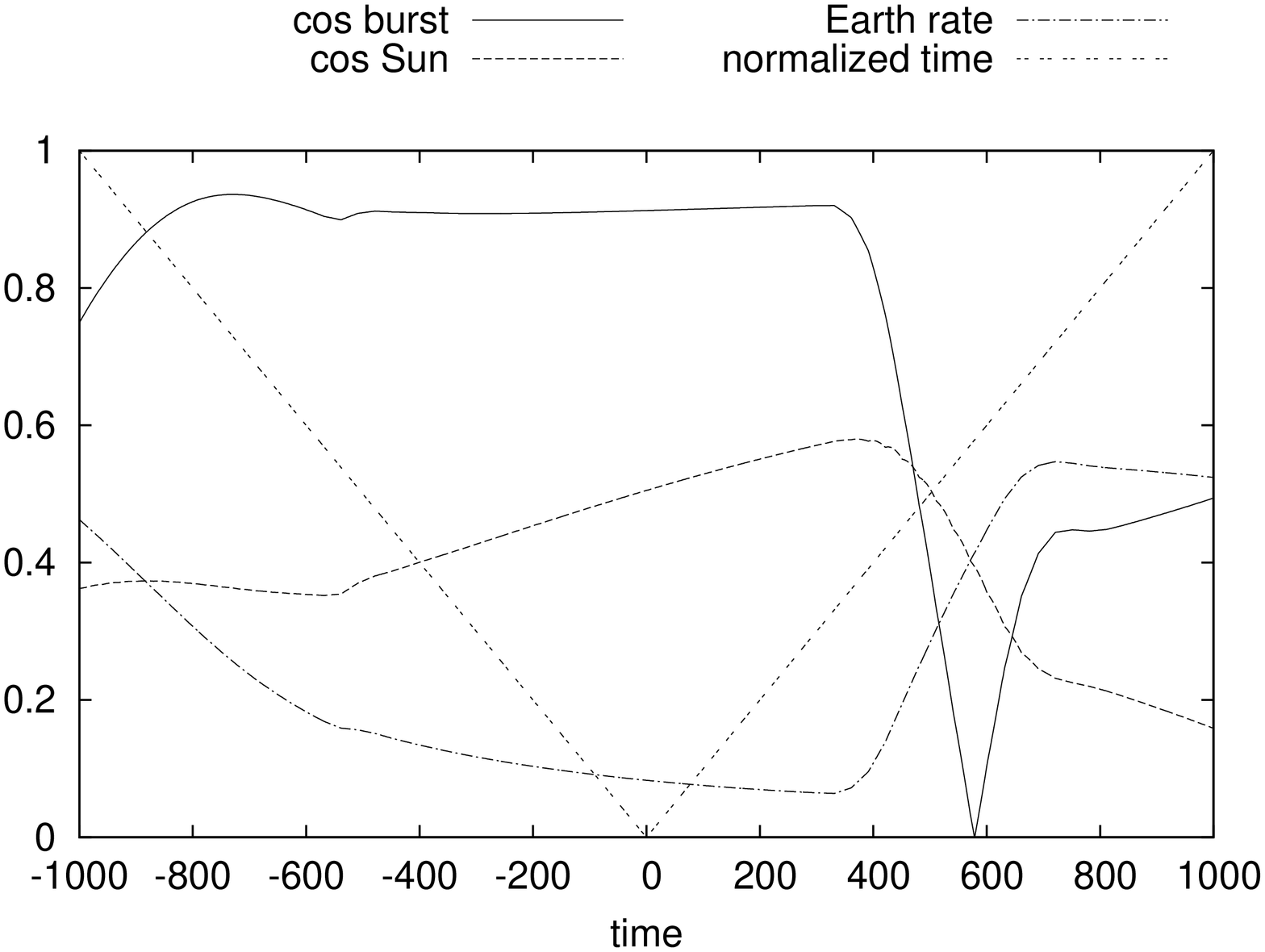}
								\includegraphics[angle=270]{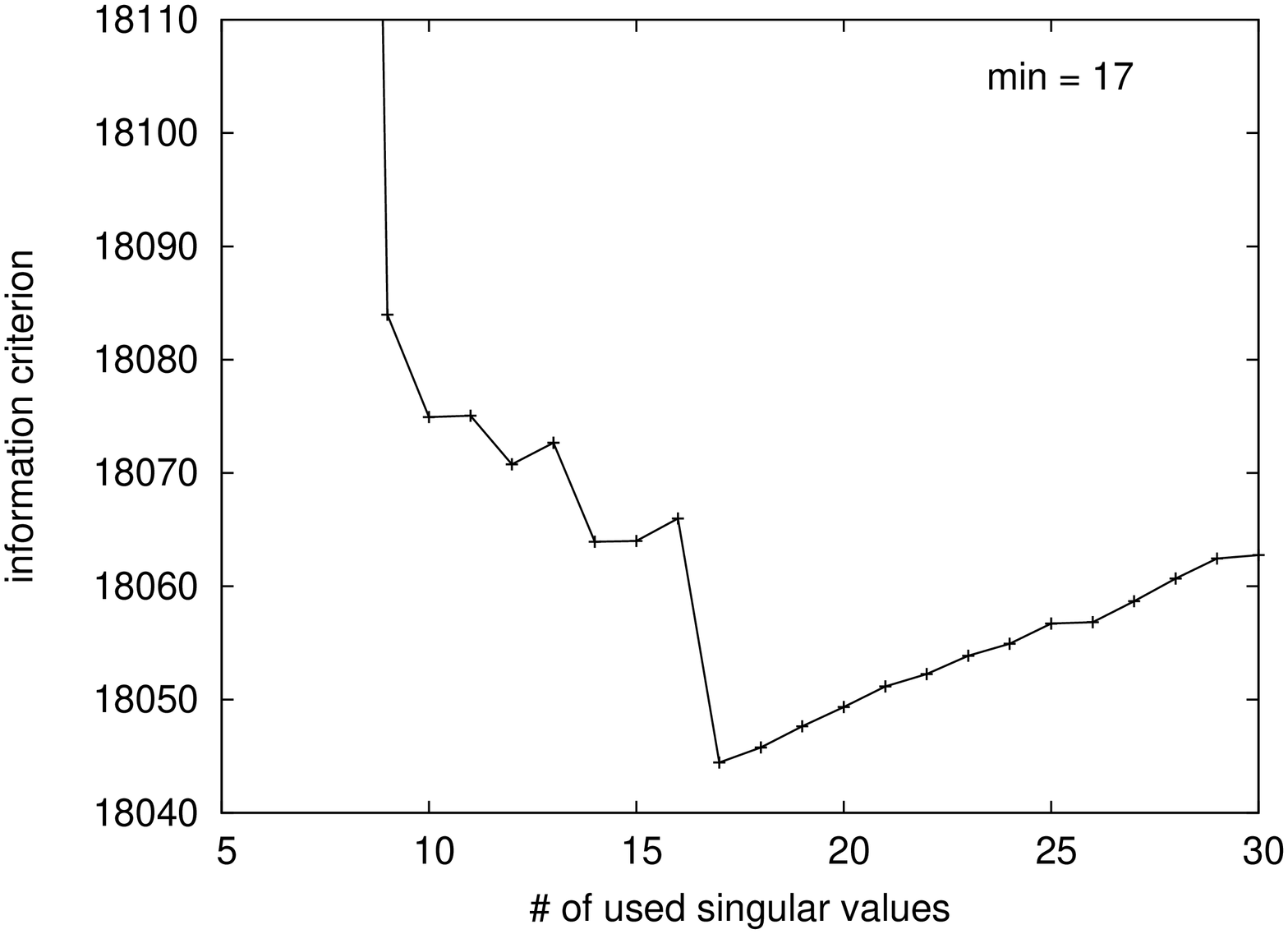}  }
		\caption{\small{
				\textsl{Top:} Lightcurve of the \textsl{Fermi}
				GRB~100130.777 as measured by the triggered
				GBM detector\,'8' and the fitted background
				with a grey line. Burst interval: [-30:90].
				\textsl{Bottom left:} Underlying variables (absolute values). See Sec.~\ref{sec:sources}.
				\textsl{Bottom right:} Akaike Information Criterion. See Sec.~\ref{sec:modelsel}. 
				}}
		\label{fig:100130777}
		\resizebox{.8\hsize}{!}{	\includegraphics[angle=270]{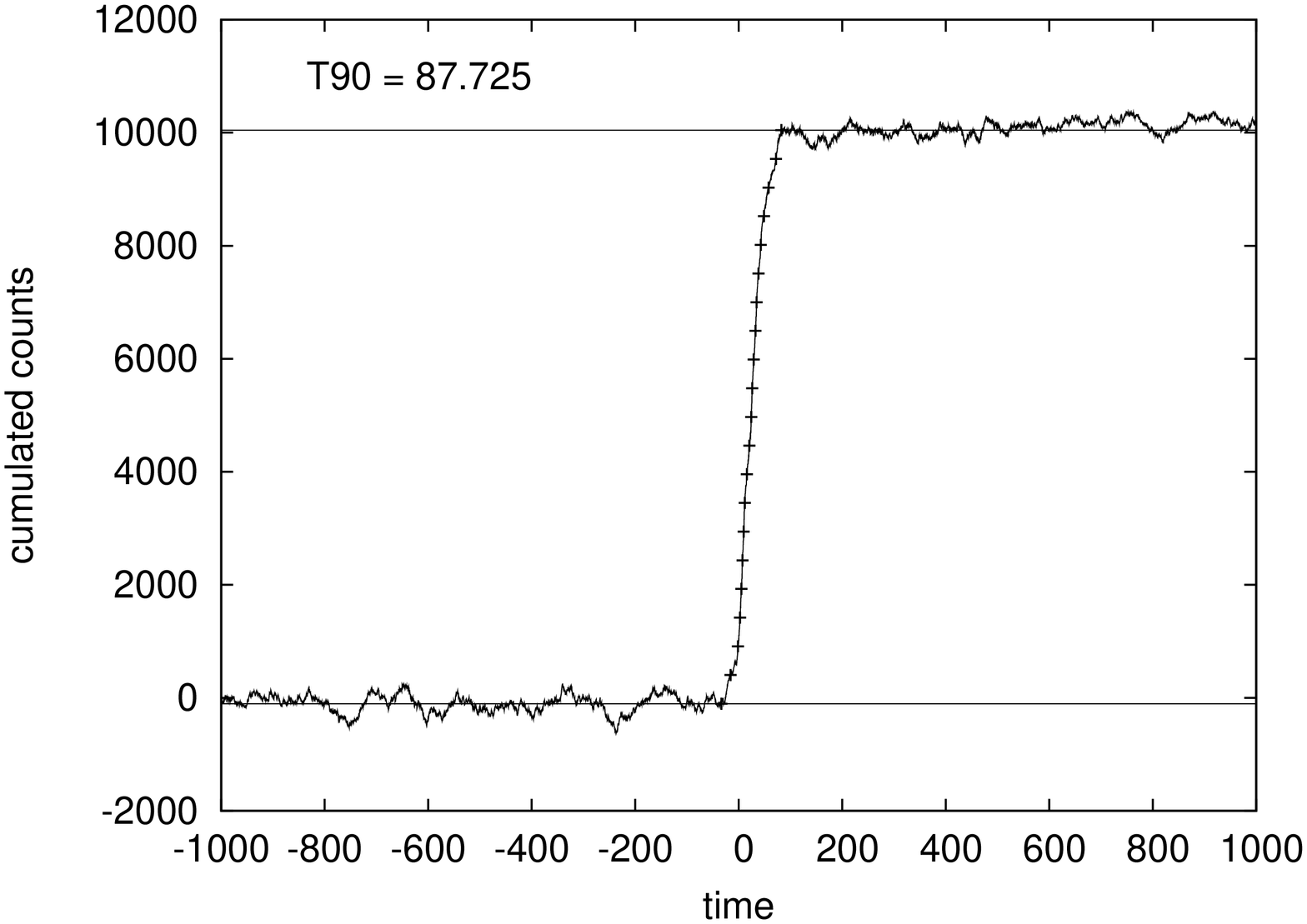}}
		\caption{\small{
				Cumulative lightcurve of GRB~100130.777.
				Horizontal lines are drawn at 0\% and 100\% of
				total cumulated counts; dots mark every 5\%.
  }}
		\label{fig:100130777INT}
\end{center}
\end{figure}
Although the background does not change extremely during the $\sim$\,80\,sec
of the burst, it is a good example to present the contribution of the
celestial position of the satellite to the actual level of the background. If
one takes a look at Fig.~\ref{fig:100130777}, one can see that the variation
in the lightcurve has a connection to the variation in the underlying
variables. 

AIC gives us a best model of 17 singular values.  After the background
subtraction, the cumulative lightcurve (Fig.~\ref{fig:100130777INT}) gives us
T$_{90}$=87.725$^{+5.311}_{-4.911}$\,sec. For error estimation, see Sec.~\ref{sec:error}.


\subsubsection{GRB~100414.097}\label{sec:Niley} 
This GRB also had an ARR event. Quoting the GCN report 10595: \textsl{'At 02:20:21.99 UT on 14 April 2010, the \textsl{Fermi} Gamma-Ray Burst Monitor
triggered and located GRB 100414A.
The \textsl{Fermi} Observatory executed a maneuver following this trigger and
tracked the burst location for the next 5 hours, subject to
Earth-angle constraints.'} \citep[10595]{GCN}

In this case, we chose to analyse a non-triggered detector (detector '5'). Because this burst was so intensive and bright, the triggered detectors show totally negligible background rate variations compared to the brightness of the burst. Since we want to demonstrate that our method works in very complicated cases as well, we analyse a lower signal-to-noise detector. Evidently, DDBF can also fit the data of the bright triggered detectors well.

\begin{figure}[h!]\begin{center}
  \resizebox{.8\hsize}{!}{\includegraphics[angle=270]{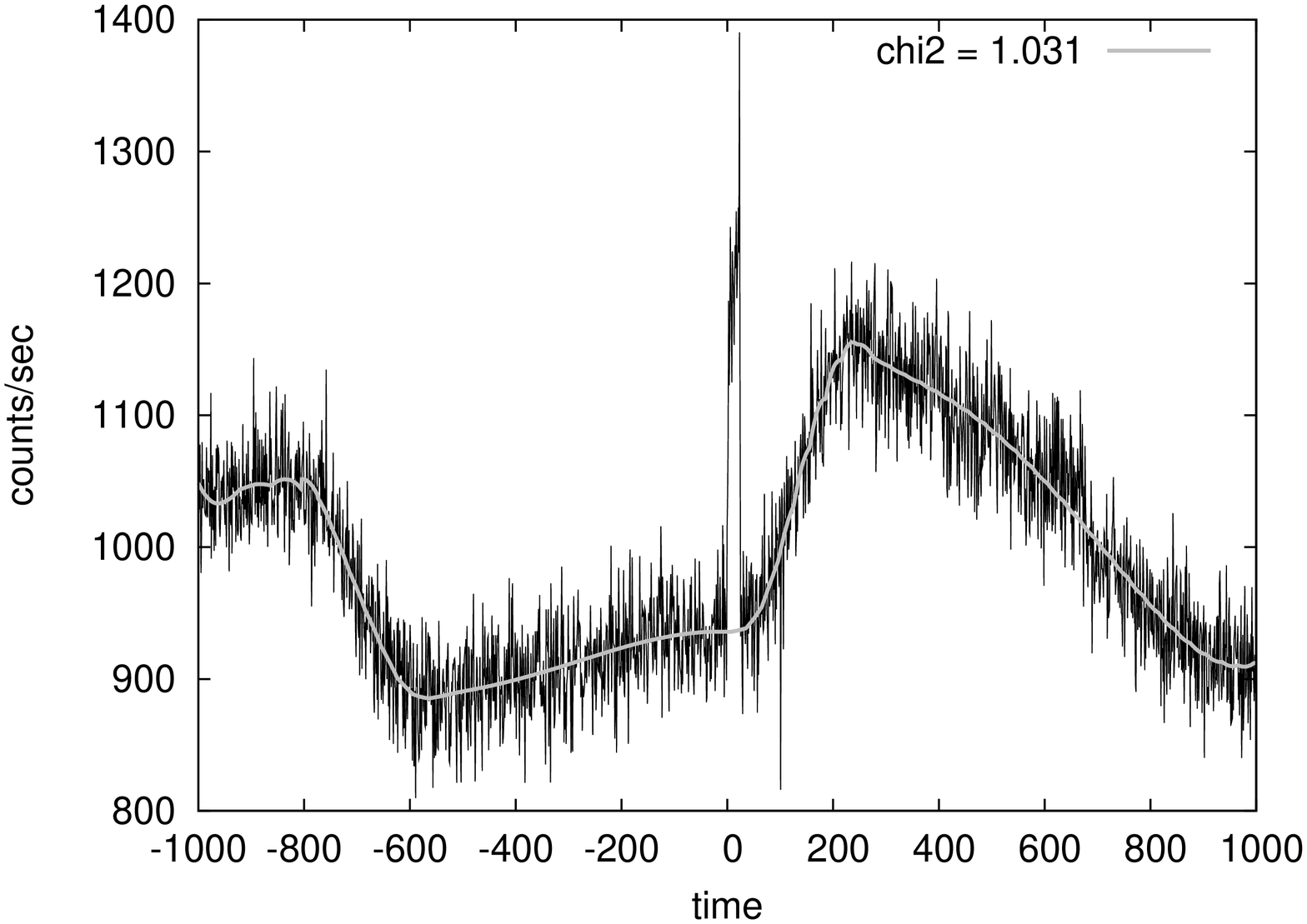}}
  \resizebox{\hsize}{!}{\includegraphics[angle=270]{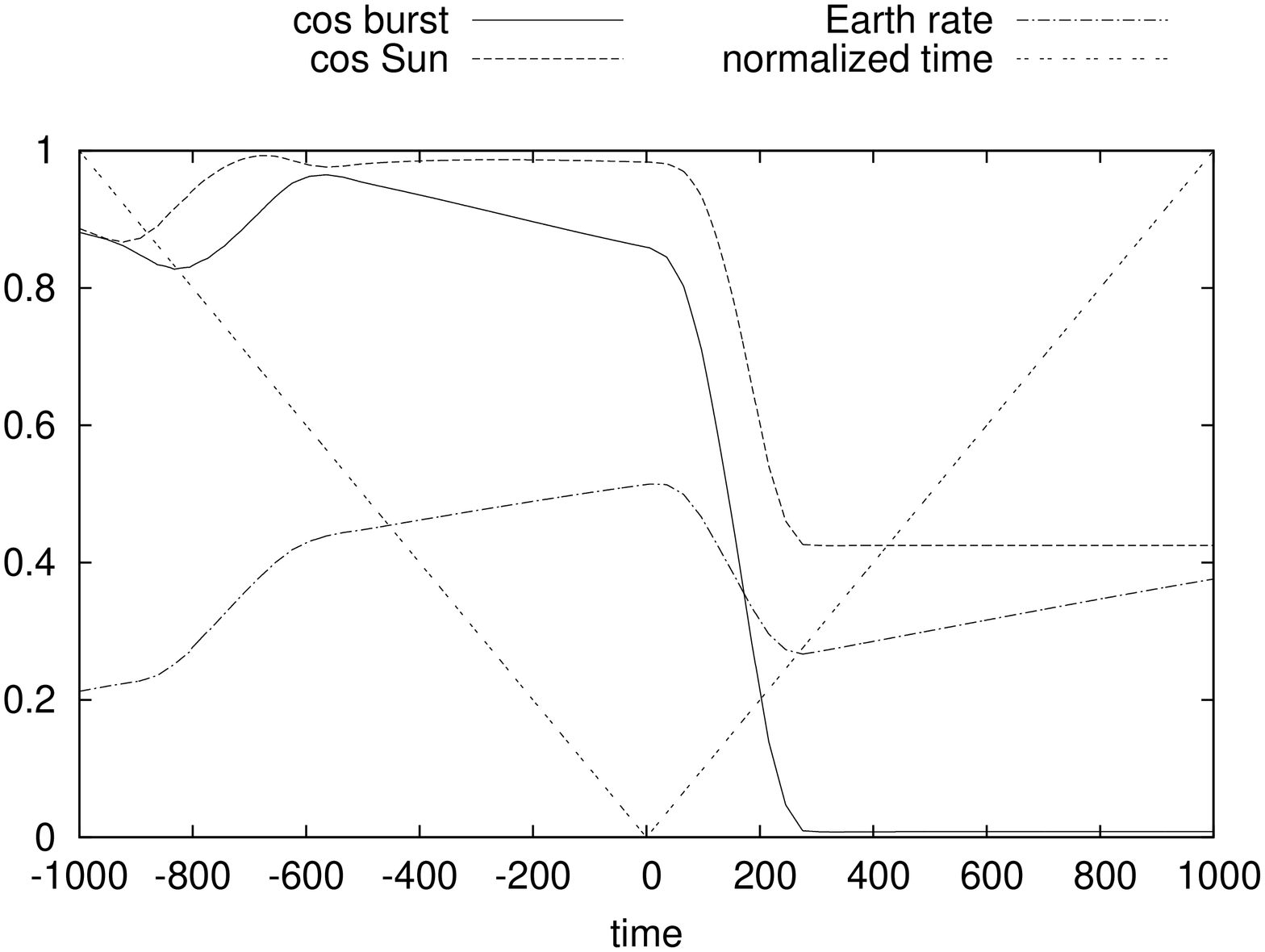}
  			\includegraphics[angle=270]{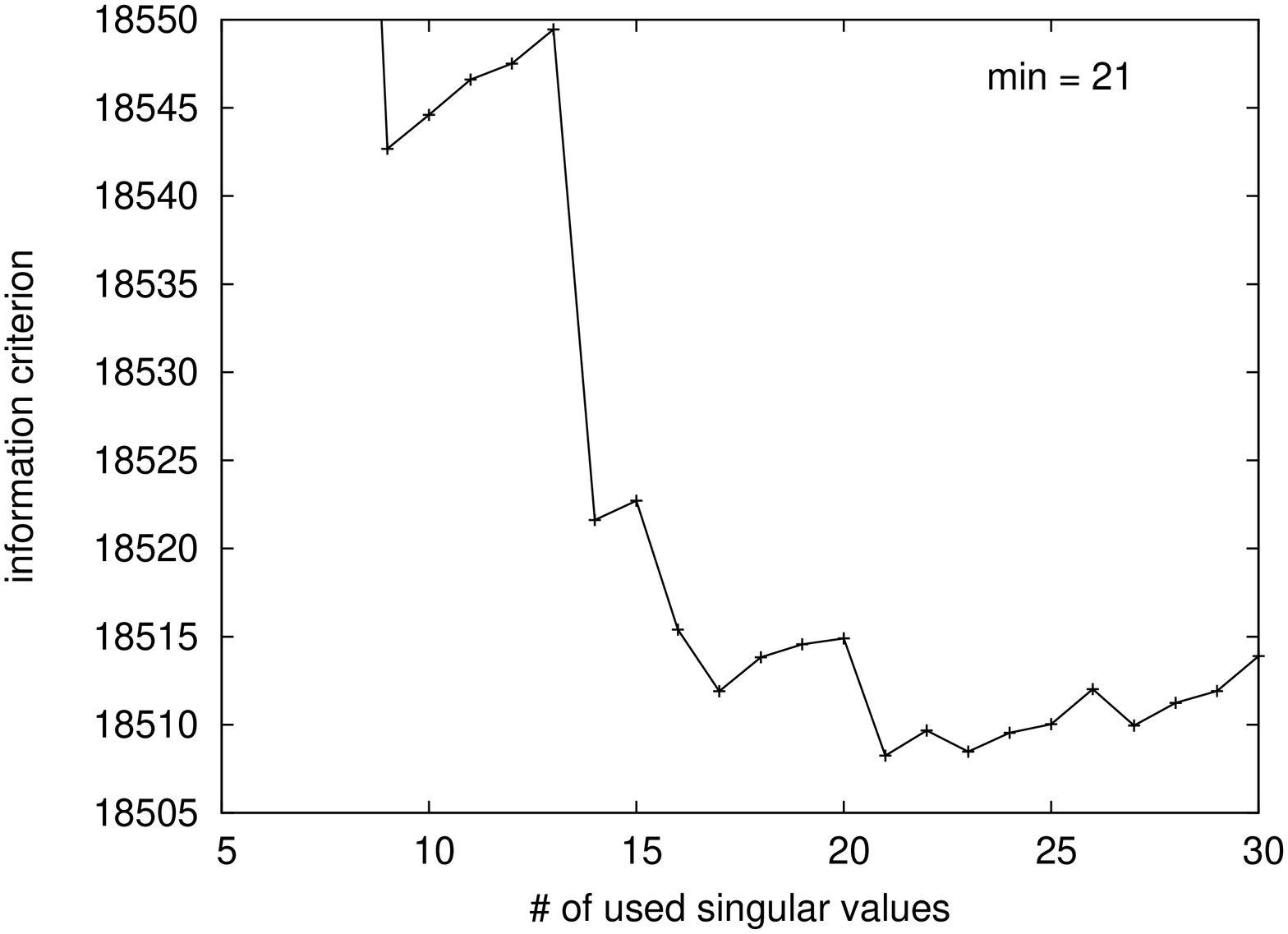}  }
  \caption{\small{
		\textsl{Top:} Lightcurve of the \textsl{Fermi} GRB~100414.097 as measured
		by the non-triggered GBM detector\,'5' and the fitted background
		with a grey line. Burst interval: [-20:30].
				\textsl{Bottom left:} Underlying variables (absolute values). See Sec.~\ref{sec:sources}.
				\textsl{Bottom right:} Akaike Information Criterion. See Sec.~\ref{sec:modelsel}.   
  }}
  \label{fig:100414097}
  \resizebox{.8\hsize}{!}{\includegraphics[angle=270]{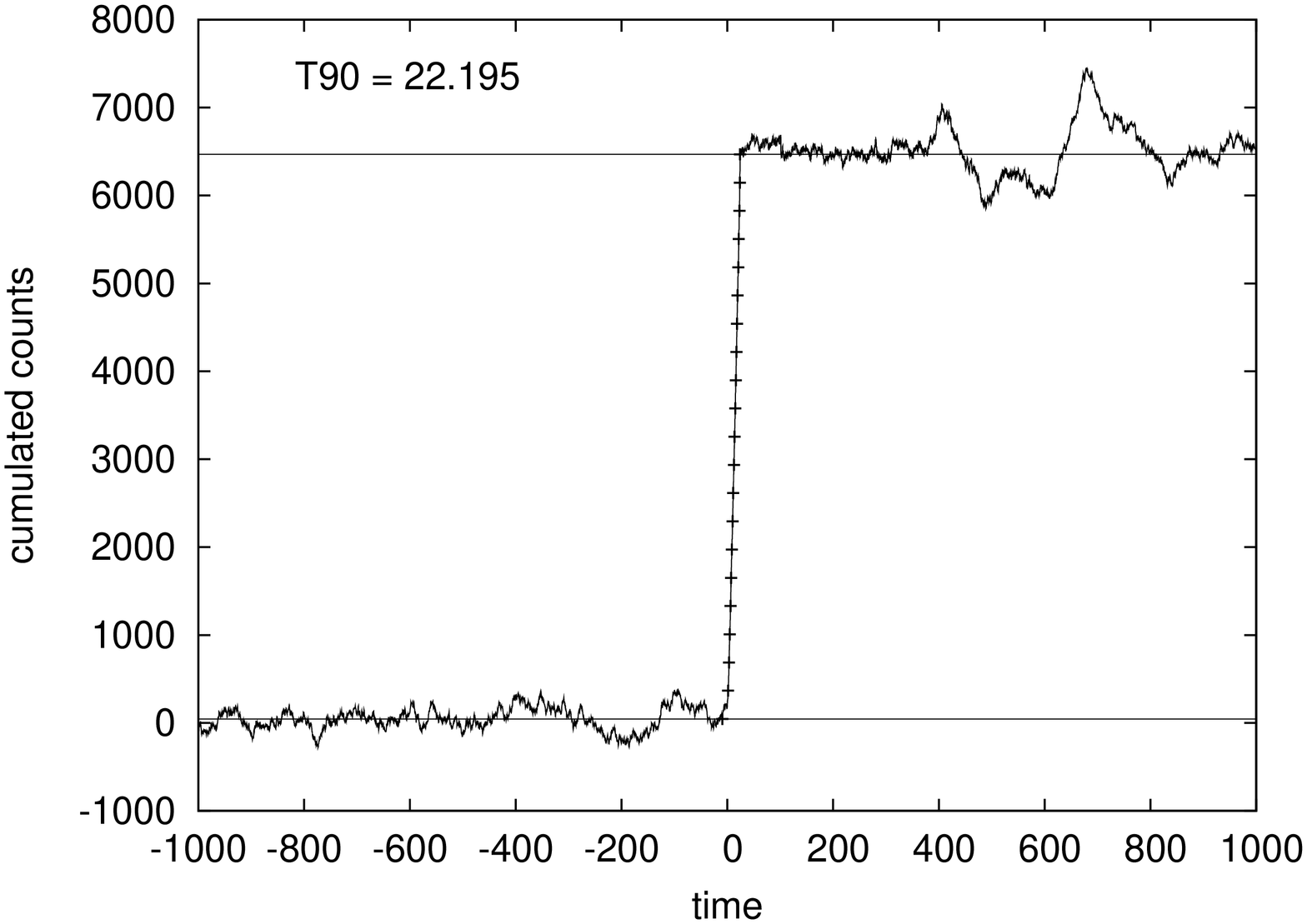} }
  \caption{\small{Cumulative lightcurve of GRB~100414.097. Horizontal lines are
  drawn at 0\% and 100\% of total cumulated counts; dots mark every 5\%. }}
  \label{fig:100414097INT}
\end{center}
\end{figure}

The GBM Catalogue reports 
a duration of T$_{90}=$26.497$\pm$2.073\,sec. 
According to the GCN 10594 and 10610, this burst also triggered the LAT and the \textsl{Suzaku Wide-band All-sky Monitor (WAM)} \citep[10594,10610]{GCN}. 

As we already mentioned above, singular values that are too high ($\gtrsim$20) deserve attention. In this case, the AIC chose 21 singular values. This 21 singular value model describes the background well. The only exception is the extra count rate around 600~sec, which is also clearly noticeable in the cumulative lightcurve. More detailed analysis of the spectral features of this event are needed to determine if this event is caused by the burst or not. Given that there were additional observations by the LAT and by the Suzaku WAM which do not report 
long emission, 
we expect that this was a local event at the GBM caused by cosmic rays or another possible transient source, which could be filtered by using different energy channels.
Our result is T$_{90}$=22.195$^{+2.149}_{-1.421}$\,sec.

%% file: error.tex
\begin{table*}[t]\begin{center}
    \hfill{}
    \caption{Final T$_{90}$ and T$_{50}$ results.}
	\begin{tabular}{|l||c||c|c|c|c||c|c|c|c|}\hline 
		Burst & sing.v. & T$_{90} (s) $ & \multicolumn{2}{c|}{Conf. int. $(s)$} & T$_{90}^{cat} (s) $ & T$_{50} (s) $ & \multicolumn{2}{c|}{Conf. int. $(s)$} & T$_{50}^{cat} (s)$ \\  \hline \hline
		090102.122 &9& 29.756  & $ {+2.971}$&${-1.198}$ & 26.624$\pm$0.810 & 10.859 & $ {+0.531}$&${-0.556}$ & 9.728$\pm$0.572\\ \hline
		090113.778 &12& 19.679  & $ {+10.883}$&${-6.421}$ & 17.408$\pm$3.238 & 6.408 & $ {+0.498}$&${-0.344}$ & 6.141$\pm$1.446\\ \hline
		090618.353 &15& 103.338 & $ {+3.842}$&${-6.725}$ & 112.386$\pm$1.086& 22.827 & $ {+2.201}$&${-1.530}$ & 23.808$\pm$0.572\\ \hline
		090828.099 &7& 63.608  & $ {+1.467}$&${-1.652}$ & 68.417$\pm$3.167& 11.100 & $ {+0.198}$&${-0.194}$ & 10.752$\pm$0.320\\ \hline
		091024.372 &26& 100.013  & $ {+7.908}$&${-4.156}$ & 93.954$\pm$5.221& 41.896 & $ {+2.987}$&${-1.731}$ & 39.937$\pm$1.056\\ \hline
		091024.380 
&7& 461.371  & $ {+48.575}$&${-71.535}$ & 450.569$\pm$2.360& 283.202 & $ {+7.360}$&${-65.306}$ & 100.610$\pm$0.923\\ \hline		
		091030.613 &14& 22.609  & $ {+13.518}$&${-4.522}$ & 19.200$\pm$0.871& 10.770 & $ {+0.388}$&${-0.424}$ & 9.472$\pm$0.345\\ \hline
		100414.097 &21& 22.195  & $ {+2.149}$&${-1.421}$ & 26.497$\pm$2.073& 11.468 & $ {+0.549}$&${-0.906}$ & 13.248$\pm$0.272\\ \hline
		100130.777 &17& 87.725  & $ {+5.311}$&${-4.911}$ & 86.018$\pm$6.988& 30.829 & $ {+1.317}$&${-1.928}$ & 34.049$\pm$1.493\\ \hline
	\end{tabular}
	\hfill{}
	\tablefoot{{Final T$_{90}$ and T$_{50}$ results, confidence intervals 	
(see Sec.~\ref{sec:error} and \citet{Monterey}), 
and the number of singular values (Sec.~\ref{sec:svd}) found with Akaike Information Criterion (Sec.~\ref{sec:modelsel}) for the bursts analysed in this paper (Sec.~\ref{sec:result}). We also show the duration value of T$_{90}^{cat}$ and T$_{50}^{cat}$ of the GBM Catalogue \citep{catalogue} for comparison (Sec.~\ref{sec:catal}).}}
	\label{tab:err}
\end{center}
\end{table*}

The DDBF method described above is too complicated to give a simple expression
for the error of T$_{90}$ using general rules of error propagation. We
therefore decided to give confidence intervals corresponding to 68\% (approximately
$1\sigma$ level).  
For this, we use Monte Carlo (MC) simulations. We simulate the data with Poisson noise: Assuming that counts are given by a Poisson process, 
we exchange our input data to one
coming from a random Poisson distribution. 
In the case of a Poisson distribution, which is parametrised
by the mean rate ($\lambda$), the expected value is given by $\lambda$. We therefore
replace each datapoint with a value drawn from a Poisson distribution with a
mean equal to the datapoint in question.	

\begin{figure}[h!]\begin{center}
  \resizebox{.8\hsize}{!}{\includegraphics{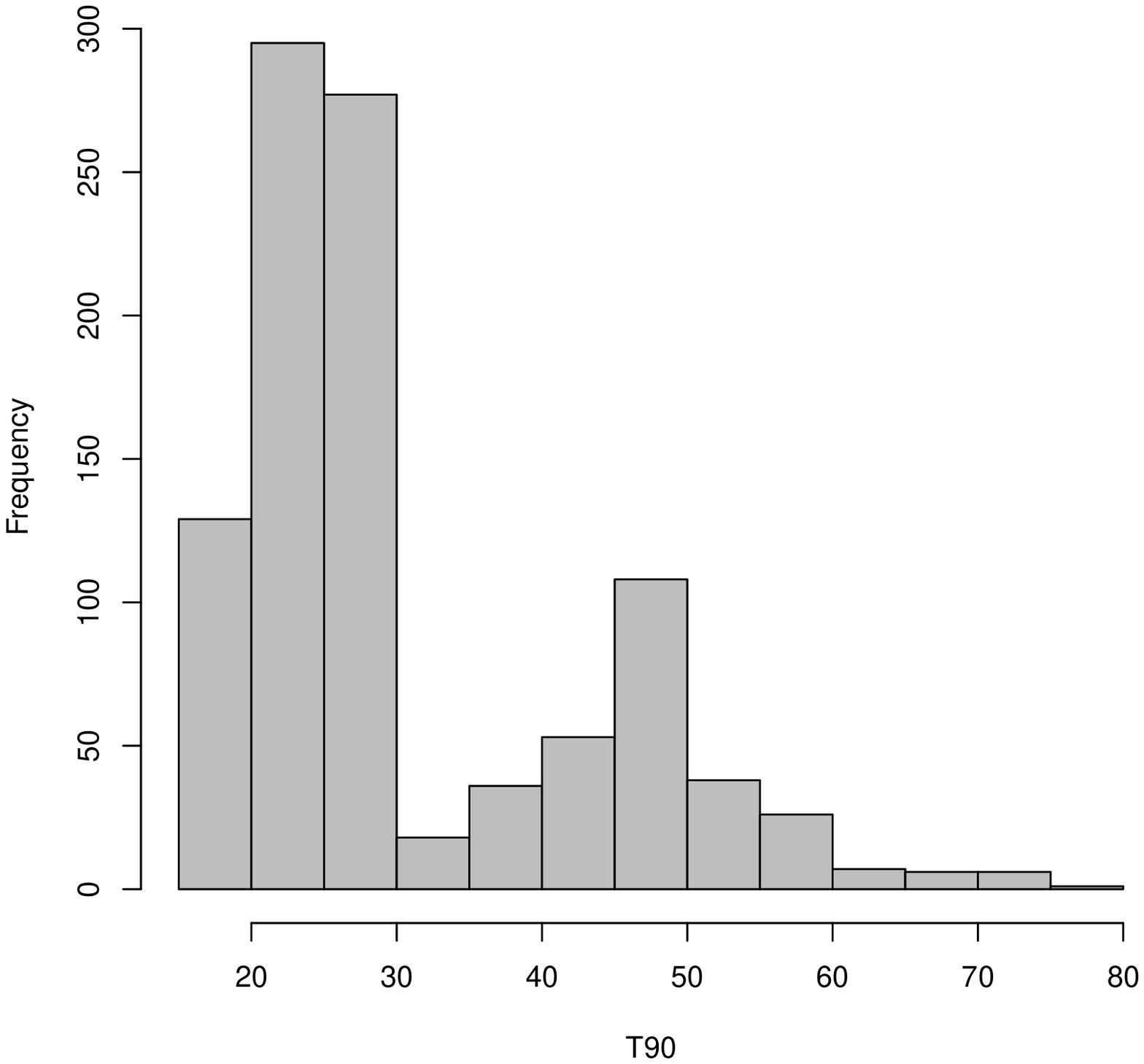}}
  \caption{\small{Distibution of the T$_{90}$ obtained from the MC simulated
  data for \textsl{Fermi} burst 091030.613   
  \citep{Monterey}.
  }}
  \label{fig:pois90}
\end{center}\end{figure}

\begin{figure}[h!]\begin{center}
  \resizebox{.8\hsize}{!}{\includegraphics{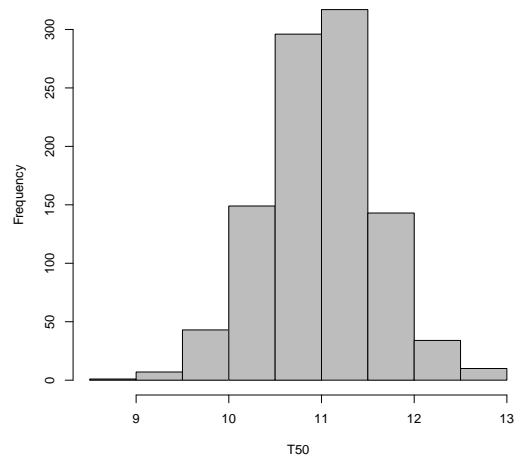}}
  \caption{\small{Distibution of the T$_{50}$ obtained from the MC simulated
  data for the \textsl{Fermi} burst 091030.613   
  \citep{Monterey}.
  }}
  \label{fig:pois50}
\end{center}\end{figure}

DDBF was repeated for 1000 MC simulated data. 
The distribution 
of the
Poisson-modified T$_{90}$ and T$_{50}$ values are shown in
Fig.~\ref{fig:pois90} and Fig.~\ref{fig:pois50} for GRB 091030.613, respectively.


Fig.~\ref{fig:pois90} shows two significant peaks around 22 and 47 seconds.
The first peak at 22 seconds corresponds to the measured T$_{90}$ value.
However, the measured T$_{90}$
value is systematically longer in some cases of the Poisson noise simulation, because this burst has a little pulse
around 47 seconds (see Figs.~\ref{fig:fit.eps}~and~\ref{fig:int.eps}), 
and T$_{90}$ is sensitive for this kind of uncertainties. 
In Fig.~\ref{fig:pois50}., there is, however, no sign of this second peak:
T$_{50}$ is more robust and less likely to be influenced by these
fluctuations \citep{Monterey}.

Final results of T$_{90}$s and T$_{50}$s with confidence intervals are given
in Table~\ref{tab:err} for the bursts mentioned in Sec.~\ref{sec:example}.

%% file: catal.tex
In Table~\ref{tab:err}, we also show the T$_{90}^{cat}$s and T$_{50}^{cat}$s of the \textsl{Fermi} GBM Catalogue \citep{catalogue} for comparison.

At this point, we need to give some notes about the differences between the method of the Catalogue and DDBF. First of all, we only used one detector when we measured the duration, whilst the Catalogue used
the sum of the brightest detectors.
	
On the other hand, there are further differences between the Catalogue's method and the DDBF. As we mentioned in Sec.~\ref{sec:result}, our method solved the problem of automatizing the identification of the 0\% and 100\% levels of cumulated counts, so the user do not need to define them by hand. This disposes of one possible error source. 

Additionally, using direction dependent variables produced the possibility of fitting the whole CTIME background (only the burst has to be taken off in the middle). This reduces the error of the user selected background intervals and, on the other hand, makes the automatic detection of a long emission possible. See Sec.~\ref{sec:features}. for more details.

With respect to the error estimation of the Catalogue,
they followed the method developed for the BATSE data by \citet{Koshut}, which uses the variance of the 0\% and 100\% levels of cumulated counts as a basis for the error estimates \citep{catalogue}. We decided to avoid this method (as we avoid the use of time-dependent polynomial methods developed for the BATSE, as seen in Sec.~\ref{sec:problem}), and give an alternative solution with Monte Carlo simulation of the data in Sec.~\ref{sec:error}. This choice is based on our belief that the DDBF is too complicated, and using the error estimation of \citet{Koshut} would underestimate the real error of our method.

Furthermore, we give different higher and lower confidence intervals. In our experience, many bursts show different amounts of uncertainties at the starting point than at the finishing point. One demonstrative example is the T$_{90}$ value of GRB~091030.613: the MC modified distribution in Fig.~\ref{fig:pois90} is clearly not symmetric. Therefore, it would be an oversimplification to give only one value as an error bar or confidence interval. For more examples, see \citet{Monterey}.
	
Given all of these facts, it follows that a comparison with the \textsl{Fermi} GBM Catalogue data is not meaningful in a quantitative sense at the moment. It is currently under way to process all \textsl{Fermi} bursts with DDBF and publish an alternative catalogue, in which we will use the combined data of the detectors. Unfortunately, we cannot say anything about the robustness of our method until we finish processing a significant number of bursts. Once it is done, we will provide an overall statistical comparison between the two dataset together with our catalogue.

%% file: discus.tex
Since the commonly used background filtering methods are not efficient 
for many cases of the \textsl{Fermi}, 
we developed a new technique based on the motion and orientation of the
satellite known as the the Direction Dependent Background Fitting (DDBF) method. 

The DDBF technique considers the position of the burst, the Sun and the Earth. Based on this information on position, we computed physically
meaningful underlying variables and fitted a four dimensional hypersurface on
the background. Singular value decomposition and Akaike information criterion
were used to reduce the number of free parameters. More research may be
required to find a more suitable model dimension reducing criterion.

The background model was subtracted from the measured data, resulting 
in
background-free lightcurves. These lightcurves can be used to perform
statistical 
surveys. We showed 
the efficiency of our DDBF method
computing durations of some very complicated cases. 
We also calculated confidence intervals for our duration values corresponding to 1$\sigma$ level.

We summarized some of the main differences between DDBF and the 
background estimation
method of the GBM Catalogue and decided not to give a quantitative comparison at this point. Our plan is to process the combined data of the detectors with DDBF for every \textsl{Fermi} burst and produce an alternative catalogue. This future work will also contain the statistically relevant comparison of the official GBM Catalogue and the DDFB Catalogue which has yet to come.

The DDBF method has the advantage of considering only variables with physical
meanings and it fits all the 2000\,sec CTIME data well as
opposed to the currently used method. 
These features are indeed necessary when
analysing long GRBs, 
where motion effects 
can influence the background rate in a very extreme way. 
Therefore, not only Sky Survey but also ARR mode
GRB's can be analysed, 
and possible long 
emission can be detected. 
	
Furthermore, there seems to be no reason why DDBF could not be used for other sources than GRBs. The method only considers the background levels before and after the event; therefore, the event itself has no influence to the resulted background model, even if it is very bright. Nevertheless, the duration can play a role in its applicability. Events that are comparably long to the 2000\,sec data file could 
be problematic.	
The DDBF is not necessary for short events, as
the effects of the motion of the spacecraft are negligible: 
One may use the time dependent polynomial fitting for short GRBs. However, DDBF is able to discover long emissions or prebursts, as we have shown in Sec.~\ref{sec:example}. Therefore, DDBF could be used to verify the final
result in the case of short bursts as well.

In summary, celestial position plays an important role in the \textsl{Fermi} data set.
If one wants to filter the background more efficiently and in a physically
more comprehensible way, one has to use 
this 
information. 
Utilizing this principle, we have created the DDBF method.	
In future work, 
DDBF will be used to create a catalogue of the durations of the \textsl{Fermi} GBM GRBs.

%% file: appendix.tex
In Sec.~\ref{sec:earth}, we defined one of the underlying variables as the
Earth-occulted sky rate -- i.e., the Earth-uncovered sky correlated to the size of the detector's Field of View (FoV). Here, we present the computations.

Let us have $R$ as the radius of Earth and $h$ as the altitude of the satellite.
(The actual $h$ during the burst is known from the LAT spacecraft data file.)
The aperture $\sigma$ of the cone constituted by the Earth-limb seen from the
board of satellite is

\begin{equation}
	\sigma=\mbox{asin} \frac{R}{R+h}.
\label{sigma}
\end{equation}

Angular dependence of the detector effective area is assumed to be constant,
so the FoV of one GBM detector is $2\pi$ sterad. However, more
precise calculations could be done knowing the real characteristics \citep{Meegan}.

When the Earth-limb is totally in the FoV, the Earth-covered area is computed by
integrating on a spherical surface as follows,
\begin{equation}
\displaystyle \Omega_{total}(\sigma)= 
\int_0^{2\pi} \int_0^{\sigma} \sin \theta \, d \theta \, d \phi  = 
2\pi\left(1 -\cos \sigma \right).
\label{Os}
\end{equation}
Eqn.~(\ref{Os}) means the solid angle of a cone of aperture $\sigma$.

\begin{figure}[h!]\begin{center}
  \resizebox{.65\hsize}{!}{\includegraphics{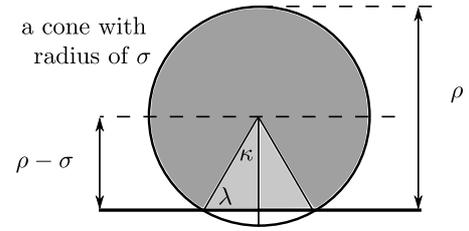}}
  \caption{Earth limb seen onboard from the \textsl{Fermi}. Detector can only see the
  coloured parts above the solid horizontal black line.}
  \label{fig:spheric.eps}
\end{center}\end{figure}

If only a fraction of the Earth-limb is in the FoV, then
$\Omega=\Omega(\sigma,\rho)$ is smaller then $\Omega_{total}$ and is a
function of the maximum altitude of the Earth-limb $\rho$ as well. In this
case, we have to separate the area in the FoV to two parts, which are marked with
light grey and dark grey on Fig.~\ref{fig:spheric.eps}. 

We can calculate the dark grey surface the same way as above. Using
$2\pi-2\kappa$, instead of $2\pi$ when integrating with respect to $\phi$, we find that
\begin{equation}
\displaystyle \Omega_{darkgrey}(\sigma,\rho)= 
2(\pi-\kappa)\left(1 -\cos \sigma \right),
\label{konkav}
\end{equation}
where $\kappa$ is a function of $\rho$ and $\sigma$.  It is easy to see that
the light grey triangle in Fig.~\ref{fig:spheric.eps} is a spherical triangle,
since its every side is a geodetic curve. Therefore, $\kappa$ can be
calculated from the Napiers pentagon:
\begin{equation}
\displaystyle 
\kappa=\acos{\left(\frac{\tan{(\rho-\sigma)}}{\tan{\sigma}}\right)}.
\label{lambda}
\end{equation}

Then, we calculate the light grey surface. The area of a spherical triangle is
given by the Girard formule:
\begin{equation}
\displaystyle 
\Omega_{lightgrey}(\sigma,\rho)= -\pi + 2\kappa + 2\lambda,
\label{Girard}
\end{equation}
where $\lambda=\acos{\left(\cos{(\rho-\sigma)}\cdot \sin{\kappa}\right)}$
from the Napiers pentagon.

Thus, the surface above the black line is the sum of the light grey and dark grey
parts:
\begin{eqnarray}
\Omega^{\sigma<\rho}(\sigma,\rho)&=&2\left[\pi - \acos{\left(\frac{\tan{(\rho-\sigma)}}{\tan{\sigma}}\right)}\right] \left(1-\cos{\sigma}\right)  \notag\\
&&-\pi + 2 \acos{\left(\frac{\tan{(\rho-\sigma)}}{\tan{\sigma}}\right)} \notag\\
&&+2\acos{\left(\cos{(\rho-\sigma)} \cdot \sin\acos{
\frac{\tan{(\rho-\sigma)}}{\tan{\sigma}}
}\right)} 
\label{eq:Osr2}
\end{eqnarray}\vspace{20pt}

Eqn.~(\ref{eq:Osr2}) has to be modified a little bit when $\rho<\sigma$: In
this case, the horizontal solid black line is \textsl{over} the half of the
circle, and the light grey triangle has to be \textsl{subtracted} from the integral
calculated from (\ref{Os}) with $2\kappa$ instead of $2\pi$:

\begin{eqnarray}
\Omega^{\rho<\sigma}(\sigma,\rho)=2\left[\acos{\left(\frac{\tan{(\sigma-\rho)}}{\tan{\sigma}}\right)}\right] \left[1-\cos{\sigma}\right] +\pi \notag\\
- 2 \acos{\left(\frac{\tan{(\sigma-\rho)}}{\tan{\sigma}}\right)} \notag\\
- 2\acos{\left(\cos{(\sigma-\rho)} \cdot \sin\acos{\left(\frac{\tan{(\sigma-\rho)}}{\tan{\sigma}}\right)}\right)}.
\label{eq:Osr1}
\end{eqnarray}\vspace{20pt}	

\begin{figure}[h!]\begin{center}
  \resizebox{.8\hsize}{!}{\includegraphics[angle=270]{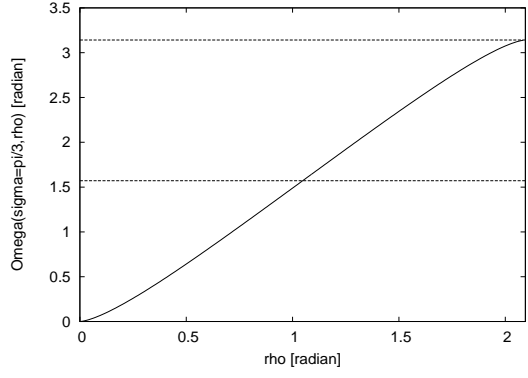}}
  \caption{Plot of equation (\ref{eq:Osr2}) and (\ref{eq:Osr1}) as a function of $\rho$ for $\sigma=\pi/3$.}
  \label{fig:plot}
\end{center}\end{figure}

We plot equation (\ref{eq:Osr2}) and (\ref{eq:Osr1}) as a function of $\rho$
for $\sigma=\pi/3$, as seen in Fig.~\ref{fig:plot}.
Equations.~(\ref{eq:Osr2})~and~(\ref{eq:Osr1}) give us equation (\ref{Os}),
when $\rho=2\sigma$, and have no meaning when $\rho < \sigma$ or $\rho >
2\sigma$.  Therefore, we define an underlying variable $x^{(3)}$ (called the
\textsl{Earth-occulted sky rate}, see Sec.~\ref{sec:earth} and \ref{sec:gls}) the
following way:

\begin{equation}
x^{(3)} = \begin{cases} 
0, & \mbox{if } \rho \leq 0; \\
\displaystyle\frac{\Omega^{\rho<\sigma}(\sigma,\rho)}{2\pi}, & \mbox{if } 0 < \rho \leq \sigma; \\
\displaystyle\frac{\Omega^{\sigma<\rho}(\sigma,\rho)}{2\pi}, & \mbox{if } \sigma < \rho < 2\sigma; \\
\displaystyle\frac{\Omega_{total}(\sigma)}{2\pi}, & \mbox{if } 2\sigma \leq \rho.
\end{cases}
\label{eq:cases}
\end{equation}

Note that we divided by $2\pi$ because we assumed that FoV of the
detector is $2\pi$ sterad. In that way, we get the rate of the Earth-limb to the
FoV. We computed expression (\ref{eq:cases}) for every second of
the lightcurve and use it as an underlying variable in Sec.~\ref{sec:earth}.